\documentclass[journal,onecolumn]{IEEEtran} 

\usepackage{mathptmx} 
\usepackage{amsmath} 
\usepackage{amssymb}
\usepackage{float}
\usepackage{xspace}
\usepackage{hhline}
\usepackage{tabularx}

\newtheorem{theorem}{Theorem}
\newtheorem{definition}{Definition} 
\newtheorem{conjecture}{Conjecture} 
\newtheorem{lemma}{Lemma}

\newtheorem{example}{Example} 
\newtheorem{fact}{Fact}
\newtheorem{prop}{Proposition}

\mathchardef\ordinarycolon\mathcode`\:  
\mathcode`\:=\string"8000
\def\vcentcolon{\mathrel{\mathop\ordinarycolon}} \begingroup
\catcode`\:=\active \lowercase{\endgroup \let :\vcentcolon }

\newcommand{\betweentable}{\vspace{0.6cm}} 

\newcommand{\ssetgen}{\ensuremath{\set{S}}} 
\newcommand{\svalgen}{\ensuremath{{s}}} 

\newcommand{\svec}{\ensuremath{{\sigma}}} 
\newcommand{\ssim}{\ensuremath{\Sigma}} 
\newcommand{\ssimuc}{\ensuremath{\Sigma^{\text{\sc uc}}}} %

\newcommand{\msetgen}{\ensuremath{\set{X}}} 
\newcommand{\mvargen}{\ensuremath{{X}}} 
\newcommand{\mvalgen}{\ensuremath{{x}}} 

\newcommand{\mvara}{\ensuremath{{X}}} 
\newcommand{\mvala}{\ensuremath{{x}}} 
\newcommand{\mvarb}{\ensuremath{{Y}}} 
\newcommand{\mvalb}{\ensuremath{{y}}} 
\newcommand{\mvec}{\ensuremath{{\pi}}} 
\newcommand{\msim}{\ensuremath{\Pi}} 
\newcommand{\msimuc}{{\ensuremath{\Pi}^{\textsc{uc}}}} 

\newcommand{\false}{\ensuremath{\mathsf F}} 
\newcommand{\true}{\ensuremath{\mathsf T}} 

\newcommand{\bayesconf}{\ensuremath{\text{\tt post-odds}}}
\newcommand{\itconf}{\ensuremath{\text{\tt bit-diff}}} 
\newcommand{\freqconf}{\ensuremath{\text{\tt p-value}}}

\newcommand{\empfreq}{\ensuremath{\mathbb P}}
\newcommand{\R}{\ensuremath{\mathbb{R}}}
\newcommand{\smfrac}[2]{\mbox{$\frac{#1}{#2}$}}
\newcommand{\ket}[1]{|#1\rangle}
\newcommand{\braket}[2]{\langle#1|#2\rangle}
\newcommand{\Exp}{\ensuremath{{\bf E}}}
\newcommand{\strength}{\ensuremath{{\mbox{\sc s}_Q^{\text{\sc uc}}}}}
\newcommand{\strengthis}{\ensuremath{{\overline{\mbox{\sc s}}_Q^{\text{\sc uc}}}}}
\newcommand{\strengthcor}{\ensuremath{{\mbox{\sc s}_Q^{\text{\sc cor}}}}}
\newcommand{\strengthuni}{\ensuremath{{\mbox{\sc s}_Q^{\text{\sc uni}}}}}
\newcommand{\priordens}{\ensuremath{{w}}}
\newcommand{\e}{\mathrm{e}} 
\renewcommand{\d}{\mathrm{d}} 
\newcommand{\xor}{\oplus}
\newcommand{\CHSH}{\textsc{CHSH}\xspace}
\newcommand{\Bell}{\textsc{Bell}\xspace}
\newcommand{\Hardy}{\textsc{Hardy}\xspace}
\newcommand{\Mermin}{\textsc{Mermin}\xspace}
\newcommand{\GHZ}{\textsc{GHZ}\xspace}
\newcommand{\partyA}{$\mathcal{A}$\xspace}
\newcommand{\partyB}{$\mathcal{B}$\xspace}

\DeclareMathAlphabet{\set}{OMS}{cmsy}{m}{n}  
\newcommand{\sups}[1]{\ensuremath{\sup_{#1}}}
\newcommand{\infm}[1]{\ensuremath{\inf_{#1}}}

\newlength{\entrywidth}
\newcolumntype{M}{>{\centering\arraybackslash $}X<{$}}
\newcolumntype{m}{>{$}c<{$}} 
\newcolumntype{v}{>{$}r<{$}} 

\newcommand{\Qtable}[4]{{
\begin{tabular}{v||m|m|}
 & a=1 & a=2 \\
Q_{ab}(X=x,Y=y) & 
\begin{tabularx}{\entrywidth}{MM} x=\true & x=\false \end{tabularx} & 
\begin{tabularx}{\entrywidth}{MM} x=\true & x=\false \end{tabularx} \\\hhline{=::==}
b=1~\begin{tabular}{m} y=\true \\y=\false\end{tabular} \hspace*{-\tabcolsep} &
 \begin{tabularx}{\entrywidth}{MM} #1 \end{tabularx} &
 \begin{tabularx}{\entrywidth}{MM} #2 \end{tabularx} \\ \hhline{-||--}
b=2~\begin{tabular}{m} y=\true \\y=\false\end{tabular}  \hspace*{-\tabcolsep} &
 \begin{tabularx}{\entrywidth}{MM} #3 \end{tabularx} &
 \begin{tabularx}{\entrywidth}{MM} #4 \end{tabularx} \\ \hhline{-||--} 
\end{tabular}}}

\newcommand{\Ptable}[4]{{
\begin{tabular}{v||m|m|}
 & a=1 & a=2 \\
P_{ab}(X=x,Y=x) & 
\begin{tabularx}{\entrywidth}{MM} x=\true & x=\false \end{tabularx} & 
\begin{tabularx}{\entrywidth}{MM} x=\true & x=\false \end{tabularx} \\\hhline{=::==}
b=1~\begin{tabular}{m} y=\true \\y=\false\end{tabular}  \hspace*{-\tabcolsep} &
 \begin{tabularx}{\entrywidth}{MM} #1 \end{tabularx} &
 \begin{tabularx}{\entrywidth}{MM} #2 \end{tabularx} \\ \hhline{-||--}
b=2~\begin{tabular}{m} y=\true \\y=\false\end{tabular}  \hspace*{-\tabcolsep} &
 \begin{tabularx}{\entrywidth}{MM} #3 \end{tabularx} &
 \begin{tabularx}{\entrywidth}{MM} #4 \end{tabularx} \\ \hhline{-||--} 
\end{tabular}}}

\newcommand{\Prtable}[4]{{
\begin{tabular}{v||m|m|}
 & a=1 & a=2 \\
\Pr_{ab}(X=x,Y=x) & 
\begin{tabularx}{\entrywidth}{MM} x=\true & x=\false \end{tabularx} & 
\begin{tabularx}{\entrywidth}{MM} x=\true & x=\false \end{tabularx} \\\hhline{=::==}
b=1~\begin{tabular}{m} y=\true \\y=\false\end{tabular}  \hspace*{-\tabcolsep} &
 \begin{tabularx}{\entrywidth}{MM} #1 \end{tabularx} &
 \begin{tabularx}{\entrywidth}{MM} #2 \end{tabularx} \\ \hhline{-||--}
b=2~\begin{tabular}{m} y=\true \\y=\false\end{tabular}  \hspace*{-\tabcolsep} &
 \begin{tabularx}{\entrywidth}{MM} #3 \end{tabularx} &
 \begin{tabularx}{\entrywidth}{MM} #4 \end{tabularx} \\ \hhline{-||--} 
\end{tabular}}}

\newlength{\merminentrywidth}
\newcommand{\MerminQtable}[9]{{
\begin{tabular}{v||m|m|m|}
 & a=1 & a=2 & a=3 \\
Q_{ab}(X=x,Y=y) & 
\begin{tabularx}{\merminentrywidth}{MM} x=\true & x=\false \end{tabularx} & 
\begin{tabularx}{\merminentrywidth}{MM} x=\true & x=\false \end{tabularx} &
\begin{tabularx}{\merminentrywidth}{MM} x=\true & x=\false \end{tabularx} \\\hhline{=::===}
b=1~\begin{tabular}{m} y=\true \\y=\false\end{tabular}  \hspace*{-\tabcolsep} &
 \begin{tabularx}{\merminentrywidth}{MM} #1 \end{tabularx} &
 \begin{tabularx}{\merminentrywidth}{MM} #2 \end{tabularx} & 
 \begin{tabularx}{\merminentrywidth}{MM} #3 \end{tabularx} \\ \hhline{-||---}
b=2~\begin{tabular}{m} y=\true \\y=\false\end{tabular}  \hspace*{-\tabcolsep} &
 \begin{tabularx}{\merminentrywidth}{MM} #4 \end{tabularx} &
 \begin{tabularx}{\merminentrywidth}{MM} #5 \end{tabularx} & 
 \begin{tabularx}{\merminentrywidth}{MM} #6 \end{tabularx} \\ \hhline{-||---}
b=3~\begin{tabular}{m} y=\true \\y=\false\end{tabular}  \hspace*{-\tabcolsep} &
 \begin{tabularx}{\merminentrywidth}{MM} #7 \end{tabularx} &
 \begin{tabularx}{\merminentrywidth}{MM} #8 \end{tabularx} & 
 \begin{tabularx}{\merminentrywidth}{MM} #9 \end{tabularx} \\ \hhline{-||---}
\end{tabular}}}

\newcommand{\MerminPtable}[9]{{
\begin{tabular}{v||m|m|m|}
 & a=1 & a=2 & a=3 \\
P_{ab}(X=x,Y=y) & 
\begin{tabularx}{\merminentrywidth}{MM} x=\true & x=\false \end{tabularx} & 
\begin{tabularx}{\merminentrywidth}{MM} x=\true & x=\false \end{tabularx} &
\begin{tabularx}{\merminentrywidth}{MM} x=\true & x=\false \end{tabularx} \\\hhline{=::===}
b=1~\begin{tabular}{m} y=\true \\y=\false\end{tabular}  \hspace*{-\tabcolsep} &
 \begin{tabularx}{\merminentrywidth}{MM} #1 \end{tabularx} &
 \begin{tabularx}{\merminentrywidth}{MM} #2 \end{tabularx} & 
 \begin{tabularx}{\merminentrywidth}{MM} #3 \end{tabularx} \\ \hhline{-||---}
b=2~\begin{tabular}{m} y=\true \\y=\false\end{tabular}  \hspace*{-\tabcolsep} &
 \begin{tabularx}{\merminentrywidth}{MM} #4 \end{tabularx} &
 \begin{tabularx}{\merminentrywidth}{MM} #5 \end{tabularx} & 
 \begin{tabularx}{\merminentrywidth}{MM} #6 \end{tabularx} \\ \hhline{-||---}
b=3~\begin{tabular}{m} y=\true \\y=\false\end{tabular}  \hspace*{-\tabcolsep} &
 \begin{tabularx}{\merminentrywidth}{MM} #7 \end{tabularx} &
 \begin{tabularx}{\merminentrywidth}{MM} #8 \end{tabularx} & 
 \begin{tabularx}{\merminentrywidth}{MM} #9 \end{tabularx} \\ \hhline{-||---}
\end{tabular}}}

\newcommand{\ssimtable}[4]{{
\begin{tabular}{v||m|m|}
\Pr(A=a,B=b)=\svec_{ab}\in\ssim & a=1 & a=2 \\ \hhline{=::==}
b=1 & #1 & #2 \\ \hhline{-||--}
b=2 & #3 & #4 \\ \hhline{-||--}
\end{tabular}}}

\newcommand{\ssimuctable}[8]{{
\begin{tabular}{v||m|m||m}
\Pr(A=a,B=b)=\svec_{ab}\in\ssimuc & a=1 & a=2 & \Pr(B=b)\\ \hhline{=::==::=}
b=1 & #1 & #2 & #3 \\ \hhline{-||--||-}
b=2 & #4 & #5 & #6 \\ \hhline{=::==::=} 
\Pr(A=a) & #7 & #8 \\
\end{tabular}}}

\begin{document}
\title{{The statistical strength of nonlocality proofs}} 
\author{Wim van Dam,~\thanks{%
Wim van Dam is with the Massachusetts Institute of Technology, 
Center for Theoretical Physics, 
77 Massachusetts Avenue,
Cambridge, MA 02139-4307,
United States of America. 
Email: \texttt{vandam@mit.edu}}
Richard~D.~Gill,~\thanks{%
Richard Gill is with the Mathematical Institute, 
University Utrecht,  
Budapestlaan 6, 
NL-3584 CD Utrecht, The Netherlands.
He is also at \textsc{Eurandom}, Eindhoven, The Netherlands. 
Email: \texttt{gill@math.uu.nl}}
Peter~D.~Gr\"unwald,~\thanks{%
Peter Gr\"unwald is with the CWI, P.O. Box 94079, NL-1090 GB, The Netherlands.
He is also at \textsc{Eurandom}, Eindhoven, The Netherlands. 
Email: \texttt{pdg@cwi.nl}}} 
\pubid{arXiv:quant-ph/0307125}

\maketitle 
\begin{abstract}
There exist numerous proofs of Bell's theorem, stating that quantum
mechanics is incompatible with local realistic theories of nature.
Here we define the strength of such nonlocality proofs in terms of the
amount of evidence against local realism provided by the corresponding
experiments.  This measure tells us how many trials of the experiment
we should perform in order to observe a substantial violation of local
realism.  Statistical considerations show that the amount of evidence
should be measured by the \emph{Kullback-Leibler} or \emph{relative
entropy} divergence between the probability distributions over the
measurement outcomes that the respective theories predict.  The
statistical strength of a nonlocality proof is thus determined by the
experimental implementation of it that maximizes the Kullback-Leibler
divergence from experimental (quantum mechanical) truth to the set of
all possible local theories.  An implementation includes a specification 
with which probabilities the different measurement settings are sampled,
and hence the maximization is done over all such setting distributions.

We analyze two versions of Bell's nonlocality proof (his original
proof and an optimized version by Peres), and proofs by
Clauser-Horne-Shimony-Holt, Hardy, Mermin, and
Greenberger-Horne-Zeilinger.  We find that the GHZ proof is at least
four and a half times stronger than all other proofs, while of the
two-party proofs, the one of CHSH is the strongest.
\end{abstract}
\begin{keywords} 
nonlocality, Bell's theorem, quantum correlations, Kullback-Leibler divergence
\end{keywords}

\tableofcontents


\newpage
\section{Introduction}\label{sec:intro}
\PARstart{A}{plethora} of proofs exist of Bell's theorem 
(``quantum mechanics
violates local realism'') encapsulated in inequalities and equalities
of which the most celebrated are those of Bell \cite{Bell64}, 
Clauser, Horne, Shimony and Holt (\CHSH) \cite{ClauserHSH69}, 
Greenberger, Horne and Zeilinger (\GHZ) \cite{GreenbergerHZ89}, 
Hardy \cite{Hardy93}, and Mermin \cite{Mermin81}.  
Competing claims exist that one proof is stronger than another. For 
instance, a proof in which quantum predictions having probabilities 
$0$ or $1$ only are involved, is often said to be more strong than 
a proof that involves quantum predictions of probabilities between 
$0$ and $1$. Other researchers argue that one should compare the
absolute differences between the probabilities that quantum 
mechanics predicts and those that are allowed by local theories.
And so on. 
The aim of this paper is to settle such questions once
and for all: we formally define the strength of a nonlocality proof
and show that our definition is the only one compatible with generally
accepted notions in information theory and theoretical statistics.

To see the connection with statistics, note first that a 
\emph{mathematical} nonlocality proof shows that the \emph{predicted}
probabilities of quantum theory are incompatible with local realism.
Such a proof can be implemented as an \emph{experimental proof}
showing that \emph{physical reality} conforms to those predictions and
hence too is incompatible with local realism. We are interested in the
strength of such experimental proofs, which should be measured in
statistical terms: how sure do we become that a certain theory is
false, after observing a certain violation from that theory, in a
certain number of experiments.

\subsection{Our Game} 
We analyze the statistics of nonlocality proofs in terms
of a two-player game. The two players are the pro-quantum theory
experimenter QM, and a pro-local realism theoretician LR. The
experimenter QM is armed with a specific proof of Bell's theorem.  A
given proof---\Bell, \CHSH, \Hardy, \Mermin, \GHZ---involves a collection
of equalities and inequalities between various experimentally
accessible probabilities. The proof specifies a given quantum state
(of a collection of entangled qubits, for instance) and experimental
settings (orientations of polarization filters or Stern-Gerlach
devices). 
All local realistic theories of LR will obey the (in)equalities,
while the observations that QM will make when performing
the experiment (assuming that quantum mechanics is true) will 
violate these (in)equalities.
The experimenter QM still has a choice of the probabilities 
with which the different combinations of settings will be
applied, in a long sequence of independent trials. In other words, he
must still decide how to allocate his resources over the different
combinations of settings.  At the same time, the local realist can
come up with all kinds of different local realistic theories,
predicting different probabilities for the outcomes given the
settings.  She might put forward different theories in response to
different specific experiments. Thus the quantum experimenter will
choose that probability distribution over his settings, for which the
\emph{best} local realistic model explains the data \emph{worst}, when
compared with the true (quantum mechanical) description.

\subsection{Quantifying Statistical Strength - Past Approaches}
How should we measure the statistical strength of a given experimental
setup?  In the past it was often simply said that
the largest deviation in the Bell
inequality is attained with such and such filter settings, and hence
the experiment which is done with these settings gives (potentially) 
the strongest proof of nonlocality.  The argument is however not
very convincing. One should take account of the statistical
variability in finite samples. The experiment that might confirm the largest
absolute deviation from local realistic theories, might be subject
to the largest standard errors, and therefore be less convincing
than an experiment where a much smaller deviation can be
proportionally much more accurately determined.

Alternatively, the argument has just been that with a large enough
sample size, even the smallest deviation between two theories can be
made firm enough.  For instance, \cite{Mermin81} has said in the
context of a particular example 
\begin{quote}\emph{``\dots to produce the conundrum it is
necessary to run the experiment sufficiently many times to establish
with overwhelming probability that the observed frequencies (which
will be close to 25\% and 75\%) are not chance fluctuations away
from expected frequencies of 33\% and 66\%.  (A million runs is
more than enough for this purpose)\dots''}
\end{quote}  
We want to replace the words
``sufficiently", ``overwhelming'', ``more than enough'' with something
more scientific.  
(See Example~\ref{ex:mermin} for our conclusion with respect to this.)
And as experiments are carried out that are harder
and harder to prepare, it becomes important to design them so that
they give conclusive results with the smallest possible sample sizes.
Initial work in this direction has been done by Peres \cite{Peres00}. 
Our approach is compatible with his, and extends it in a number of 
directions---see Section~\ref{sec:peres}. 

\subsection{Quantifying Statistical Strength - Our Approach}
We measure statistical strength using an information-theoretic
quantification, namely the Kullback-Leibler (KL) divergence (also
known as \emph{information deficiency} or \emph{relative entropy}
\cite{CoverT91}). We show (Appendix~\ref{app:formal}) that for large
samples, all reasonable definitions of statistical strength that can
be found in the statistical and information-theoretic literature
essentially coincide with our measure. For a given type of experiment,
we consider the game in which the experimenter wants to maximize the
divergence while the local theorist looks for theories that minimize
it.  The experimenter's game space is the collection of probability
distributions over joint settings, which we call in the sequel, for
short, ``setting distributions''.  (More properly, these are ``joint
setting distributions''.)  The local realist's game space is the space
of local realistic theories.  This game defines an experiment, such
that each trial (assuming quantum mechanics is true) provides on
average, the maximal support for quantum theory against the best
explanation that local realism can provide, at that setting
distribution.

\subsection{Our Results - Numerical}
We determined the statistical strength of five two-party proofs:
Bell's original proof and Peres' optimized variant of it, and the
proofs of \CHSH, Hardy, and Mermin.  Among these, \CHSH turns out to
be the strongest by far.  We also determined the strength of the
three-party \GHZ proof.  Contrary to what has sometimes been claimed
(see Section~\ref{sec:discussion}), even the \GHZ experiment has to be
repeated a fair number of times before a substantial violation of
local realism is likely to be observed. Nevertheless, it is about
$4.5$ times stronger than the \CHSH experiment, meaning that, in order
to observe the same support for QM and against LR, the \CHSH
experiment has to be run about $4.5$ times as often as the \GHZ
experiment.
\subsection{Our Results - Mathematical}
To find the (joint) setting distribution that optimizes the strength
of a nonlocality proof is a highly nontrivial computation. In the
second part of this paper, we prove several mathematical properties of
our notion of statistical strength. These provide insights in the
relation between local realist and quantum distributions that are
interesting in their own right. They also imply that determining
statistical strength of a given nonlocality proof may be viewed as a
convex optimization problem which can be solved numerically. We also
provide a game-theoretic analysis involving minimax and maximin KL
divergences. This analysis allows us to shortcut the computations in
some important special cases.

\subsection{Organization of This Paper}
Section~\ref{sec:formal} gives a formal definition of what we mean by
a nonlocality proof and the corresponding experiment, as well as 
the notation that we will use throughout the article. 
The kinds of nonlocality proofs that this article
analyzes are described in Section~\ref{sec:nonlocalityproofs}, using
the \CHSH proof as a specific example; the other proofs are described
in more detail in Appendices~\ref{app:nonlocalityarguments}
and~\ref{app:nonlocalityproofs}. The definition of the `statistical
strength of a nonlocality proof' is presented in Section~\ref{sec:kl},
along with some standard facts about the Kullback-Leibler divergence 
and its role in hypothesis testing.  With this definition, we are 
able to calculate the strengths of various nonlocality proofs. 
The results of these calculations for six well-known proofs 
are listed in Section~\ref{sec:results} 
(with additional details again in Appendix~\ref{app:nonlocalityproofs}).  
The results are interpreted,
discussed and compared in Section~\ref{sec:discussion}, which also
contains four conjectures.
Section~\ref{sec:computing} constitutes the second, more mathematical
part of the paper. It presents the mathematical results that allow us
to compute statistical strength efficiently.

We defer all issues that require knowledge of the mathematical
aspects of quantum mechanics to the appendices.  There we provide more
detailed information about the nonlocality proofs we analyzed, the
relation of Kullback-Leibler divergence to hypothesis testing, 
and the proofs of the theorems we present in the main text.

Depending on their interests, readers might want to skip 
certain sections of this, admittedly, long paper.  
Only the first six sections are crucial; all other parts 
provide background information of some sort or the other.


\section{Formal Setup} \label{sec:formal}
A basic nonlocality proof (``quantum mechanics violates local
realism'') has the following ingredients.  There are two parties
\partyA and \partyB, who can each dispose over one of two
entangled qubits.  They may each choose out of two different
measurement settings.  In each trial of the experiment, 
\partyA and \partyB randomly sample from the four different 
joint settings and each observe one of two different binary
outcomes, say ``$\false$'' (false) and ``$\true$'' (true).  Quantum
mechanics enables us to compute the joint probability distribution of
the outcomes, as a function of the measurement settings and of the
joint state of the two qubits.  Thus possible design choices are: the
state of the qubits, the values of the settings; and the probability
distribution over the settings. More complicated experiments may
involve more parties, more settings, and more outcomes. Such a
generalized setting is formalized in Appendix~\ref{app:beyond2x2x2}.
In the main text, we focus on the basic $2\times 2\times 2$ case
(which stands for `$2$ parties $\times$ $2$ measurement settings per
party $\times$ $2$ outcomes per measurement setting').
Below we introduce notation for all ingredients involved in
nonlocality proofs.

\subsection{Distribution of Settings} \label{sec:distributionsettings}
The random variable $A$ denotes the measurement setting of party 
\partyA and the random variable $B$ denotes the measurement setting of party
\partyB.  Both $A$ and $B$ take values in $\{1,2\}$. The experimenter QM
will decide on the distribution $\svec$ of $(A,B)$, giving the
probabilities (and, after many trials of the experiment, the
frequencies) with which each (joint) measurement setting is
sampled. This \emph{setting distribution} $\svec$ is identified with 
its probability vector $\svec := (\svec_{11},\svec_{12},\svec_{21},\svec_{22}) \in
\ssim$, and $\ssim$ is the unit simplex in $\R^4$ defined by
\begin{IEEEeqnarray}{rCl}
\ssim &:=& 
\big\{ (\svec_{11},\svec_{12},\svec_{21},\svec_{22}) \mid
\sum_{a,b \in \{1,2\}} \svec_{ab} = 1, \text{for all $a,b:\ $}
\svec_{ab} \geq 0 \big\}.
\end{IEEEeqnarray} 
We use $\ssimuc$ to denote the set of vectors representing 
\emph{uncorrelated} distributions in $\ssim$. 
Formally, $\svec \in \ssimuc$ if and only if 
$\svec_{ab} = (\svec_{a1}+\svec_{a2})(\svec_{1b}+\svec_{2b})$
for all $a,b \in \{1,2\}$.

\subsection{Measurement Outcomes}
The random variable $\mvara$ denotes the measurement outcome of party 
\partyA and the random variable $\mvarb$ denotes that
of party \partyB. Both $\mvara$ and $\mvarb$ take values in $\{\true,\false\}$;
$\false$ standing for `false' and $\true$ standing for `true'. Thus, the
statement `$\mvara = \false, \mvarb = \true$' and describes the event that
party \partyA observed $\false$ and party \partyB observed $\true$.

The distribution of $(\mvara,\mvarb)$ depends on the chosen setting
$(a,b) \in \{1,2\}^2$.  The state of the entangled qubits together
with the measurement settings determines four conditional
distributions $Q_{11}, Q_{12}, Q_{21}, Q_{22}$ for $(\mvara,\mvarb)$,
one for each joint measurement setting, where $Q_{ab}$ is the distribution
of $(\mvara,\mvarb)$ given that measurement setting $(a,b)$ has been
chosen. For example, $Q_{ab}(\mvara = \false, \mvarb = \true)$,
abbreviated to $Q_{ab}(\false,\true)$, denotes the probability that
party \partyA observes $\false$ and party \partyB observes
$\true$, given that the device of \partyA is in setting $a$ and
the device of \partyB is in setting $b$.
According to QM, the total outcome $(\mvara,\mvarb,A,B)$ 
of a single trial is then distributed as $Q_{\svec}$, defined by
$  Q_{\svec}(\mvara=\mvala, \mvarb = \mvalb,A=a,B=b) := 
 \svec_{ab} Q_{ab}(\mvara=\mvala, \mvarb= \mvalb)$.

\subsection{Definition of a Nonlocality Proof and Corresponding Nonlocality 
Experiments}
\label{sec:defnon}
A \emph{nonlocality proof} for $2$ parties, $2$ measurement settings
per party, and $2$ outcomes per measurement, is identified with an
entangled quantum state of two qubits (realized, by, e.g., two
photons) and two measurement devices (e.g., polarization filters)
which each can be used in one of two different measurement settings
(polarization angles).  Everything about the quantum state, the
measurement devices, and their settings that is relevant for the
probability distribution of outcomes of the corresponding experiment
can be summarized by the four distributions $Q_{ab}$ of
$(\mvara,\mvarb)$, one for each (joint) setting $(a, b) \in
\{1,2\}^2$.  Henceforth, we will simply \emph{identify} a $2\times
2\times 2$ nonlocality proof with the vector of distributions $Q :=
(Q_{11},Q_{12},Q_{21},Q_{22})$.

This definition can easily be extended in an entirely straightforward 
manner to a situation with more than two parties, 
two settings per party, or two outcomes per setting. 
In Appendix~\ref{app:beyond2x2x2} we provide a 
formal definition of the general case where the numbers of parties,
settings, and outcomes are arbitrary.

We call a nonlocality proof $Q = (Q_{11},Q_{12},Q_{21},Q_{22})$
\emph{proper} if and only if it violates local realism, i.e.\ if there
exists no local realist distribution $\mvec$ (as defined below) 
such that $P_{ab;\mvec}(\cdot ) 
= Q_{ab}(\cdot)$ for all $(a,b) \in \{1,2\}^2$.

For the corresponding $2\times 2\times 2$ nonlocality 
\emph{experiment} we have to specify the setting distribution 
$\svec$ with which the experimenter QM samples the different 
settings $(a,b)$.  Thus, for a single nonlocality proof $Q$, QM 
can use different experiments (different in $\svec$) to verify 
Nature's nonlocality.  Each experiment consists of a series of trials, 
where---per trial---the event 
$(\mvara,\mvarb,A,B)$ occurs with probability  
$Q_{\svec}(\mvara=\mvala, \mvarb = \mvalb,A=a,B=b) = 
 \svec_{ab} Q_{ab}(\mvara=\mvala, \mvarb= \mvalb)$.

\subsection{Local Realist Theories}
The local realist (LR) may provide any `local' theory she likes to
explain the results of the experiments. 

Under such a theory it is possible to talk about ``the outcome that
\partyA would have observed, if she had used setting $1$'',
independently of which setting was used by \partyB and indeed of
whether or not \partyA actually did use setting $1$ or $2$. Thus
we have four binary random variables, which we will call $\mvara_1$,
$\mvara_2$, $\mvarb_1$ and $\mvarb_2$. As before, variables named $X$
correspond to \partyA's observations, and variables named $Y$
correspond to \partyB's observations.  According to LR, each
experiment determines values for the four random variables
$(\mvara_1,\mvara_2,\mvarb_1,\mvarb_2)$.  For $a \in \{1,2\}$,
$\mvara_a \in \{\true,\false\}$ denotes the outcome that party
\partyA would have observed if the measurement setting of
\partyA had been $a$. Similarly, for $b \in \{1,2\}$, $\mvarb_b
\in \{ \true,\false\}$ denotes the outcome that party \partyB
would have observed if the measurement setting of \partyB had
been $b$.

A local theory $\mvec$ may be viewed as a probability distribution for
$(\mvara_1,\mvara_2,\mvarb_1,\mvarb_2)$. Formally, we define $\mvec$
as a $16$-dimensional probability vector with indices 
$(\mvala_1,\mvala_2,\mvalb_1,\mvalb_2) \in \{\true,\false\}^4$. 
By definition,
$P_{\mvec}(\mvara_1 = \mvala_1, \mvara_2 = \mvala_2, 
\mvarb_1 = \mvalb_1, \mvarb_2 = \mvalb_2) := 
\mvec_{\mvala_1 \mvala_2 \mvalb_1 \mvalb_2}$.
For example, $\mvec_{\false \false \false \false}$ denotes LR's probability
that, in all possible measurement settings, \partyA and \partyB would both
have observed $\false$.  The set of local theories can thus be
identified with the unit simplex in $\R^{16}$, which we will
denote by $\msim$.

Recall that the quantum state of the entangled qubits determines four
distributions over measurement outcomes $Q_{ab}(\mvara = \cdot, \mvarb
= \cdot)$, one for each joint setting $(a,b) \in \{1,2\}^2$.
Similarly, each LR theory $\mvec \in \msim$ determines four
distributions $P_{ab ; \mvec}(\mvara = \cdot, \mvarb = \cdot )$. These
are the  distributions, according to the local realist theory
$\mvec$, of the random variables $(\mvara,\mvarb)$ given that setting
$(a,b)$ has been chosen.
Thus, the value $P_{ab ; \mvec}(\mvara = \cdot, \mvarb = \cdot )$ 
is defined as the sum of four terms:
\begin{IEEEeqnarray}{rCl} \label{eq:condlr}
P_{ab ; \mvec}(\mvara = \mvala, \mvarb = \mvalb )
& := &
\sum_{\substack{\mvala_1,\mvala_2,\mvalb_1,\mvalb_2 \in \{\true,\false\}\\
\mvala_a = \mvala; \mvalb_b = \mvalb}}{\mvec_{\mvala_1 \mvala_2 \mvalb_1\mvalb_2}}.
\end{IEEEeqnarray}
We suppose that LR does not dispute the actual setting distribution
$\sigma$ which is used in the experiment, she only disputes the
probability distributions of the measurement outcomes given the
settings.
According to LR therefore, the outcome of a single trial is
distributed as $P_{\svec; \mvec}$ defined by
$ P_{\svec; \mvec}(\mvara = \mvala, \mvarb = \mvalb,A=a,B=b) :=
 \svec_{ab} P_{ab; \mvec}(\mvara = \mvala, \mvarb = \mvalb)$.


\section{The Nonlocality Proofs}
\label{sec:nonlocalityproofs}
In this section we briefly describe the five (or six, since we have two
versions of Bell's proof) celebrated nonlocality
proofs for which we will compute the statistical strength. 
In Appendix~\ref{app:nonlocalityproofs}, we provide further details about
the entangled quantum states that give rise to the violations of the
various inequalities. 

Let us interpret the measurement outcomes $\false$ and $\true$ in
terms of Boolean logic, i.e.\ $\false$ is ``false'' and $\true$ is
``true''. We can then use Boolean expressions such as $\mvara_2 \&
\mvarb_2$, which evaluates to true whenever both $\mvara_2$ and
$\mvarb_2$ evaluate to `true', i.e.\ when both $\mvara_2 = \true$ and
$\mvarb_2 =\true$.  We derive the proofs by applying the rule that if
the event $X=\true$ implies the event $Y=\true$ (in short ``$X\implies
Y$''), then $\Pr(X)\leq\Pr(Y)$.  In similar vein, we will use rules
like $\Pr(X\vee Y)\leq \Pr(X)+\Pr(Y)$ and
$1-\Pr(\neg X)-\Pr(\neg Y)\leq 1-\Pr(\neg X \vee \neg Y)=\Pr(X\&
Y)$.

As an aside we want to mention that the proofs of Bell, CHSH and Hardy
all contain the following argument, which can be traced back to the
nineteenth century logician George Boole (1815--1864) \cite{Boole54}.
Consider four events such that $\neg B \cap \neg C \cap \neg D
\implies \neg A$. Then it follows that $A\implies B \cup C \cup
D$. And from this, it follows that $\Pr(A)\leq
\Pr(B)+\Pr(C)+\Pr(D)$. In the \CHSH argument and the Bell argument,
the events concern the equality or inequality of one of the $\mvara_i$
with one of the $\mvarb_j$.  In the Hardy argument, the events concern
the joint equality or inequality of one of the $\mvara_i$, one of the
$\mvarb_j$, and a specific value $\false$ or $\true$.

\begin{example}[The CHSH Argument] \label{ex:CHSH}
For the \CHSH argument one notes that the implication 
\begin{IEEEeqnarray}{rCl}
[(\mvara_1 = \mvarb_1)  \&   (\mvara_1 = \mvarb_2)  \&  (\mvara_2 = \mvarb_1)] 
& \implies & 
(\mvara_2 = \mvarb_2)
\end{IEEEeqnarray}
is logically true, and hence $(\mvara_2 \ne \mvarb_2)
\implies [(\mvara_1 \ne \mvarb_1) \vee (\mvara_1 \ne \mvarb_2) \vee 
(\mvara_2 \ne \mvarb_1)]$ holds.
As a result, local realism implies the following ``\CHSH inequality''
\begin{IEEEeqnarray}{rCl}\label{e:chsh}
\Pr(\mvara_2 \ne \mvarb_2) 
&\leq & 
\Pr(\mvara_1 \ne \mvarb_1)+ \Pr(\mvara_1 \ne \mvarb_2) + \Pr(\mvara_2 \ne \mvarb_1),
\end{IEEEeqnarray}
which can be violated by many choices of settings and states under
quantum theory. 
As a specific example, CHSH identified quantum states and settings
such that  the first probability equals (approximately)  $0.85$
while the three probabilities on the right are each (approximately)
$0.15$, thus clearly violating (\ref{e:chsh}).  In full detail,  
the probability distribution that corresponds to \CHSH's 
proof is as follows
\begin{IEEEeqnarray}{c}\label{eq:chshtable}
\begin{tabular}{r||cc|cc|}
$\Pr$ & $\mvara_1 = \true$ & $\mvara_1 = \false$ & $\mvara_2 = \true$ & $\mvara_2 = \false$
 \\\hhline{=::====} 
$\mvarb_1=\true$  & $0.4267766953$ & $0.0732233047$ & $0.4267766953$ & $0.0732233047$ \\ 
$\mvarb_1=\false$ & $0.0732233047$ & $0.4267766953$ & $0.0732233047$ & $0.4267766953$ \\\hline
$\mvarb_2=\true$  & $0.4267766953$ & $0.0732233047$ & $0.0732233047$ & $0.4267766953$ \\ 
$\mvarb_2=\false$ & $0.0732233047$ & $0.4267766953$ & $0.4267766953$ & $0.0732233047$ \\
\hhline{-||----}
\end{tabular}
\end{IEEEeqnarray}

In Appendix~\ref{app:CHSH} we explain how to arrive at this table.
The table lists the $4$ conditional distributions $Q = (Q_{11},Q_{12},Q_{21},Q_{22})$ 
defined in Section~\ref{sec:defnon}, and thus uniquely determines the nonlocality proof $Q$. 
As an example of how to read the table, note that $\Pr(\mvara_2 
\neq \mvarb_2)$ is given by 
\begin{IEEEeqnarray*}{rCcCcCl}
\Pr(\mvara_2 \neq \mvarb_2) 
&=& 
\Pr(\mvara_2=\true  \&  \mvarb_2=\false) + \Pr(\mvara_2=\false \& \mvarb_2=\true) 
&=& 
0.4267766953 + 0.4267766953 
&\approx &
0.85,
\end{IEEEeqnarray*}
showing that the expression on the left in (\ref{e:chsh}) is approximately 
$0.85$. That on the right evaluates to approximately $0.45$.
\end{example}
The other nonlocality proofs are derived in a similar manner: one
shows that according to any and all local realist theories, the random
variables $\mvara_1,\mvara_2,\mvarb_1,\mvarb_2$ must satisfy certain
logical constraints and hence probabilistic (in)equalities. One then
shows that these constraints or (in)equalities can be violated by
certain quantum mechanical states and settings, giving rise to a table
of probabilities of observations similar to
(\ref{eq:chshtable}). Details are given in
Appendix~\ref{app:nonlocalityarguments}.


\section{Kullback-Leibler Divergence and Statistical Strength}\label{sec:kl}
\subsection{Kullback-Leibler Divergence}
In this section we formally define our notion of `statistical strength
of a nonlocality proof'. The notion will be based on the KL
divergence, an information theoretic quantity which we now
introduce. Let $\set{Z}$ be an arbitrary finite set.  For a
distribution $Q$ over $\set{Z}$, $Q(z)$ denotes the probability of
event $\{ z \}$.  For two (arbitrary) distributions $Q$ and $P$
defined over $\set{Z}$, the Kullback-Leibler (KL) divergence from
$Q$ to $P$ is defined as
\begin{IEEEeqnarray}{rCl}\label{eq:kl}
D(Q\|P) 
&:=& 
\sum_{z \in \set{Z}}{Q(z) \log \frac{ Q(z)}{P(z)}}
\end{IEEEeqnarray}
where the logarithm is taken here, as in the rest of the paper, to
base $2$.  We use the conventions that, for $y > 0$, $y \log 0 :=
\infty$, and $0 \log 0 := \lim_{y \downarrow 0} y \log y = 0$.

The KL divergence is also known as relative entropy, cross-entropy,
information deficiency or $I$-divergence. Introduced in
\cite{KullbackL51}, KL divergence has become a central notion in
information theory, statistics and large deviation theory. A good
reference is \cite{CoverT91}.  It is straightforward to show (using
concavity of the logarithm and Jensen's inequality) that $D(Q \| P)
\geq 0$ with equality if and only if $P=Q$; in this sense, KL
divergence behaves like a distance. However, in general $D(P\|Q) \neq
D(Q\|P)$, so formally $D(\cdot\| \cdot)$ is not a distance.
(See the examples in Appendix~\ref{app:klvariation} for a 
clarification of this asymmetry.)

KL divergence expresses the average disbelief in $P$, when observing
random outcomes $Z$ from $Q$. Thus occasionally (with respect to $Q$)
one observes an outcome $Z$ that is more likely under $P$ than
$Q$, but on average (with respect to $Q$), the outcomes are more
likely under $Q$ than $P$, expressed by the fact that $D(Q\|P)
\geq 0$.  In the Appendices~\ref{app:klproperties} 
and \ref{app:klvariation} we provide several
properties and examples of the KL divergence.

KL divergence has several different interpretations and applications.
Below we focus on the interpretation we are concerned with in this
paper: KL divergence as a measure of `statistical closeness' in the
context of statistical hypothesis testing.

\subsubsection{KL Divergence and Statistical Strength in Simple Hypothesis
  Testing} Let $Z_1, Z_2,\dots$ be a sequence of random variables
independently generated either by some distribution $P$ or by some
distribution $Q$ with $Q \neq P$. Suppose we are given a sample
(sequence of outcomes) $z_1,\dots, z_n$. We want to perform
a statistical test in order to find out whether the sample is from $P$
or $Q$. Suppose that the sample is, in fact, generated by $Q$ (`$Q$ is
true'). Then, given enough data, the data will with very high ($Q$-)
probability be overwhelmingly more likely according to $Q$ than
according to $P$. That is, the data strongly suggest that they were
sampled from $Q$ rather than $P$. The `statistical distance' between
$P$ and $Q$ indicates \emph{how strongly} or, equivalently, \emph{how
convincingly} data that are generated by $Q$ will suggest that they
are from $Q$ rather than $P$. It turns out that this notion of
`statistical distance' between two distributions is precisely captured
by the KL divergence $D(Q\|P)$, which can be interpreted
as \emph{ the average amount of support in favor of $Q$ and against
$P$ \emph{per trial}}. The larger the KL divergence, the larger the
amount of support per trial. It turns out that
\begin{enumerate}
\item For a fixed sample size $n$, the larger $D(Q\|P)$, the more
  support there will be in the sample $z_1,\dots, z_n$ for $Q$ versus
  $P$ (with high probability under $Q$) .
\item For a pre-determined fixed level of support in favor of $Q$
  against $P$ (equivalently, level of `confidence' in $Q$, level of
  `convincingness' of $Q$), we have that the larger $D(Q\|P)$, the
  smaller the sample size before this level of support is achieved
  (with high probability under $Q$).
\item If, based on observed data $z_1,\dots, z_n$, an experimenter
  decides that $Q$ rather than $P$ must have generated the data, then,
  the larger $D(Q\|P)$, the larger the \emph{confidence} the
  experimenter should have in this decision (with high probability
  under $Q$).
\end{enumerate}
What exactly do we mean by `level of support/convincingness'? Different approaches to
statistical inference define this notion in a different
manner. Nevertheless, \emph{for large samples}, all definitions of support one finds in the
literature become equivalent, and are determined by the KL
divergence up to lower order terms in the exponent.

For example, in the Bayesian approach to statistical inference, the
statistician assigns initial, so-called \emph{prior} probabilities to
the hypotheses `$Q$ generated the data' and `$P$ generated the data',
reflecting the fact that he does not know which of the two
distributions in fact did generate the data. For example, he may
assign probability $1/2$ to each hypothesis. Then given data $Z_1, Z_2,
\ldots, Z_n$, he can compute the \emph{posterior} probabilities of
the two hypotheses, conditioned on this data. It turns out that,
if $Q$ actually generated the data, and the prior probabilities on
both $P$ and $Q$ are nonzero, then 
the posterior odds that $P$ rather than $Q$
generated the data typically behaves as $2^{- n D(Q \| P) + o(n)}$.
Thus, the larger  
the KL divergence $D(Q \| P)$, the larger the odds in favour of $Q$
and therefore, 
the larger the confidence in
the decision `$Q$ generated the data'. Different approaches to hypothesis testing provide
somewhat different measures of `level of support' such as \emph{p-values} 
and \emph{code length difference}. But, as we show
in Appendix~\ref{app:kl}, these different measures of support can also be related
to the KL divergence. In the appendix we also give a
more intuitive and informal
explanation of how KL divergence is related to these notions, and
we explain why, contrary to what has sometimes been implicitly
assumed, \emph{absolute deviations} between probabilities can be quite
\emph{bad} indicators of statistical distance.

\subsubsection{KL Divergence and Statistical Strength in Composite 
Hypothesis Testing} 
Observing a sample generated by $Q$ or $P$ and trying to
infer whether it was generated by $Q$ or $P$ is called
\emph{hypothesis testing} in the statistical literature. A hypothesis
is \emph{simple} if it consists of a single probability
distribution. A hypothesis is called \emph{composite} if it consists
of a \emph{set} of distributions. The composite hypothesis 
`$\set{P}$' should be interpreted as `there exists a $P \in \set{P}$ that
generated the data'. Above, we related the KL divergence to
statistical strength when testing two simple hypotheses against each
other. Yet in most practical applications (and in this paper) the aim
is to test two hypotheses, at least one of which is composite. For
concreteness, suppose we want to test the distribution $Q$ against the
set of distributions $\set{P}$.
In this case, under some regularity conditions on $\set{P}$ and
$\set{Z}$, the element $P \in \set{P}$ that is \emph{closest} in
statistical divergence to $Q$ determines the statistical strength of
the test of $Q$ against $\set{P}$. Formally, for a set of
distributions $\set{P}$ on $\set{Z}$ we define (as is customary,
\cite{CoverT91})
\begin{IEEEeqnarray}{rCl}
D(Q \| \set{P})
& :=& 
\inf_{P \in \set{P}}{D(Q \| P)}.
\end{IEEEeqnarray}
Analogously to $D(Q \| P)$, $D(Q \| \set{P})$ may be interpreted as
the \emph{average amount of support in favor of $Q$ and against 
$\set{P}$ per trial}, if data are generated according to $Q$.

In our case, QM claims that data are generated by some distribution
$Q_{\svec}$. LR claims that data are generated by some $P \in 
\set{P}_{\svec}$, where $\set{P}_{\svec} := \{ P_{\svec; \mvec} \; : \;
\mvec \in \msim \}$.  Here $Q_{\svec}$ corresponds to a nonlocality
proof equipped with setting distribution $\svec$, 
and $\set{P}_{\svec}$ is  the set of probability distributions
of all possible local theories with the same $\sigma$ --- 
see  Section~\ref{sec:formal}.
QM and LR agree to test the hypothesis $Q_{\svec}$ against 
$\set{P}_{\svec}$. QM, who knows that data are really generated according
to $Q_{\svec}$, wants to select $\svec$ in such a way that the
average amount of support in favor of $Q$ and against $\set{P}$ is
maximized. We shall argue in Section~\ref{sec:discussion} 
that QM should restrict himself to uncorrelated settings. 
The previous discussion then suggests that he should 
pick the $\svec \in \ssimuc$ that \emph{maximizes statistical strength}
$D(Q_{\svec} \| \set{P}_{\svec})$. In Appendix~\ref{app:kl} we show that 
this is (in some sense) also the optimal choice according to statistical
theory. Thus we define the
statistical strength of $Q$ as $\sup_{\svec \in \ssimuc} 
D(Q_{\svec} \| \set{P}_{\svec})$, but we also present results for two
alternative classes of setting distributions.

\subsection{Formal Definition of Statistical Strength}\label{sec:strengthdef}
We define `the statistical strength of nonlocality proof $Q$'
in three different manners, depending on the freedom that we allow QM
in determining the sampling probabilities of the different
measurement settings.

\begin{definition}[Strength for Uniform Settings] 
When each measurement setting is sampled with equal probability,
the resulting strength $\strengthuni$ is defined by 
\begin{IEEEeqnarray}{rCcCl}\label{eq:strengthuni}
\strengthuni 
&:=& 
D(Q_{\svec^\circ} \| \set{P}_{\svec^\circ}) 
&=&  
\infm{\mvec\in\msim}{D(Q_{\svec^\circ} \| P_{\svec^\circ,\mvec})},
\end{IEEEeqnarray} 
where $\svec^\circ$ denotes the uniform distribution over the settings.
\end{definition}
\begin{definition}[Strength for Uncorrelated Settings] 
When the experimenter QM is allowed to choose any  distribution 
on measurement settings, as long as the distribution for each party is
uncorrelated with the distributions of the other parties, 
the resulting strength $\strength$ is defined by 
\begin{IEEEeqnarray}{rCcCl}\label{eq:strength4}
\strength 
&:=& 
\sups{\svec\in\ssimuc}{D(Q_{\svec} \| \set{P}_{\svec} )}
&=& 
\sups{\svec\in\ssimuc}{\infm{\mvec\in\msim}{D(Q_{\svec} \| P_{\svec,\mvec})}},
\end{IEEEeqnarray} 
where $\svec\in\ssimuc$ denotes the use of uncorrelated settings.
\end{definition}
\begin{definition}[Strength for Correlated Settings] 
When the experimeniter QM is allowed to choose 
any distribution on measurement settings 
(including correlated distributions), 
the resulting strength $\strengthcor$ is defined by 
\begin{IEEEeqnarray}{rCcCl}\label{eq:strengthcor}
\strengthcor 
&:=& 
\sups{\svec\in\ssim}{D(Q_{\svec} \| \set{P}_{\svec} )} 
&=& 
\sups{\svec\in\ssim}{\infm{\mvec\in\msim}{D(Q_{\svec} \| P_{\svec,\mvec})}},
\end{IEEEeqnarray} 
where $\svec\in\ssim$ denoted the use of correlated settings. 
\end{definition}
Throughout the remainder of the paper, we sometimes abbreviate the subscript 
$\svec \in \ssimuc$ to $\ssimuc$, and $\mvec \in \msim$ to $\msim$. 

In Section~\ref{sec:properties} we list
some essential topological and analytical properties of our three notions of strength.
For now, we only need the following reassuring fact
(Theorem~\ref{thm:supinf}, Section~\ref{sec:properties}, part 2(c)):
\begin{fact} \label{fact}
$\strengthuni \leq \strength \leq \strengthcor$. 
Moreover, $\strengthuni > 0$ if and only if $Q$ is a proper
nonlocality proof.
\end{fact}

As we explain in Section~\ref{sec:discussion}, we regard the
definition $\strength$ allowing maximization over uncorrelated
distributions as the `right' one. Henceforth, whenever we speak of
`statistical strength' without further qualification, we refer to
$\strength$. Nevertheless, to facilitate comparisons, we list our
results also for the two alternative definitions of statistical
strength.


\section{The Results}\label{sec:results}
The following table summarizes the statistical strengths of the various
nonlocality proofs. Note that the numbers in the \emph{middle} column
correspond to the `right' definition  $\strength$, 
which considers uncorrelated distributions for the measurement settings. 
\begin{IEEEeqnarray}{c}\label{eq:THEtable} 
\begin{tabular}{r||ccc|}
\text{Strength} & 
\text{Uniform Settings} \strengthuni & 
\text{Uncorrelated Settings} \strength &
\text{Correlated Settings} \strengthcor \\\hhline{=::===}
\text{Original \Bell}  & 0.0141597409 & 0.0158003672 & 0.0169800305 \\ 
\text{Optimized \Bell} & 0.0177632822 & 0.0191506613 & 0.0211293952 \\ 
\CHSH & 0.0462738469 & 0.0462738469 & 0.0462738469 \\ 
\text{\Hardy} & 0.0278585182 & 0.0279816333 & 0.0280347655 \\
\text{\Mermin} & 0.0157895843 & 0.0191506613 & 0.0211293952 \\
\GHZ & 0.2075187496 & 0.2075187496 & 0.4150374993 \\\hhline{-||---}
\end{tabular}
\end{IEEEeqnarray}

\begin{example}[The CHSH Results]
To help interpret the table, we continue our Example~\ref{ex:CHSH}
on \CHSH.  The entry in the first (`uniform') column for the CHSH
proof was obtained as follows.  $\svec$ was set to the uniform
distribution $\svec^\circ = (1/4,1/4,1/4,1/4)$. $Q$ was set to the
values in Table~(\ref{eq:chshtable}), resulting in a joint
distribution $Q_{\svec^\circ}$ on measurement settings and outcomes.
$Q_{\svec^\circ}$ was used to determine the local theory 
${\mvec}^*\in\msim$ that obtains the minimum in 
\begin{IEEEeqnarray*}{c}
\infm{\mvec\in\msim} D(Q_{\svec^\circ} \| {P}_{\svec^\circ, \mvec}).
\end{IEEEeqnarray*}
The resulting $\mvec^*$ can be found numerically. The corresponding
$P_{ab ; \mvec}$ distributions is given in Table~\ref{tb:bestclassCHSH}.
\begin{table}
\centering
\caption{Best Classical Theory for CHSH} 
\label{tb:bestclassCHSH}
\Ptable%
{0.375 & 0.125 \\ 0.125 & 0.375} {0.375 & 0.125 \\ 0.125 & 0.375}
{0.375 & 0.125 \\ 0.125 & 0.375} {0.125 & 0.375 \\ 0.375 & 0.125}
\end{table}

The KL divergence between $Q_{\svec^\circ}$ and $P_{\svec^\circ,
{\mvec}^*}$  can now be calculated. It is equal to $0.0462738469$, the left-most
entry in Table~(\ref{eq:THEtable}) in the CHSH-row. To get the
rightmost entry in this row, we performed the same computation for
\emph{all} $\svec \in \ssim$ (we will explain later how to do this
efficiently). We found that the resulting KL divergence
$D(Q_{\svec},P_{\svec,{\mvec}^*})$ (where ${\mvec}^*$ depends on
$\svec$) was, in fact, maximized for $\svec = \svec^\circ$: there was no gain in trying any 
other value for $\svec$. Thus, the rightmost column is equal to the leftmost column. 
Finally, Fact~\ref{fact} 
above implies that the middle column entry must be in between the leftmost
and the rightmost, explaining the entry in the middle column.
The corresponding analysis for the other nonlocality proofs is done in
Appendix~\ref{app:nonlocalityproofs}. 
\end{example}

\begin{example}[Mermin's ``a million runs'']\label{ex:mermin}
We recall Mermin's quote from the Introduction where he says that 
``a million runs'' of his experiment should be enought to convince 
us that ``the observed frequencies \dots are not chance fluctations''.  
We now can put numbers to this.  

Assuming that we perform Mermin's experiment with the optimized, 
uncorrelated settings, we should get a strength of 1,000,000 $\times$ 
0.0191506613 $\approx$ 19,150.  This means that after the million runs of 
the experiment, the likelihood of local realism still being true is 
comparable with the likelihood of a coin being fair after 19,150 tosses  
when the outcome was ``tails'' all the time.
\end{example}

>From Table~(\ref{eq:THEtable}) we  see that in the two-party setting, the 
nonlocality proof of \CHSH is much stronger than those of Bell, Hardy or Mermin,
and that this optimal strength is obtained for uniform measurement
settings.  Furthermore it is clear that the three-party proof
of \GHZ is an four and a half times stronger than all the two-party proofs.

We also note that the nonlocality proof of Mermin---in the case
of non-uniform settings---is equally strong as the optimized 
version of Bell's proof.  The setting distributions tables
in Appendix~\ref{app:mermin} shows why this is the case:
the optimal setting distribution for Mermin exclude one setting 
on $A$'s side, and one setting on $B$'s side, thus reducing Mermin's 
proof to that of Bell.  One can view this is as an
example of how a proof that is easier to understand (Mermin) is 
not necessarily stronger than one that has more subtle arguments
(Bell).

We also see that in general, except for \CHSH's proof,
uniform setting distributions do not give the optimal strength
of a nonlocality proof.  Rather, the experimenter obtains more
evidence for the nonlocality of nature by employing sampling
freqencies that are biased towards those settings that are 
more relevant for the nonlocality proof. 


\section{Interpretation and Discussion} \label{sec:discussion}
\subsection{Which nonlocality proof is strongest and what does it mean?}
\subsubsection{Caveat: statistical strength is not the whole story}
First of all, we stress that statistical strength is by no means the
only factor in determining the `goodness' of a nonlocality proof and
its corresponding experiment.  Various other aspects also come into
play, such as: how easy is it to prepare certain types of particles in
certain states?  Can we arrange to have the time and spatial
separations which are necessary to make the results convincing? Can we
implement the necessary random changes in settings per trial, quickly
enough? Our notion of strength neglects all these important practical
aspects.
\subsubsection{Comparing GHZ and CHSH}
\GHZ is the clear winner among all proofs that we investigated, being
about $4.5$ times stronger than CHSH, the strongest two-party proof
that we found. This means that, to obtain a given level of support for
QM and against LR, the optimal CHSH experiment has to be repeated
about $4.5$ times as often as the optimal \GHZ experiment.

On the other hand, the \GHZ proof is much harder to prepare
experimentally. In light of the reasoning above, and assuming that
both \CHSH and \GHZ can be given a convincing experimental
implementation, it may be the case that repeating the \CHSH
experiment $4.5 \times n$ times is much cheaper than repeating \GHZ
$n$ times.

\subsubsection{Nonlocality `without inequality'?}
The \GHZ proof was the first of a new class of proofs of Bell's
theorem, ``without inequalities''.  It specifies a state and
collection of settings, such that all QM probabilities are zero or
one, while this is impossible under LR. The QM probabilities involved
are just the probabilities of the four events in
Equation~(\ref{eq:GHZ}), Appendix~\ref{app:GHZ}. The fact that all
these must be either 0 or 1 has led some to claim that the
corresponding experiment has to performed only once in order to rule
out local realism\footnote{Quoting \cite{Peres00}, ``The list of
authors [claiming that a single experiment is sufficient to invalidate
local realism] is too long to be given explicitly, and it would be
unfair to give only a partial list.''}.  As has been observed before
\cite{Peres00}, this is not the case. This can be seen immediately if
we let LR adopt the uniform distribution on all possible
observations. Then, although QM is correct, no matter how often the
experiment is repeated, the resulting sequence of observations does
not have zero probability under LR's local theory --- simply because
\emph{no} sequence of observations has probability 0 under LR's
theory.  We can only decide that LR is wrong on a statistical basis:
the observations are \emph{much more likely} under QM than under
LR. This happens even if, instead of using the uniform distribution,
LR uses the local theory that is closest in KL divergence to the $Q$
induced by the \GHZ scenario. The reason is that there exists a
positive $\epsilon$ such that any local realist theory which comes
within $\epsilon$ of all the equalities but one, is forced to deviate
by more than $\epsilon$ in the last. Thus, accompanying the \GHZ style
proof without inequalities, is an implied 
\emph{inequality}, and it is this latter inequality that can be tested
experimentally.
\subsection{Is our definition of statistical strength the right one?}
We can think of two objections against our definition of statistical
strength. First, we may wonder whether the KL divergence is really the
right measure to use. Second, assuming that KL divergence is the right
measure, is our game-theoretic setting justified? We treat both issues in turn.
\subsubsection{Is Kullback-Leibler divergence justified?}
We can see two possible objections against KL divergence: (1)
different statistical paradigms such as the `Bayesian' and
`frequentist' paradigm define `amount of support' in different manners
(Appendix~\ref{app:formal}); (2) `asymptopia': KL divergence is an
inherently asymptotic notion.

These two objections are inextricably intertwined: there exists no
non-asymptotic measure which would (a) be acceptable to all
statisticians; (b) would not depend on prior considerations, such as a
`prior distribution' for the distributions involved in the Bayesian
framework, and a pre-set significance level in the frequentist
framework.  Thus, since we consider it most important to arrive at a
generally acceptable and objective measure, we decided to opt for the
KL divergence. We add here that even though this notion is asymptotic,
it can be used to provide numerical bounds on the actual,
non-asymptotic amount of support provided on each trial, both in
Bayesian and in frequentist terms.  We have not pursued this option
any further here.
\subsubsection{Game-Theoretic Justification}
There remains the question of whether to prefer $\strengthuni$,
$\strengthcor$ or, as we do, $\strength$.  The problem with
$\strengthuni$ is that, for any given combination of nonlocality proof
$Q$ and local theory $\mvec$, different settings may provide, on
average, more information about the nonlocality of nature than others.
This is evident from Table~(\ref{eq:THEtable}). We see no reason
for the experimenter not to exploit this.

On the other hand, allowing QM to use \emph{correlated} distributions
makes QM's case much weaker: LR might now argue that there is some
hidden communication between the parties. Since QM's goal is to
provide an experiment that is as convincing as possible to LR, we
do not allow for this situation. Thus, among the three definitions considered,
$\strength$ seems to be the most reasonable one.  Nevertheless, one
may still argue that \emph{none} of the three definitions of strength
are correct: they all seem unfavourable to QM, since we allow LR to
adjust his theory to whatever frequency of measurement settings QM is
going to use. In contrast, our definition does \emph{not} allow 
QM to adjust his setting distribution to LR's choice
(which would lead to strength defined as $\inf \sup$ rather than  $\sup \inf$).
The reason  why we favour LR in this way is that the quantum experimenters QM  
should try to convince LR that nature is nonlocal \emph{in a setting about 
which LR cannot complain}. Thus, if
LR wants to entertain several local theories at the same time, 
or wants to have a look at the probabilities $\svec_{ab}$ before the 
experiment is conducted, QM should allow him to do so---he will 
still be able to convince LR, even
though he may need to repeat the experiment a few more times.
Nevertheless, in developing clever strategies for computing
$\strength$, it turns out to be useful to investigate the $\inf \sup$
scenario in more detail. This is done in Section~\ref{sec:game}.

Summarizing, our approach is highly nonsymmetric between quantum
mechanics and local realism. There is only one quantum theory, and QM
believes in it, but he must arm himself against any and all local
realists.
\footnote{Some readers might wonder what would happen if
one would replace the  $D(Q\|P)$ in our analysis by $D(P\|Q)$.
In short, $D(P\|Q)$ quantifies how strongly the predictions of
quantum mechanics disagree with the outcomes of a classical
system $P$.  Hence such an analysis would be useful if one has
to prove that the statistics of a local realistic experiment (say, a
network of classically communicating computers) are not in
correspondance with the predictions of quantum mechanics.
The minimax solution of the game based on $D(P\|Q)$ provides a value of $Q$
that QM should specify as part of a challenge to LR to reproduce quantum
predictions with LR's theory. With this challenge, the computer simulation
using LR's theory can be run in as short as possible amount of time,
before giving sufficient evidence that LR has failed.}

\subsection{Related Work by Peres} \label{sec:peres}
Earlier work in our direction has been done by Peres \cite{Peres00} who
adopts a Bayesian type of approach. Peres implicitly
uses the same definition of strength of nonlocality proofs as we do
here, after merging equal probability joint outcomes of the experiment.
Our work extends his in several ways; most importantly, we allow the
experimentalist to optimize her experimental settings, whereas
Peres assumes particular (usually uniform) distributions over the settings. 
Peres determines LR's best theory by an inspired guess.
The proofs he considers have so many symmetries, that the
best LR theory has the same equal probability joint outcomes as the
QM experiment, the reduced experiment is binary,
and his guess always gives the right answer. His strategy would not
work for, e.g., the Hardy proof.

Peres starts out with a nonlocality proof $Q_\svec$ to be tested
against local theory $P_{\svec; \mvec}$, for some fixed distribution
$\svec$. Peres then defines the \emph{confidence depressing factor for
  $n$ trials}. In fact, Peres rediscovers the notion of KL divergence, 
since a straightforward calculation shows that for large $n$,
\begin{equation}
D(Q_\svec \| P_{\svec; \mvec}) = \frac{1}{n} 
\log(\texttt{confidence depressing factor}).
\end{equation}
For any given large $n$, the larger the confidence depressing factor
for $n$, the more evidence against $P_{\svec; \mvec}$ we are likely to
get on the basis of $n$ trials. Thus, when comparing a fixed quantum
experiment (with fixed $\svec$) $Q_\svec$ to a fixed local theory
$P_{\svec; \mvec}$, Peres' notion of strength is equivalent to ours .
Peres then goes on to say that, when comparing a fixed quantum
experiment $Q_\svec$ to the corresponding set of \emph{all} local
theories $\set{P}_{\svec}$, we may expect that LR will choose the
local theory with the least confidence depressing factor, i.e. the
smallest KL divergence to $Q_\svec$. Thus, whenever Peres chooses
uniform $\svec$, his notion of strength corresponds to our
$\strengthuni$, represented in the first column of
Table~\ref{eq:THEtable}.
In practice, Peres chooses an intuitive $\svec$, which is usually,
but not always uniform in our sense.  For example, in the \GHZ
scenario, Peres implicitly assumes that only those measurement
settings are used that correspond to the probabilities (all 0 or 1)
appearing in the \GHZ-inequality (\ref{eq:GHZ}),
Appendix~\ref{app:GHZ}. Thus, his experiment corresponds to a uniform
distribution on those four settings. Interestingly, such a
distribution on settings is \emph{not} allowed under our definition
of strength $\strength$, since it makes the probability of the setting
at party A dependent on (correlated with) the other settings. This
explains that Peres obtains a larger strength for \GHZ than we do: he
obtains $\log 0.75^{-n} = 0.4150\ldots$, which corresponds to our
$\strengthcor$: the uniform distribution on the restricted set of
settings appearing in the \GHZ proof turns out to be the optimum over
all distributions on measurement settings.

Our approach may be viewed as an extension of Peres' in several ways. 
First, we relate his confidence
depressing factor to the Kullback-Leibler divergence and show that
this is the right measure to use not just from a Bayesian point of
view, but also from an information-theoretic point of view and the
standard, `orthodox' frequentist statistics point of view. Second, we
extend his analysis to non-uniform distributions $\svec$ over measurement
settings and show that in some cases, substantial statistical strength
can be gained if QM uses non-uniform sampling distributions. Third, we
give a game-theoretic treatment of the maximization of $\svec$ and 
develop the necessary mathematical tools to enable fast computations
of statistical strength. Fourth, 
whereas he finds the best LR theory by cleverly guessing, we show 
the search for this theory can be performed automatically.

\subsection{Future Extensions and Conjectures}

The purpose of our paper has been to objectively compare the
statistical strength of existing proofs of Bell's theorem. The tools
we have developed, can be used in many further ways.

Firstly, one can take a given quantum state, and ask the
question, what is the best experiment which can be done with it.
This leads to a measure of statistical nonlocality of a given
joint state, whereby one is optimizing (in the outer optimization)
not just over setting distributions, but also over the settings
themselves, and even over the number of settings.

Secondly, one can take a given experimental type, for 
instance: the $2\times2\times2$ type, and ask what is
the best state, settings, and setting distribution for that
type of experiment? This comes down to replacing
the outer optimization over setting distributions, with
an optimization over states, settings, and setting distribution.

Using numerical optimization, we were able to analyse
a number of situations, leading to the following conjectures.
\begin{conjecture}
Among all $2\times2\times2$ proofs, and
allowing correlated setting distributions, CHSH is best.
\end{conjecture}
\begin{conjecture}
Among all $3\times2\times2$ proofs, and
allowing correlated setting distributions, GHZ is best.
\end{conjecture}
\begin{conjecture}
The best experiment with the Bell singlet state
is the CHSH experiment.
\end{conjecture}
In \cite{Acin02} Ac\'{\i}n et al.\ investigated the natural generalization of 
CHSH type experiments to qutrits.  Their main interest was the resistance 
of a given experiment to noise, and to their surprise they discovered in 
the $2\times 2\times 3$ case, 
that a less entangled state was more resistant to noise than the maximally
entangled state.  After some preliminary investigations, we found that 
that a similar experiment with an \emph{even less entangled} state
gives a stronger nonlocality experiment. 
\begin{conjecture}\label{conj:2x2x3}
The strongest possible $2\times 2\times 3$ nonlocality proof 
has statistical strength $0.077$, and it uses the bipartite state 
$\approx 0.6475\ket{1,1}+0.6475\ket{2,2}+0.4019\ket{3,3}$.
\end{conjecture}
If true, this conjecture is in remarkable contrast with what appears 
to be the strongest possible $2\times 2\times 3$ nonlocality proof 
that uses the maximally entangled state $(\ket{1,1}+\ket{2,2}+\ket{3,3})/\sqrt{3}$,
which has a statistical strength of only $0.058$.

Conjecture~\ref{conj:2x2x3} 
suggests that it is not always the case that a quantum state
with more `entropy of entanglement' \cite{Bennett96} will always
give a stronger nonlocality proof.  
Rather, it seems that entanglement and statistical nonlocality are different
quantities. One possibility however is that the counterintuitive
results just mentioned would disappear if one could do joint
measurements on several pairs of entangled qubits, qutrits,
or whatever. A regularized measure of nonlocality of a given
state, would be the limit for $k\to\infty$, of the strength of the
best experiment based on $k$ copies of the state (where
the parties are allowed to make joint measurements on
$k$ systems at the same time), divided by $k$. One may
conjecture for instance that the best experiment based on
two copies of the Bell singlet state is more than twice as
good as the best experiment based on single states.
That would be a form of ``superadditivity of nonlocality'',
quite in line with other forms of superadditivity which is
known to follow from entanglement.

\begin{conjecture}
There is an experiment on pairs of Bell
singlets, of the $2\times4\times4$ type, more than 
twice as strong as CHSH, and involving joint measurements
on the pairs.
\end{conjecture}

\section{Mathematical and Computational Properties of Statistical Strength}
\label{sec:computing}
Having presented and discussed the strengths of various nonlocality
proofs, we now turn to the second, more mathematical part of the
paper. We first prove several mathematical properties of our three
variations of statistical strength. Some of these are interesting in
their own right, giving new insights in the relation between
distributions predicted by quantum theory and local realist
approximations of it. But their main purpose is to help us compute
$\strength$. 
\subsection{Basic Properties}
\label{sec:properties}
We proceed to list some
essential properties of $\strengthuni, 
\strength$ and $\strengthcor$.  We say that ``nonlocality
proof $Q$ is \emph{absolutely continuous with respect to local realist
theory $\mvec$}'' \cite{Feller68b} if and only if for all 
$a,b \in \{1,2\}, \mvala, \mvalb \in \{\true,\false\}$,
it holds that if $Q_{ab}(\mvala,\mvalb) > 0$ then 
$P_{ab ; \mvec}(\mvala,\mvalb) > 0$.
\begin{theorem} \label{thm:supinf}
\newcounter{cr:supinf}
\setcounter{cr:supinf}{\value{theorem}}
Let $Q$ be a given (not necessarily $2\times 2\times 2$) nonlocality
proof and $\msim$ the corresponding set of local realist theories.
\begin{enumerate}
\item Let $U(\svec,\mvec) := D(Q_{\svec} \| P_{\svec;\mvec})$, then:
\begin{enumerate}
\item For a $2\times 2\times 2$ proof, we have that
\begin{IEEEeqnarray}{rCl}\label{eq:avkl}
U(\svec,\mvec) 
&=& 
\sum_{a,b \in \{1,2\}}{\svec_{ab} D(Q_{ab}(\cdot ) \| P_{ab; \mvec}(\cdot ))}.
\end{IEEEeqnarray} 
Hence, the KL divergence $D(Q_{\svec} \| P_{\svec;\mvec})$ may
alternatively be viewed as the average KL divergence between the
conditional distributions of $(X,Y)$ given the setting $(A,B)$, where 
the average is over the setting. For a generalized nonlocality proof, 
the analogous generalization of Equation~(\ref{eq:avkl}) holds.
\item For fixed $\svec$, $U(\svec,\mvec)$ is convex and lower
  semicontinuous on $\msim$, and continuous and differentiable on the
  interior of $\msim$.
\item If $Q$ is absolutely continuous with respect to some fixed
  $\mvec$, then $U(\svec,\mvec)$ is linear in $\svec$.
\end{enumerate}
\item Let
\begin{IEEEeqnarray}{rCl}\label{eq:uinf}
U(\svec)
& :=& 
\infm{\msim}{U(\svec,\mvec)},
\end{IEEEeqnarray} 
then
\begin{enumerate}
\item For all $\svec \in \ssim$, the infimum in Equation~(\ref{eq:uinf})
is achieved for some $\mvec^*$.
\item The function $U(\svec)$ is nonnegative, bounded, concave and
continuous on $\svec$.
\item If $Q$ is not a proper nonlocality proof, then for all $\svec
  \in \ssim, U(\svec) = 0$. If $Q$ is a proper nonlocality proof, then
  $U(\svec) > 0$ for \emph{all} $\svec$ in the interior of $\ssim$.
\item For a 2 party, 2 measurement settings per party nonlocality
  proof, we further have that, even if $Q$ is proper, then still
  $U(\svec) = 0$ for all $\svec$ on the boundary of $\ssim$.
\end{enumerate}
\item Suppose that $\svec$ is in the interior of $\ssim$, then:
\begin{enumerate}
\item Let $Q$ be a $2\times 2\times 2$ nonlocality proof. Suppose that
  $Q$ is non-trivial in the sense that, for some $a,b$, $Q_{ab}$ is
  not a point mass (i.e.\ $0 < Q_{ab}(\mvala,\mvalb) < 1$ for some
  $\mvala,\mvalb$). Then $\mvec^* \in \msim$ achieves the infimum in
  Equation~(\ref{eq:uinf}) if and only if the following 16
  (in)equalities hold:
\begin{IEEEeqnarray}{rCl}\label{eq:em}
\sum_{a,b \in \{1,2\}} \svec_{ab} \frac{ Q_{ab}(\mvala_a, \mvalb_b ) }
{ P_{ab ; \mvec^*}(\mvala_a, \mvalb_b )} 
&=& 
1
\end{IEEEeqnarray}
for all $(\mvala_1,\mvala_2,\mvalb_1,\mvalb_2) \in \{\true,\false\}^4$
such that $\mvec^*_{\mvala_1,\mvala_2,\mvalb_1,\mvalb_2} > 0$, and
\begin{IEEEeqnarray}{rCl} \label{eq:emb}
\sum_{a,b \in \{1,2\}} \svec_{ab} \frac{ Q_{ab}(\mvala_a, \mvalb_b ) }
{ P_{ab ; \mvec^*}(\mvala_a, \mvalb_b )} 
&\leq& 
1
\end{IEEEeqnarray}
for all $(\mvala_1,\mvala_2,\mvalb_1,\mvalb_2) \in \{\true,\false \}^4$
such that $\mvec^*_{\mvala_1,\mvala_2,\mvalb_1,\mvalb_2} = 0$.

For generalized nonlocality proofs, $\mvec^* \in \msim$ achieves
Equation~(\ref{eq:uinf}) if and only if the corresponding analogues of
Equations~(\ref{eq:em}) and (\ref{eq:emb}) both hold.
\item Suppose that $\mvec^*$ and $\mvec^\circ$ both achieve the
infimum in Equation~(\ref{eq:uinf}).  Then for all $\mvala, \mvalb \in
\{\true,\false\}$, $a,b \in \{1,2\}$ with $Q_{ab}(\mvala,\mvalb) > 0$,
we have
$P_{ab; \mvec^*}(\mvala,\mvalb) = P_{ab ; \mvec^\circ}(\mvala, \mvalb) > 0$.
In words, $\mvec^*$ and $\mvec^\circ$ coincide in every measurement
setting for every measurement outcome that has positive probability
according to $Q_{\svec}$, and $Q$ is absolutely continuous with
respect to $\mvec^*$ and $\mvec^\circ$.
\end{enumerate}
\end{enumerate}
\end{theorem}
The proof of this theorem is in Appendix~\ref{app:prf:supinf}.

In general, $\infm{\msim} U(\svec,\mvec)$ may be achieved for several,
different $\mvec$. By part 2 of the theorem, these must induce the
same four marginal distributions $P_{ab ; \mvec}$.  It also follows
directly from part 2 of the theorem that, for $2\times 2\times 2$
proofs, $\strength := \sups{\ssimuc} U(\svec)$ is achieved
for some $\svec^* \in \ssimuc$, where $\svec^*_{ab} > 0$ for all $a,b
\in \{1,2\}$.
\subsection{Game-Theoretic Considerations}
\label{sec:game}
The following considerations will enable us to compute $\strength$
very efficiently in some special cases, most notably the \CHSH proof.

We consider the following variation of our basic scenario. Suppose
that, before the experiments are actually conducted, LR has to decide
on a \emph{single} local theory $\mvec_0$ (rather than the set
$\msim$) as an explanation of the outcomes that will be observed. QM
then gets to see this $\mvec$, and can choose $\svec$ depending on the
$\mvec_0$ that has been chosen. Since QM wants to maximize the
strength of the experiment, he will pick the $\svec$ achieving
$\sups{\ssimuc} D(Q_{\svec} \| P_{\svec ; \mvec_0})$. In
such a scenario, the `best' LR theory, minimizing statistical
strength, is the LR theory $\mvec_0$ that minimizes, over $\mvec \in
\msim$, $\sups{\ssimuc} D(Q_{\svec} \| P_{\svec ;
\mvec})$. Thus, in this slightly different setup, the statistical
strength is determined by
\begin{IEEEeqnarray}{rCl}\label{eq:infsupstrength}
\strengthis 
& := & 
\infm{\msim}{\sups{\ssimuc}{D(Q_{\svec} \| P_{\svec ; \mvec})}}
\end{IEEEeqnarray}
rather than $\strength := \sups{\ssimuc} \infm{\msim} 
D(Q_{\svec} \| P_{\svec ; \mvec})$. Below we show that
$\strengthis \geq \strength$. As we already argued in
Section~\ref{sec:discussion}, 
we consider the definition
$\strength$ to be preferable over $\strengthis$. 
Nevertheless, it is useful to investigate under what conditions
$\strength = \strengthis$. Von Neumann's famous minimax
theorem of game theory \cite{VonNeumann28} suggests that
\begin{IEEEeqnarray}{rCl}\label{eq:neumann}
\sups{\ssim^*}{\infm{\msim}{D(Q_{\svec} \| P_{\svec ; \mvec})}}
& =& 
\infm{\msim}{\sups{\ssim^*}{D(Q_{\svec} \| P_{\svec ; \mvec})}},
\end{IEEEeqnarray}
if $\ssim^*$ is a convex subset of $\ssim$. Indeed,
Theorem~\ref{thm:spcor} below shows that Equation~(\ref{eq:neumann})
holds if we take $\ssim^* = \ssim$. Unfortunately, $\ssimuc$ is
\emph{not} convex, and Equation~(\ref{eq:neumann}) does not hold in
general for $\ssim^* = \ssimuc$, whence in general $\strength \neq
\strengthis$. Nevertheless, Theorem~\ref{thm:2x2} provides some
conditions under which Equation~(\ref{eq:neumann}) does hold with
$\ssim^* = \ssimuc$. In Section~\ref{sec:chshgame} we put this fact to
use in computing $\strength$ for the \CHSH nonlocality proof.  But
before presenting Theorems~\ref{thm:spcor} and~\ref{thm:2x2}, we first
need to introduce some game-theoretic terminology.

\subsubsection{Game-Theoretic Definitions}
\begin{definition}[Statistical Game \cite{Ferguson67}] 
A \emph{statistical game} is a triplet $(A,B,L)$
where $A$ and $B$ are arbitrary sets and $L: A \times B \rightarrow \R
\cup \{ - \infty, \infty\}$ is a \emph{loss function}. 
If 
\begin{IEEEeqnarray}{rCl}
\sup_{\alpha \in A} \inf_{\beta \in B} L(\alpha,\beta) 
& =& 
\inf_{\beta\in B} \sup_{\alpha \in A} L(\alpha,\beta),
\end{IEEEeqnarray}
we  say that the game has \emph{value} $V$ with 
\begin{IEEEeqnarray}{rCl}
V 
& := & 
\sup_{\alpha \in A} \inf_{\beta \in B} L(\alpha,\beta).
\end{IEEEeqnarray}
If for some $(\alpha^*,\beta^*) \in A \times B$ we have
\begin{IEEEeqnarray*}{rL}
\text{For all $\alpha \in A$:~} & L(\alpha,\beta^*) \leq L(\alpha^*,\beta^*) \\ 
\text{For all $\beta \in B$:~} & L(\alpha^*,\beta) \geq L(\alpha^*,\beta^*) 
\end{IEEEeqnarray*}
then we call $(\alpha^*,\beta^*)$ a \emph{saddle point} of the game.
It is easily seen (Proposition~\ref{prop:game},
Appendix~\ref{app:theoremproofs}) that, if $\alpha^\circ$ achieves
$\sup_{\alpha \in A} \inf_{\beta \in B} L(\alpha,\beta)$ and
$\beta^\circ$ achieves $\inf_{\beta \in B} L(\alpha,\beta)$ and the
game has value $V$, then $(\alpha^\circ,\beta^\circ)$ is a
saddle point and $L(\alpha^\circ,\beta^\circ) = V$.
\end{definition}

\begin{definition}[Correlated Game]
With each nonlocality proof we associate a corresponding
\emph{correlated game,} which is just the statistical game defined by
the triple $(\ssim,\msim,U)$, where $U: \ssim \times \msim \rightarrow
\R \cup \{ \infty \}$ is defined by
\begin{IEEEeqnarray}{rCl}
U(\svec,\mvec) 
&:=& 
D(Q_\svec \| P_{\svec;\mvec}).
\end{IEEEeqnarray}
By the definition above, if this game has a value then
it is equal to  $V$ defined by 
\begin{IEEEeqnarray}{rCcCl}
V
&:=& 
\infm{\msim} \sups{\ssim} U(\svec,\mvec) 
 & = & 
\sups{\ssim}\infm{\msim} U(\svec,\mvec).
\end{IEEEeqnarray}
We call the game \emph{correlated} because we allow distributions
${\svec}$ over measurement settings to be such that the probability
that party $A$ is in setting $a$ is correlated with (is dependent on)
the setting $b$ of party $B$.
The fact that each correlated game has a well defined value
is made specific in Theorem~\ref{thm:spcor} below. 
\end{definition}

\begin{definition}[Uncorrelated Game]
Recall that we use $\ssimuc$ to denote the set of vectors representing
uncorrelated distributions in $\ssim$.  With each nonlocality proof
we can associate the game $(\ssimuc,\msim,U)$ which we call the
corresponding \emph{uncorrelated game}.
\end{definition}

\subsubsection{Game-Theoretic, Saddle Point Theorems}
\begin{theorem}[Saddle point for Potentially Correlated Settings]
\label{thm:spcor}
\newcounter{cr:spcor}
\setcounter{cr:spcor}{\value{theorem}}
For every (generalized) nonlocality proof, the correlated game
$(\msim,\ssim,U)$ corresponding to it has a finite value, i.e.\ there
exist a $0 \leq V < \infty$ with
$\infm{\msim} \sups{\ssim} U(\svec,\mvec) = V =  
\sups{\ssim}\infm{\msim} U(\svec,\mvec)$.
The infimum on the left is achieved for some $\mvec^* \in \msim$; the
supremum on the right is achieved for some $\svec^*$ in $\ssim$, so
that $(\mvec^*, \svec^*)$ is a saddle point.
\end{theorem}
The proof of this theorem is in Appendix~\ref{app:prf:spcor}.

In the information-theoretic literature, several well-known minimax
and saddle point theorems involving the Kullback-Leibler divergence
exist; we mention \cite{Haussler97,Topsoe79}. However, all these deal
with settings that are substantially different from ours.

In the case where there are two parties and two measurement settings
per party, we can say a lot more.
\begin{theorem} [Saddle point for $2\times 2\times N$ Nonlocality Proofs]\label{thm:2x2}
\newcounter{cr:2x2}
\setcounter{cr:2x2}{\value{theorem}}
Fix any proper nonlocality proof based on 2 parties with 2
measurement settings per party and let $(\ssim,\msim,U)$ and
$(\ssimuc,\msim,U)$ be the corresponding correlated and uncorrelated
games, then:
\begin{enumerate}
\item The correlated game has a saddle point with value $V > 0$. Moreover,
\begin{IEEEeqnarray}{rCcCl}\label{eq:ucltcor}
\sups{\ssimuc} \infm{\msim} U(\svec,\mvec) 
&\leq & 
\sups{\ssim}  \infm{\msim} U(\svec,\mvec) 
&=&
 V, \\ 
\label{eq:uciscor}
\infm{\msim} \sups{\ssimuc} U(\svec,\mvec) 
&=& 
\infm{\msim}  \sups{\ssim} U(\svec,\mvec) 
&=&
V.
\end{IEEEeqnarray}
\item Let 
\begin{IEEEeqnarray}{rCl}
{\msim^*} 
&:=& \{ \mvec : \mvec \text{\ achieves\ } \infm{\msim} \sups{ \ssim} U(\svec,\mvec) \},\\
{\msimuc}^* 
&:=& 
\{ \mvec : \mvec \text{\ achieves\ } 
\infm{\msim} \sups{ \ssimuc} U(\svec,\mvec) \},
\end{IEEEeqnarray}
then
\begin{enumerate}
\item $\msim^*$ is non-empty.
\item $\msim^* = \msimuc^*$.
\item All $\mvec^* \in \msim^*$ are `equalizer strategies',
i.e.\  for all $\svec \in \ssim$ we have the equality 
$U(\svec,\mvec^*) = V$.
\end{enumerate}
\item The uncorrelated game has a saddle point if and only if there
exists $(\mvec^*,\svec^*)$, with $\svec^* \in \ssimuc$, such that
\begin{enumerate}
\item $\mvec^*$ achieves $\infm{\msim} U(\svec^*,\mvec)$.
\item $\mvec^*$ is an equalizer strategy.
\end{enumerate}
If such $(\svec^*,\mvec^*)$ exists, it is a saddle point.
\end{enumerate}
\end{theorem}
The proof of this theorem is in Appendix~\ref{app:prf:2x2}.

\subsection{Computing Statistical Strength} \label{sec:chshgame}
We are now armed with the mathematical tools needed to compute
statistical strength.  By convexity of $U(\svec,\mvec)$ in $\mvec$, we
see that for fixed $\svec$, determining 
$D(Q_{\svec} \| \set{P}_{\svec}) = \infm{\msim} U(\svec,\mvec)$ is
a convex optimization problem, which suggests that numerical
optimization is computationally feasible. Interestingly, it turns out that computing
$\infm{\msim} U(\svec,\mvec)$ is formally equivalent to
computing the maximum likelihood in a well-known statistical missing
data problem. 
Indeed, we obtained our results by using a `vertex direction
algorithm' \cite{GroeneboomJW}, a clever
numerical optimization algorithm specifically designed for
statistical missing data problems.

By concavity of $U(\svec)$ as defined in Theorem~\ref{thm:supinf}, we
see that determining $\strengthcor$ is a concave optimization problem.
Thus, numerical optimization can again be performed. There are some
difficulties in computing the measure $\strength$, since the set
$\ssimuc$ over which we maximize is not convex. Nevertheless, for the
small problems (few parties, particles, measurement settings) we
consider here it can be done.

In some special cases, including \CHSH, we can do all the calculations 
\emph{by hand} and do not have to resort to numerical optimization. 
We do this by  making an educated
guess of the $\svec^*$ achieving $\sups{\ssimuc} D(Q_{\svec}
\| \set{P}_{\svec})$, and then verify our guess using
Theorem~\ref{thm:supinf} and the game-theoretic tools developed in Theorem~\ref{thm:2x2}.
This can best be illustrated using \CHSH as an example. 
\begin{example}[CHSH, continued]
Consider the \CHSH nonlocality argument. The quantum distributions $Q$,
given in the table in Section~\ref{sec:nonlocalityproofs} have
traditionally been compared with the local theory $\tilde{\mvec}$
defined by
\begin{IEEEeqnarray}{rCcCcCcCcCcCcCcCl}
\tilde{\mvec}_{\false \false \false \false}
&=&
\tilde{\mvec}_{\true \true \true\true}
&=&
\tilde{\mvec}_{\false \false \false \true}
&=& 
\tilde{\mvec}_{\true \true\true \false}
&=& 
\tilde{\mvec}_{\false \false \true \false}
&=& 
\tilde{\mvec}_{\true\true \false \true}
&=& 
\tilde{\mvec}_{\true \false \false \true}
&=&
\tilde{\mvec}_{\false \true \true \false} 
&=&
\smfrac{1}{8},
\end{IEEEeqnarray}
and $\tilde{\mvec}_{\mvala_1 \mvala_2 \mvalb_1 \mvalb_2}=0$ otherwise.
This gives rise to the following probability table:
\begin{IEEEeqnarray}{c}
\begin{tabular}{r||rr|rr|}
$P_{ab ; \tilde{\mvec}}$ & $\mvara_1=\true$ & $\mvara_1=\false$ & 
$\mvara_2=\true$ & $\mvara_2=\false$ \\\hhline{=::====} 
$\mvarb_1 = \true$  & $0.375$ & $0.125$ & $0.375$ & $0.125$ \\ 
$\mvarb_1 = \false$ & $0.125$ & $0.375$ & $0.125$ & $0.375$ \\\hhline{-||----} 
$\mvarb_2 = \true$  & $0.375$ & $0.125$ & $0.125$ & $0.375$ \\
$\mvarb_2 = \false$ & $0.125$ & $0.375$ & $0.375$ & $0.125$ \\\hhline{-||----} 
\end{tabular}
\end{IEEEeqnarray}
There exists no local theory which has uniformly smaller absolute
deviations from the quantum probabilities in all four tables. Even
though, in general, absolute deviations are not a good indicator of
statistical strength, based on the fact that all four tables `look the
same', we may still \emph{guess} that, in this particular case, for
uniform measurement settings $\tilde{\svec}_{ab} = 1/4$, $a,b \in
\{1,2\}$, the optimal local realist theory is given by the
$\tilde{\mvec}$ defined above. We can now use
Theorem~\ref{thm:supinf}, part 3(a) to check our guess. Checking the
16 equations~(\ref{eq:em}) and (\ref{eq:emb}) shows that our guess was
correct: $\tilde{\mvec}$ achieves $\inf U(\svec,\mvec)$ for the
uniform measurement settings $\tilde{\svec}$. It is clear that
$\tilde{\mvec}$ is an equalizer strategy and that $\tilde{\svec}$ is
uncorrelated.  But now Theorem~\ref{thm:2x2}, part (3) tells us that
$(\tilde{\svec},\tilde{\mvec})$ is a saddle point in the uncorrelated
game. This shows that $\tilde{\svec}$ achieves $\sups{\ssimuc}
\infm{\msim} D(Q_{\svec} \| {P}_{\svec})$. Therefore, the statistical
strength of the \CHSH nonlocality proof must be given by
\begin{IEEEeqnarray}{rCcCl}
\strength 
&=& 
\sups{\ssimuc}{\infm{\msim}{D(Q_{\svec} \| {P}_{\svec})}} 
&=& 
D(Q_{\tilde{\svec}} \| P_{\tilde{\svec} ; \tilde{\mvec}}),
\end{IEEEeqnarray}
which is straightforward to evaluate.
\end{example}


\section{Acknowledgments}
The authors are very grateful to Piet Groeneboom for providing us with 
the software \cite{GroeneboomJW} 
needed to compute $\inf_{P \in \set{P}} D(Q \| P)$, and helping 
us to adjust the program  for our purposes.  

The authors would like to thank \textsc{Eurandom} for financial support.  
Part of this research was performed while Peter Gr\"unwald was visiting the
University of California at Santa Cruz (UCSC).  The visit was funded
by NWO (the Netherlands Organization for Scientific Research) and the
Applied Math and Computer Science Departments of UCSC.  
Wim van Dam's work is supported in part by funds provided by the
U.S.\ Department of Energy (DOE) and cooperative research
agreement DF-FC02-94ER40818, by a CMI postdoctoral fellowship,
and by an earlier HP/MSRI fellowship.
Richard Gill's research
was partially funded by project RESQ (IST-2001-37559) of the IST-FET
programme of the European Union.
Richard Gill is grateful for the hospitality of the Quantum Probability group
at the department of mathematics of the University of Greifswald,
Germany. His research there was supported by
European Commission grant HPRN-CT-2002-00279, RTN QP-Applications.

\appendices
\section{Beyond 2$\times$2$\times$2: General Case of Nonlocality Proofs}
\label{app:beyond2x2x2} 
Here we extend the $2\times 2\times 2$ setting to more than two
parties, settings and outcomes.  A general \emph{nonlocality proof}
is defined as a tuple $(k,\ssetgen,\msetgen,Q)$ where
\begin{enumerate}
\item $k$ is the number of parties,
\item $\ssetgen := \ssetgen_1 \times \dots \times \ssetgen_k$ is the
set of possible measurement settings.
\begin{enumerate}
\item $\ssetgen_j := \{1,2,\dots, N^{\svalgen}_j\}$ is the set of
measurement settings for party $j$.
\item $N^{\svalgen}_j$ is the number of settings of party $j$.
\end{enumerate}
\item $\msetgen := \msetgen_1 \times \dots \times \msetgen_k$ is the
  set of possible measurement outcomes.
\begin{enumerate}
\item $\msetgen_j := \msetgen_{(j,1)} \times \dots \times
\msetgen_{(j,N_j^{\svalgen})}$ is the set of measurement outcomes for
party $j$.
\item $\msetgen_{(j,\svalgen)} := \{1,2,\dots,
  N^{\mvalgen}_{(j,\svalgen)}\}$ is the set of measurement outcomes
  for party $j$ \emph{when party $j$ chooses setting $\svalgen$}.
\item $N^{\mvalgen}_{(j,\svalgen)}$ is the number of measurement
  outcomes for party $j$ when party $j$ chooses setting $\svalgen$.
\item $(\mvargen_1,\dots, \mvargen_k)$ are the random variables
  indicating the outcome at parties $1,2,\dots, k$.
\end{enumerate}
\item $Q := (Q_{\svalgen_1 \dots \svalgen_k} : (\svalgen_1,\dots,
  \svalgen_k) \in \ssetgen)$ is a list of all the distributions
  $Q_{\svalgen_1 \dots \svalgen_k}(\mvargen_1 = \cdot,\dots,
  \mvargen_k =\cdot )$, one for each joint measurement setting
  $(\svalgen_1,\dots, \svalgen_k) \in \ssetgen$.  These are the
  distributions on outcomes induced by the state that the quantum
  experimenter's entangled states are in.
\end{enumerate}
To each nonlocality proof $(k,\ssetgen,\msetgen,Q)$ there corresponds
a set of local realist distributions $\msim$. Each such distribution
is identified with its probability vector $\mvec$. Formally, $\mvec$
is a distribution for the tuple of random variables
\begin{IEEEeqnarray}{c}
\begin{array}{ccc}
\mvargen_{(1,1)} & \cdots & \mvargen_{(1,N^{\svalgen}_1)}\\ 
\vdots & \ddots & \vdots \\ 
\mvargen_{(k,1)} & \cdots & \mvargen_{(k,N^{\svalgen}_k)}
\end{array}
\end{IEEEeqnarray}
Here $\mvargen_{(j,\svalgen)}$ denotes LR's distribution of $Z_j$ when
party $j$'s measurement device is in setting $\svalgen$.

Once again, we call a nonlocality proof \emph{proper} if and only if
it violates local realism, i.e.\ if there exists no local realist
distribution $\mvec$ such that $P_{\svalgen_1 \dots \svalgen_k ;
\mvec}(\cdot ) = Q_{\svalgen_1 \dots \svalgen_k}(\cdot )$ for all
$(\svalgen_1,\dots, \svalgen_k) \in \ssetgen$.

The definition of statistical strength remains unchanged.


\section{The Nonlocality Arguments} \label{app:nonlocalityarguments}
In this Appendix we present the inequalities and logical constraints
that must hold under local realism yet can be violated under quantum
mechanics. The specific quantum states chosen to violate these
inequalities, as well as the closest possible (in the KL divergence
sense) local theories are listed in Appendix~\ref{app:nonlocalityproofs}.
\subsection{Arguments of Bell and CHSH}
CHSH's argument was described in Example~\ref{ex:CHSH}. 
By exactly the same line of reasoning as used in obtaining the CHSH
inequality (\ref{e:chsh}), one also obtains Bell's inequality
\begin{IEEEeqnarray}{rCl}
\Pr(X_1=Y_1) 
& \leq & 
\Pr(X_2\neq Y_2) + \Pr(X_2\neq Y_1) + \Pr(X_1+Y_2).
\end{IEEEeqnarray}
See Sections~\ref{sec:bellor} and \ref{sec:bellop} for how this inequality can be violated.

\subsection{Hardy's Argument} 
Hardy noted the following: if $(\mvara_2 \& \mvarb_2)$ is true, and
$(\mvara_2 \implies \mvarb_1)$ is true, and $(\mvarb_2 \implies
\mvara_1)$ is true, then $(\mvara_1 \& \mvarb_1)$ is true. Thus
$(\mvara_2 \& \mvarb_2)$ implies: $\neg(\mvara_2 \implies \mvarb_1)$
or $\neg(\mvarb_2 \implies \mvara_1)$ or $(\mvara_1 \& \mvarb_1)$.
Therefore 
\begin{IEEEeqnarray}{rCl}\label{e:hardy}
\Pr(\mvara_2 \& \mvarb_2 ) 
&\leq & 
\Pr(\mvara_2 \& \neg \mvarb_1) +
\Pr(\neg \mvara_1 \& \mvarb_2 )+ \Pr (\mvara_1 \& \mvarb_1).
\end{IEEEeqnarray}
On the other hand, according to quantum mechanics it is possible that
the first probability is positive, in particular, equals $0.09$, while
the three other probabilities here are all zero.  
See Section~\ref{sec:hardy} for the precise  probabilities.

\subsection{Mermin's Argument}
Mermin's argument uses three settings on both sides of the
two parties, thus giving the set of six events 
$\{X_1,Y_1,X_2,Y_2,X_3,Y_3\}$.  First, observe that 
the three equalities in $(X_1=Y_1)\& (X_2=Y_2)\& (X_3=Y_3)$ 
implies at least one of the three statements in
$((X_1=Y_2) \&  (X_2=Y_1))\vee ((X_1=Y_3) \& 
(X_3=Y_1))\vee ((X_2=Y_3) \&  (X_3=Y_2))$. 
By the standard arguments that we used before, we see that 
\begin{IEEEeqnarray*}{rCl}
1-\Pr(X_1\neq Y_1) -\Pr(X_2\neq Y_2) -\Pr(X_3\neq Y_3) 
&\leq&  \Pr((X_1=Y_1)\& (X_2=Y_2)\& (X_3=Y_3)),
\end{IEEEeqnarray*}
and that 
\begin{IEEEeqnarray*}{rCl}
\Pr\left(\begin{array}{c}
((X_1=Y_2) \&  (X_2=Y_1))\\\vee\\ ((X_1=Y_3) \&  (X_3=Y_1))\\\vee\\ ((X_2=Y_3) 
\&  (X_3=Y_2))
\end{array} \right)
& \leq & 
\left(\begin{array}{c}
\Pr((X_1=Y_2) \&  (X_2=Y_1))\\+\\ \Pr((X_1=Y_3) \&  
(X_3=Y_1))\\+\\ \Pr((X_2=Y_3) \&  (X_3=Y_2))
\end{array}
\right)\\
&\leq &  \frac{1}{2}\left(\begin{array}{c}
\Pr(X_1=Y_2) +\Pr(X_2=Y_1)\\ + \\ \Pr(X_1=Y_3) +\Pr (X_3=Y_1)
\\+\\  \Pr(X_2=Y_3) +\Pr(X_3=Y_2)
\end{array}
\right).
\end{IEEEeqnarray*}
As a result we have the `Mermin inequality'
\begin{IEEEeqnarray*}{rCl}
1 & \leq & 
\sum_{i=1}^3{\Pr(X_i\neq Y_i)}+\frac{1}{2}
\sum_{\substack{i,j=1\\i\neq j}}^{3}{\Pr(X_i=Y_j)}, 
\end{IEEEeqnarray*}
which gets violated by a state and measurement setting
that has probabilities $\Pr(X_i\neq Y_i)=0$ and 
$\Pr(X_i=Y_j)=\smfrac{1}{4}$ for $i\neq j$
(see Section~\ref{app:mermin} in the appendix).

\subsection{GHZ's Argument} \label{app:GHZ}
Starting with \cite{GreenbergerHZ89}, \GHZ, proofs against local
realism have been based on systems of three or more qubits, on systems
of higher-dimensional quantum systems, and on larger sets of
measurements (settings) per particle. 
Each time we are allowed to search over a wider space we may be able 
to obtain stronger nonlocality proofs,
though each time the actual experiment may become harder to set up in
the laboratory.  

Let $\xor$ denote the exclusive or operation such that 
$X\xor Y$ is true if and only if $X\neq Y$.  
Then the following implication must hold 
\begin{IEEEeqnarray}{rCl}
((X_1 \xor Y_2 = Z_2) \&  
(X_2 \xor Y_1 = Z_2)  \&  
(X_2 \xor Y_2 = Z_1))  
&\implies &
(X_1 \xor Y_1 = Z_1).
\end{IEEEeqnarray}
Now, by considering the contrapositive, we get 
\begin{IEEEeqnarray}{rCl}
\Pr(X_1\xor Y_1 \neq Z_1) 
& \leq & 
\Pr((X_1 \xor Y_2\neq Z_2) \vee (X_2 \xor Y_1 \neq Z_2) \vee (X_2 \xor Y_2 \neq Z_1)).
\end{IEEEeqnarray}
And because $\Pr(X\xor Y\neq Z) = \Pr(X\xor Y\xor Z)$,
this gives us
\GHZ's inequality:
\begin{IEEEeqnarray}{rCl}
\label{eq:GHZ}
\Pr(X_1\xor Y_1 \xor Z_1) 
& \leq & 
\Pr(X_1 \xor Y_2\xor Z_2) +
\Pr(X_2 \xor Y_1 \xor Z_2) +
\Pr(X_2 \xor Y_2 \xor Z_1).
\end{IEEEeqnarray}
This inequality can be violated by a three way entangled state 
and measurement settings that give 
$\Pr(X_1\xor Y_1 \xor Z_1)=1$ and 
$\Pr(X_1 \xor Y_2\xor Z_2) =
\Pr(X_2 \xor Y_1 \xor Z_2) =
\Pr(X_2 \xor Y_2 \xor Z_1)=0$.
The details of this proof are in Section~\ref{app:ghz}.


\section{The Nonlocality Proofs, Their Optimal Setting Distributions 
and Best Classical Theories} 
\label{app:nonlocalityproofs}
In this appendix we list the nonlocality proofs of 
Bell, an optimized version of Bell, CHSH, Hardy,  
Mermin and \GHZ and their solutions. 
The proofs themselves are described by a multipartite quantum state
and the measurement bases $\ket{m_\cdot^\cdot}$ of the parties.  
Because all bases are two dimensional in the proofs below, it 
is sufficient to only describe the vector $\ket{m_\cdot^\cdot}$,
where it is understood that the other basis vector $(\ket{\perp 
m^\cdot_\cdot})$ is the orthogonal one. 
Because of its frequent use, we define for the whole appendix the 
rotated vector $\ket{R(\phi)} := \cos(\phi)\ket{0}+\sin(\phi)\ket{1}$.
A \emph{measurement setting} refers to the bases that parties use
during a trial of the experiment.  All proofs, except Mermin's,
have two different settings per party (in \Mermin they have three).  

Given the state and the measurement bases, the proof is summarized
in a table of probabilities of the possible measurement outcomes.
Here we list these probabilities conditionally on the 
specific measurement settings. 
For example, for Bell's original nonlocality proof, which uses 
the state $\ket{\Psi}:=\smfrac{1}{\sqrt{2}}(\ket{0_A0_B}+\ket{1_A1_B})$ 
and the measurement vectors $\ket{X=\true}_{a=1}:=\ket{R(0)}$ and 
$\ket{Y=\true}_{b=1}:=\ket{R(\smfrac{\pi}{8})}$, we list the 
probability $Q_{11}(X=\true,Y=\true)=|\braket{\Psi}{X=\true,Y=\true}_{a=1,b=1}|^2
\approx 0.4268$ in the table.

As discussed in the article (Section~\ref{sec:strengthdef}), the
strength of a nonlocality proof will depend on the probabilities
$\svec$ with which the parties use the different measurement
settings. Recall that we defined three different notions of strength,
depending on how these probabilities are determined: uniform settings
($\strengthuni$), uncorrelated settings ($\strength$) and correlated
settings ($\strengthcor$).  For both the correlated and the
uncorrelated settings, the parties can optimize their setting 
distributions to get the strongest possible statistics to prove the
nonlocality of their measurement outcomes.  We list these optimal
distributions below where, for example, $\Pr(a=1) = \svec_{10} +
\svec_{11}$ stands for the probability that party $A$ uses the
measurement basis $\{\ket{(X=\true|a=1)},\ket{(X=\false|a=1)}\}$
and $\Pr(a=1,b=2) = \svec_{ab}$ is the probability that $A$ uses the
basis $\{\ket{(X=\true|a=1)},\ket{(X=\false|a=1)}\}$ while $B$ uses
the basis $\{\ket{(Y=\true|b=2)},\ket{(Y=\false|b=2)}\}$, etc.

Associated with these optimal distributions there is an optimal local
realist theory $\mvec\in\msim$ (see Section~\ref{sec:strengthdef}).
The probabilities for such optimal classical theories are listed below
as well and should be compared with the tables of the nonlocality
proofs.  Combining these data tables for each proof and each scenario
we obtain the strengths that were listed in Section~\ref{sec:results}.

\subsection{Original Bell}\label{sec:bellor}
For Bell's proof of nonlocality we have to make a distinction between
the original version, which Bell described \cite{Bell64}, and the
optimized version, which is described by Peres in \cite{Peres95}.

First we discuss Bell's original proof.
Take the bipartite state 
$\smfrac{1}{\sqrt{2}}\ket{0_A0_B} + \smfrac{1}{\sqrt{2}}\ket{1_A 1_B}$,
and the measurement settings
\begin{IEEEeqnarray*}{rClCrCl}
\ket{X=\true}_{a=1} &:= &\ket{R(0)}& \text{ and }& 
\ket{X=\true}_{a=2} &:=& \ket{R(\smfrac{\pi}{8})} \\ 
\ket{Y=\true}_{b=1} &:=& \ket{R(\smfrac{\pi}{8})}&\text{ and }& 
\ket{Y=\true}_{b=2} &:=& \ket{R(\smfrac{\pi}{4})}
\end{IEEEeqnarray*} 

With these settings, quantum mechanics predicts 
the conditional probabilities of Table~\ref{tb:orb_QP} 
(where $\smfrac{1}{4}+\smfrac{1}{8}\sqrt{2}\approx 0.4267766953$ and
$\smfrac{1}{4}-\smfrac{1}{8}\sqrt{2} \approx 0.0732233047$).

\subsubsection*{(1) Uniform Settings, Original Bell}
When the two parties use uniform distributions for their settings, 
the optimal classical theory is the one described 
in Table~\ref{tb:orb_usP}.
The corresponding KL distance is $0.0141597409$.

\subsubsection*{(2) Uncorrelated Settings, Original Bell}
The optimized, uncorrelated setting distribution 
is described in Table~\ref{tb:orb_uncsF}.
The probabilities of the best classical theory for this 
uncorrelated setting distribution are those in Table~\ref{tb:orb_uncsP}.
The KL distance for Bell's original proof, with uncorrelated
measurement settings is $0.0158003672$.

\subsubsection*{(3) Correlated Settings, Original Bell}
The optimized, correlated setting distribution 
is described in Table~\ref{tb:orb_corsF}.
The probabilities of the best classical theory for this 
distribution are described in Table~\ref{tb:orb_corsP}. 
The corresponding KL distance is $0.0169800305$.

\begin{table}
\centering
\caption{Quantum Predictions Original Bell} 
\label{tb:orb_QP}
\Qtable%
{0.4267766953 & 0.0732233047 \\ 0.0732233047  & 0.4267766953}
{0.5 & 0 \\  0 & 0.5}
{0.25 & 0.25 \\ 0.25 & 0.25}
{0.4267766953 & 0.0732233047 \\ 0.0732233047 & 0.4267766953}

\betweentable

\caption{Best Classical Theory for 
Uniform Settings Original Bell. 
KL Distance: $0.0141597409$.}
\label{tb:orb_usP}
\Ptable%
{0.3970311357 & 0.1029688643 \\ 0.1029688643 & 0.3970311357}
{0.5000000000 & 0.0000000000 \\ 0.0000000000 & 0.5000000000}
{0.2940622714 & 0.2059377286 \\ 0.2059377286 & 0.2940622714}
{0.3970311357 & 0.1029688643 \\ 0.1029688643 & 0.3970311357} 

\betweentable

\caption{Optimized Uncorrelated Setting Distribution Original Bell}
\label{tb:orb_uncsF}
\ssimuctable
{0.2316110419}{0.1327830656}{0.3643941076}
{0.4039948505}{0.2316110419}{0.6356058924}
{0.6356058924}{0.3643941076}

\betweentable

\caption{Best Classical Theory for 
Uncorrelated Settings Original Bell. 
KL Distance: $0.0158003672$.}
\label{tb:orb_uncsP}
\Ptable%
{0.3901023259 & 0.1098976741 \\ 0.1098976741 & 0.3901023259}
{0.5000000000 & 0.0000000000 \\ 0.0000000000 & 0.5000000000} 
{0.2802046519 & 0.2197953481 \\0.2197953481 & 0.2802046519}
{0.3901023259 & 0.1098976741 \\ 0.1098976741 & 0.3901023259} 
\betweentable
\caption{Optimized Correlated Setting Distribution Original Bell}
\label{tb:orb_corsF}
\ssimtable{0.2836084841}{0.1020773549}{0.3307056768}{0.2836084841}
\betweentable
\caption{Best Classical Theory for
Correlated Settings Original Bell.
KL Distance: $0.0169800305$.}
\label{tb:orb_corsP}
\Ptable%
{0.3969913979 & 0.1030086021 \\ 0.1030086021 & 0.3969913979}
{0.4941498806 & 0.0058501194 \\ 0.0058501194 & 0.4941498806}
{0.2881326764 & 0.2118673236 \\ 0.2118673236 & 0.2881326764} 
{0.3969913979 & 0.1030086021 \\ 0.1030086021 & 0.3969913979}
\end{table}

\subsection{Optimized Bell} \label{sec:bellop}
Take the bipartite state 
$\smfrac{1}{\sqrt{2}}\ket{0_A0_B} + \smfrac{1}{\sqrt{2}}\ket{1_A 1_B}$,
and the measurement settings
\begin{IEEEeqnarray*}{rClCrCl}
\ket{X=\true}_{a=1} &:=& \ket{R(0)}
& \text{ and }& 
\ket{X=\true}_{a=2} &:=& \ket{R(\smfrac{\pi}{6})}\\ 
\ket{Y=\true}_{b=1} &:=& \ket{R(0)}
& \text{ and }&
\ket{Y=\true}_{b=2} &:=& \ket{R(\smfrac{\pi}{3})}.
\end{IEEEeqnarray*} 

With these settings, quantum mechanics predicts 
the conditional probabilities of Table~\ref{tb:opb_QP}.

\subsubsection*{(1) Uniform Settings, Optimized Bell}
For the uniform setting distribution
the best classical approximation is given in Table~\ref{tb:opb_usP},
which gives a KL distance of $0.0177632822$.

\subsubsection*{(2) Uncorrelated Settings, Optimized Bell}
The optimal, uncorrelated setting distribution is 
given in Table~\ref{tb:opb_uncsF}.
The probabilities of the best classical theory for this distribution  
are those of Table~\ref{tb:opb_uncsP}.
The corresponding KL distance is $0.0191506613$.  

\subsubsection*{(3) Correlated Settings, Optimized Bell}
The optimal correlated setting distribution is 
given in Table~\ref{tb:opb_corsF}.
The probabilities of the best classical theory for this distrubtion is 
given in Table~\ref{tb:opb_corsP}. The corresponding KL distance is $0.0211293952$. 

\begin{table}
\centering
\caption{Quantum Predictions Optimized Bell}
\label{tb:opb_QP}
\Qtable%
{0.5 & 0 \\ 0 & 0.5}
{0.375 & 0.125 \\ 0.125 & 0.375}
{0.125 &  0.375 \\ 0.375 & 0.125}
{0.375 & 0.125 \\ 0.125 & 0.375}

\betweentable

\caption{Best Classical Theory for 
Uniform Settings Optimized Bell.
KL Distance: $0.0177632822$.}
\label{tb:opb_usP}
\Ptable%
{0.5000000000 & 0.0000000000 \\ 0.0000000000 & 0.5000000000}
{0.3333333333 & 0.1666666667 \\ 0.1666666667 & 0.3333333333} 
{0.1666666667 & 0.3333333333 \\ 0.3333333333 & 0.1666666667}
{0.3333333333 & 0.1666666667 \\ 0.1666666667 & 0.3333333333}

\betweentable

\caption{Optimized Uncorrelated Setting Distribution Optimized Bell} 
\label{tb:opb_uncsF}
\ssimuctable%
{0.1497077788}{0.2372131160}{0.3869208948}
{0.2372131160}{0.3758659893}{0.6130791052}
{0.3869208948}{0.6130791052}

\betweentable

\caption{Best Classical Theory for 
Uncorrelated Settings Optimized Bell.
KL Distance: $0.0191506613$.}
\label{tb:opb_uncsP}
\Ptable%
{0.5000000000 & 0.0000000000 \\ 0.0000000000 & 0.5000000000} 
{0.3267978563 & 0.1732021436 \\ 0.1732021436 & 0.3267978563}
{0.1732021436 & 0.3267978563 \\ 0.3267978563 & 0.1732021436}
{0.3464042873 & 0.1535957127 \\ 0.1535957127 & 0.3464042873} 

\betweentable

\caption{Optimized Correlated Setting Distribution Optimized Bell}
\label{tb:opb_corsF}
\ssimtable{0.1046493146}{0.2984502285}{0.2984502285}{0.2984502285}

\betweentable

\caption{Best Classical Theory for 
Correlated Settings Optimized Bell.
KL Distance: $0.0211293952$.}
\label{tb:opb_corsP}
\Ptable%
{0.4927305107 & 0.0072694892 \\ 0.0072694892 & 0.4927305107}
{0.3357564964 & 0.1642435036 \\ 0.1642435036 & 0.3357564964}
{0.1642435036 & 0.3357564964 \\ 0.3357564964 & 0.1642435036}
{0.3357564964 & 0.1642435036 \\ 0.1642435036 & 0.3357564964} 
\end{table}

\subsection{CHSH}\label{app:CHSH}
The bipartite state $\smfrac{1}{\sqrt{2}}\ket{0_A0_B} +
\smfrac{1}{\sqrt{2}}\ket{1_A 1_B}$.
$A$'s and $B$'s measurement settings are:
\begin{IEEEeqnarray}{rClCrCl}
\ket{X=\true}_{a=1} &:=& \ket{R(0)} &\mbox{ and }& 
\ket{X=\true}_{a=2} &:=& \ket{R(\smfrac{\pi}{4})}, \\ 
\ket{Y=\true}_{b=1} &:=& \ket{R(\smfrac{\pi}{8})} &\mbox{ and }& 
\ket{Y=\true}_{b=2} &:=& \ket{R(-\smfrac{\pi}{8})}.
\end{IEEEeqnarray}
With these settings, quantum mechanics predicts 
the conditional probabilities of Table~\ref{tb:chsh_QP}
(with $\smfrac{1}{4}+\smfrac{1}{8}\sqrt{2}\approx 0.4267766953$ and
$\smfrac{1}{4}-\smfrac{1}{8}\sqrt{2} \approx 0.0732233047$).

\subsubsection*{Uniform, Uncorrelated and Correlated Settings, \CHSH}
The optimal measurement settings is the uniform settings, where both
$A$ and $B$ perform uses one of the two measurements with probability
$0.5$ (that is $\svec_{ab} = 0.25$)

The optimal classical theory in this scenario has the probabilities
of Table~\ref{tb:chsh_usP}.
\begin{table}
\centering
\caption{Quantum Predictions CHSH} 
\label{tb:chsh_QP}
\Qtable%
{0.4267766953 & 0.0732233047 \\ 0.0732233047 & 0.4267766953} 
{0.4267766953 & 0.0732233047 \\ 0.0732233047 & 0.4267766953}
{0.4267766953 & 0.0732233047 \\ 0.0732233047 & 0.4267766953}
{0.0732233047 & 0.4267766953 \\ 0.4267766953 & 0.0732233047}
\betweentable
\caption{Best Classical Theory for 
Uniform Settings CHSH. 
KL Distance: $0.0462738469$.}
\label{tb:chsh_usP}
\Ptable%
{0.375 & 0.125 \\ 0.125 & 0.375} 
{0.375 & 0.125 \\ 0.125 & 0.375} 
{0.375 & 0.125 \\ 0.125 & 0.375} 
{0.125 & 0.375 \\ 0.375 & 0.125} 
\end{table}

\subsection{Hardy}\label{sec:hardy}
The bipartite state $\alpha\ket{0_A0_B} -\beta\ket{1_A 1_B}$,
with $\alpha := \smfrac{1}{2}\sqrt{2+2\sqrt{-13+6\sqrt{5}}} \approx
0.907$ and $\beta := \sqrt{1-\alpha^2} \approx 0.421$ 
(such that indeed $\alpha^2+\beta^2=1$).  
$A$'s and
$B$'s measurement settings are now identical and given by:
\begin{IEEEeqnarray}{rCcCl}
\ket{X=\true}_{a=1} &=& 
\ket{Y=\true}_{b=1} & := & 
\sqrt{\frac{\beta}{\alpha+\beta}}\ket{0} + \sqrt{\frac{\alpha}{\alpha+\beta}}\ket{1},\\ 
\ket{X=\true}_{a=2} &=& 
\ket{Y=\true}_{b=2}  & := & 
-\sqrt{\frac{\beta^3}{\alpha^3+\beta^3}}%
\ket{0}+\sqrt{\frac{\alpha^3}{\alpha^3+\beta^3}}\ket{1}.
\end{IEEEeqnarray}
With these settings, quantum mechanics predicts 
the conditional probabilities of Table~\ref{tb:hardy_QP}.

\subsubsection*{(1) Uniform Settings, Hardy}
For uniform measurement settings, the best classical theory to
describe the quantum mechanical statistics is given in 
Table~\ref{tb:hardy_usP},  with KL divergence: $0.0278585182$.

\subsubsection*{(2) Uncorrelated Settings, Hardy}
The optimized uncorrelated setting distribution is given in 
Table~\ref{tb:hardy_uncsF}.
The probabilities of the best classical theory for this distribution 
are in Table~\ref{tb:hardy_uncsP}.
The corresponding KL distance is $0.0279816333$.

\subsubsection*{(3) Correlated Settings, Hardy}
The optimized correlated setting distribution is given in 
Table~\ref{tb:hardy_corsF}.
The probabilities of the best classical theory for this distribution  
 are described in Table~\ref{tb:hardy_corsP}.
The corresponding KL distance is $0.0280347655$.

\begin{table}
\centering
\caption{Quantum Predictions Hardy} 
\label{tb:hardy_QP}
\Qtable%
{0 & 0.38196601125 \\ 0.38196601125 & 0.23606797750} 
{0.23606797750 & 0.14589803375 \\  0 & 0.61803398875} 
{0.23606797750 & 0 \\ 0.14589803375 & 0.61803398875} 
{0.09016994375 & 0.14589803375 \\ 0.14589803375 & 0.61803398875}

\betweentable

\caption{Best Classical Theory for 
Uniform Settings Hardy. 
KL Distance: $0.0278585182$.}
\label{tb:hardy_usP}
\Ptable%
{0.0338829434 & 0.3543640363 \\ 0.3543640363 & 0.2573889840} 
{0.2190090188 & 0.1692379609 \\ 0.0075052045 & 0.6042478158}
{0.2190090188 & 0.0075052045 \\ 0.1692379609 & 0.6042478158}
{0.0488933524 & 0.1776208709 \\ 0.1776208709 & 0.5958649058} 

\betweentable

\caption{Optimized Uncorrelated Setting Distribution Hardy}
\label{tb:hardy_uncsF}
\ssimuctable%
{0.2603092699}{0.2498958554}{0.5102051253} 
{0.2498958554}{0.2398990193}{0.4897948747}
{0.5102051253}{0.4897948747}

\betweentable

\caption{Best Classical Theory for 
Uncorrelated Settings Hardy. 
KL Distance: $0.0279816333$.}
\label{tb:hardy_uncsP}
\Ptable%
{0.0198831449 & 0.3612213769 \\ 0.3612213769 & 0.2576741013} 
{0.2143180373 & 0.1667864844 \\ 0.0141212511 & 0.6047742271}
{0.2143180373 & 0.0141212511 \\ 0.1667864844 & 0.6047742271}
{0.0481256471 & 0.1803136414 \\ 0.1803136414 & 0.5912470702}

\betweentable

\caption{Optimized Correlated Setting Distribution Hardy}
\label{tb:hardy_corsF}
\ssimtable{0.2562288294}{0.2431695652}{0.2431695652}{0.2574320402}

\betweentable

\caption{Best Classical Theory for 
Correlated Settings Hardy. 
KL Distance: $0.0280347655$.}
\label{tb:hardy_corsP}
\Ptable%
{0.0173443545 & 0.3620376608 \\ 0.3620376608 & 0.2585803238} 
{0.2123471649 & 0.1670348504 \\ 0.0165954828 & 0.6040225019} 
{0.2123471649 & 0.0165954828 \\ 0.1670348504 & 0.6040225019}
{0.0505353201 & 0.1784073276 \\ 0.1784073276 & 0.5926500247} 
\end{table}

\subsection{Mermin}\label{app:mermin}
In \cite{Mermin81}, we find the following nonlocality proof with two
parties, three measurement settings, and two possible outcomes.  Let
the entangled state be
$\smfrac{1}{\sqrt{2}}(\ket{0_A0_B}+\ket{1_A1_B})$, and the
measurement settings:
\begin{IEEEeqnarray*}{rCcCl}
\ket{X=\true}_{a=1}&=&\ket{Y=\true}_{b=1} & := & \ket{0},\\ 
\ket{X=\true}_{a=2}&=&\ket{Y=\true}_{b=2} & :=  & \ket{R(\smfrac{2}{3}\pi)},\\ 
\ket{X=\true}_{a=3}&=&\ket{Y=\true}_{b=3} & := &  \ket{R(\smfrac{4}{3}\pi)}.
\end{IEEEeqnarray*}
With these settings, quantum mechanics predicts the conditional 
probabilities of Table~\ref{tb:mermin_QP}.

\subsubsection*{(1) Uniform Settings, Mermin}
The probabilities of the best classical theory for the
uniform measurement settings is give  in Table~\ref{tb:mermin_usP}.
\subsubsection*{(2) Uncorrelated Settings, Mermin}
The optimal uncorrelated setting distribution is given in 
Table~\ref{tb:mermin_uncsF}.
The probabilities of the best classical theory for this distribution 
is in Table~\ref{tb:mermin_uncsP}.
\subsubsection*{(3) Correlated Settings, Mermin}
The optimal correlated setting distribution is given in 
Table~\ref{tb:mermin_corsF} (note that there are also other optimal 
distributions).
The probabilities of the best classical theory for this specific distribution 
are described in Table~\ref{tb:mermin_corsP}.
The corresponding KL distance is $0.0211293952$.

\begin{table}
\centering
\caption{Quantum Predictions Mermin}
\label{tb:mermin_QP}
\MerminQtable%
{0.5 & 0 \\ 0 & 0.5}
{0.125 & 0.375 \\ 0.375 & 0.125}
{0.125 & 0.375 \\ 0.375 & 0.125}
{0.125 & 0.375 \\ 0.375 & 0.125}  
{0.5 & 0 \\ 0 & 0.5}
{0.125 & 0.375 \\ 0.375 & 0.125}
{0.125 & 0.375 \\ 0.375 & 0.125}  
{0.125 & 0.375 \\ 0.375 & 0.125}
{0.5 & 0 \\ 0 & 0.5}

\betweentable

\caption{Best Classical Theory for 
Uniform Settings Mermin. 
KL distance: $0.0157895843$.}
\label{tb:mermin_usP}
\MerminPtable%
{0.50000 & 0.00000 \\ 0.00000 & 0.50000}
{0.16667 & 0.33333 \\ 0.33333 & 0.16667}
{0.16667 & 0.33333 \\ 0.33333 & 0.16667} 
{0.16667 & 0.33333 \\ 0.33333 & 0.16667}
{0.50000 & 0.00000 \\ 0.00000 & 0.50000} 
{0.16667 & 0.33333 \\ 0.33333 & 0.16667}
{0.16667 & 0.33333 \\ 0.33333 & 0.16667}  
{0.16667 & 0.33333 \\ 0.33333 & 0.16667}
{0.50000 & 0.00000 \\ 0.00000 & 0.50000}

\betweentable

\caption{Optimized Uncorrelated Setting Distribution Mermin}
\label{tb:mermin_uncsF}
\begin{tabular}{v||m|m|m||m}
\svec_{ab}\in\ssimuc & a=1 & a=2 & a=3 & \Pr(B=b)\\ \hhline{=::===::=}
b=1 & 0.1497077711 & 0 & 0.2372131137 & 0.3869208848 \\ \hhline{-||---||-}
b=2 & 0.2372131137 & 0 & 0.3758660015 & 0.6130791152 \\ \hhline{-||---||-}
b=3 & 0 & 0 & 0 & 0 \\ \hhline{=::===::=} 
\Pr(A=a) & 0.3869208848 & 0 & 0.6130791152 
\end{tabular}

\betweentable

\caption{Best Classical Theory for 
Uncorrelated Settings Mermin. 
KL Distance: $0.0191506613$.}
\label{tb:mermin_uncsP}
\MerminPtable%
{0.50000 & 0.00000 \\ 0.00000 & 0.50000} 
{0.50000 & 0.00000 \\ 0.50000 & 0.00000}
{0.17320 & 0.32680 \\ 0.32680 & 0.17320}
{0.17320 & 0.32680 \\ 0.32680 & 0.17320} 
{0.50000 & 0.00000 \\ 0.50000 & 0.00000}
{0.15360 & 0.34640 \\ 0.34640 & 0.15360}
{0.50000 & 0.50000 \\ 0.00000 & 0.00000}  
{1.00000 & 0.00000 \\ 0.00000 & 0.00000}
{0.50000 & 0.50000 \\ 0.00000 & 0.00000}

\betweentable

\caption{Optimized Correlated Setting Distribution Mermin}
\label{tb:mermin_corsF}
\begin{tabular}{v||m|m|m|}
\Pr(A=a,B=b)=\svec_{ab}\in\ssim & a=1 & a=2 & a=3 \\ \hhline{=::===}
b=1 & 0.1046493071 & 0 & 0.2984502310 \\ \hhline{-||---}
b=2 & 0.2984502310 & 0 & 0.2984502310\\ \hhline{-||---}
b=3 & 0 & 0 & 0 \\ \hhline{-||---}
\end{tabular}

\betweentable

\caption{Best Classical Theory for 
Correlated Settings Mermin. 
KL Distance: $0.0211293952$.}
\label{tb:mermin_corsP}
\MerminPtable%
{0.49273 & 0.00727 \\ 0.00727 & 0.49273}
{0.50000 & 0.00000 \\ 0.50000 & 0.00000}
{0.16424 & 0.33576 \\ 0.33576 & 0.16424}
{0.16424 & 0.33576 \\ 0.33576 & 0.16424} 
{0.50000 & 0.00000 \\ 0.50000 & 0.00000} 
{0.16424 & 0.33576 \\ 0.33576 & 0.16424}
{0.50000 & 0.50000 \\ 0.00000 & 0.00000}
{1.00000 & 0.00000 \\ 0.00000 & 0.00000}
{0.50000 & 0.50000 \\ 0.00000 & 0.00000}
\end{table}

\subsection{GHZ}\label{app:ghz}
The tripartite state
$\smfrac{1}{\sqrt{2}}\ket{0_A0_B0_C}+\smfrac{1}{\sqrt{2}}\ket{1_A1_B1_C}$.
The settings for all three parties are identical:
\begin{IEEEeqnarray}{rCcCcCl}
\ket{X=\true}_{a=1} &=& \ket{Y=\true}_{b=1} &=& \ket{Z=\true}_{c=1} & := & 
\smfrac{1}{\sqrt{2}}\ket{0}+\smfrac{1}{\sqrt{2}}\ket{1},\\ 
\ket{X=\true}_{a=2} &=& \ket{Y=\true}_{b=2} &=& \ket{Z=\true}_{c=2} & := & 
\smfrac{1}{\sqrt{2}}\ket{0}+\smfrac{\mathrm{i}}{\sqrt{2}}\ket{1}.
\end{IEEEeqnarray}
With these settings, quantum mechanics predicts 
the conditional probabilities of Table~\ref{tb:ghz_QP}.

\subsubsection*{(1) Uniform and Uncorrelated Settings, GHZ}
For all three settings, the best possible classical statistics that 
approximate the \GHZ experiment is that of Table~\ref{tb:ghz_usP}.
The optimal uncorrelated setting is the uniform settings that samples
all eight measurement settings with equal probability.  The
corresponding KL divergence is: $0.2075187496$.

\subsubsection*{(2) Correlated Settings, GHZ}
The optimal correlated setting samples with equal probability those
four settings that yield the $(0.125,0)$ outcome probabilities (those
are the settings where an even number of the measurements are measuring
along the $m_1$ axis).  The KL divergence in this setting is twice
that of the previous uniform setting: $0.4150374993$.

\begin{table}
\centering
\caption{Quantum Predictions GHZ}
\label{tb:ghz_QP}
\newlength{\ghzentrywidth}
\setlength{\ghzentrywidth}{3cm}
{\newlength{\hshlength}
\setlength{\hshlength}{\tabcolsep} 
\newcommand{\hsh}{\hspace*{\hshlength}}
\setlength{\tabcolsep}{0pt}
\begin{tabular}{>{$}r<{$}||>{$}c<{$}|>{$}c<{$}|}
 & a=1 & a=2 \\
\hsh Q_{abc}(X=x,Y=y,Z=z)\hsh & 
\begin{tabularx}{\ghzentrywidth}{MM} x=\true & x=\false \end{tabularx} & 
\begin{tabularx}{\ghzentrywidth}{MM} x=\true & x=\false \end{tabularx} \\\hhline{=::==}
\hsh c=1\hsh \begin{tabular}{>{$}c<{$}} 
   \hsh b=1 \hsh \begin{tabular}{>{$}c<{$}} 
         \hsh z=\true \hsh\begin{tabular}{>{$}c<{$}} 
             \hsh y=\true\hsh\\
             \hsh y=\false\hsh 
             \end{tabular} \\ 
         \hsh z=\false\hsh \begin{tabular}{>{$}c<{$}} 
           \hsh y=\true\hsh \\
            \hsh y=\false\hsh
             \end{tabular}
           \end{tabular} \\ \hline
   \hsh b=2 \hsh \begin{tabular}{>{$}c<{$}} 
        \hsh z=\true\hsh \begin{tabular}{>{$}c<{$}} 
            \hsh y=\true\hsh \\
            \hsh y=\false\hsh
             \end{tabular} \\ 
        \hsh z=\false\hsh \begin{tabular}{>{$}c<{$}} 
            \hsh y=\true\hsh \\
            \hsh y=\false\hsh
             \end{tabular}
           \end{tabular} 
      \end{tabular} & 
\begin{tabularx}{\ghzentrywidth}{MM} 
0.25 & 0 \\
0    & 0.25 \\ 
0    & 0.25 \\
0.25 & 0 \\ \hline 
0.125 & 0.125 \\
0.125 & 0.125 \\ 
0.125 & 0.125 \\
0.125 & 0.125 \\  
\end{tabularx} 
& 
\begin{tabularx}{\ghzentrywidth}{MM} 
0.125  & 0.125  \\
0.125  & 0.125  \\ 
0.125  & 0.125  \\
0.125  & 0.125  \\ \hline 
0 & 0.25 \\
0.25 & 0 \\ 
0.25 & 0 \\
0 & 0.25 \\  
\end{tabularx} 
\\ \hhline{-||--}
\hsh c=2 \hsh \begin{tabular}{>{$}c<{$}} 
   \hsh b=1\hsh \begin{tabular}{>{$}c<{$}} 
        \hsh z=\true\hsh \begin{tabular}{>{$}c<{$}} 
            \hsh y=\true\hsh \\
            \hsh y=\false\hsh
             \end{tabular} \\ 
        \hsh z=\false\hsh \begin{tabular}{>{$}c<{$}} 
            \hsh y=\true\hsh \\
            \hsh y=\false\hsh
             \end{tabular}
           \end{tabular} \\ \hline
   \hsh b=2\hsh \begin{tabular}{>{$}c<{$}} 
         \hsh z=\true\hsh \begin{tabular}{>{$}c<{$}} 
            \hsh y=\true\hsh \\
            \hsh y=\false\hsh
             \end{tabular} \\ 
         \hsh z=\false\hsh \begin{tabular}{>{$}c<{$}} 
            \hsh y=\true\hsh \\
            \hsh y=\false\hsh
             \end{tabular}
           \end{tabular} 
      \end{tabular} & 
\begin{tabularx}{\ghzentrywidth}{MM} 
0.125 & 0.125 \\
0.125 & 0.125 \\ 
0.125 & 0.125 \\
0.125 & 0.125 \\ \hline 
0 & 0.25 \\
0.25 & 0 \\ 
0.25 & 0 \\
0 & 0.25 \\  
\end{tabularx} 
& 
\begin{tabularx}{\ghzentrywidth}{MM} 
0 & 0.25 \\
0.25 & 0 \\ 
0.25 & 0 \\
0 & 0.25 \\ \hline 
0.125 & 0.125 \\
0.125 & 0.125 \\ 
0.125 & 0.125 \\
0.125 & 0.125\\  
\end{tabularx} 
\\ \hhline{-||--}
\end{tabular}}

\betweentable

\caption{Best Classical Theory for
Uniform Settings GHZ.
KL Distance: $0.2075187496$.}
\label{tb:ghz_usP}
\setlength{\ghzentrywidth}{3cm}
{%
\setlength{\hshlength}{\tabcolsep} 
\newcommand{\hsh}{\hspace*{\hshlength}}
\setlength{\tabcolsep}{0pt}
\begin{tabular}{>{$}r<{$}||m|m|}
 & a=1 & a=2 \\
\hsh P_{abc}(X=x,Y=y,Z=z) \hsh & 
\begin{tabularx}{\ghzentrywidth}{MM} x=\true & x=\false \end{tabularx} & 
\begin{tabularx}{\ghzentrywidth}{MM} x=\true & x=\false \end{tabularx} \\\hhline{=::==}
\hsh c=1\hsh  \begin{tabular}{>{$}c<{$}} 
   \hsh b=1\hsh  \begin{tabular}{>{$}c<{$}} 
         \hsh z=\true\hsh  \begin{tabular}{>{$}c<{$}} 
             \hsh y=\true\hsh  \\
             \hsh y=\false\hsh  
             \end{tabular} \\ 
         \hsh z=\false\hsh  \begin{tabular}{>{$}c<{$}} 
             \hsh y=\true\hsh  \\
             \hsh y=\false\hsh 
             \end{tabular}
           \end{tabular} \\ \hline
   \hsh b=2\hsh  \begin{tabular}{>{$}c<{$}} 
         \hsh z=\true\hsh  \begin{tabular}{>{$}c<{$}} 
             \hsh y=\true\hsh  \\
             \hsh y=\false\hsh 
             \end{tabular} \\ 
         \hsh z=\false\hsh  \begin{tabular}{>{$}c<{$}} 
             \hsh y=\true\hsh  \\
             \hsh y=\false\hsh 
             \end{tabular}
           \end{tabular} 
      \end{tabular} & 
\begin{tabularx}{\ghzentrywidth}{MM} 
0.1875 & 0.0625 \\
0.0625   & 0.1875 \\ 
0.0625   & 0.1875 \\
0.1875 & 0.0625 \\ \hline 
0.125 & 0.125 \\
0.125 & 0.125 \\ 
0.125 & 0.125 \\
0.125 & 0.125 \\  
\end{tabularx} 
& 
\begin{tabularx}{\ghzentrywidth}{MM} 
0.125  & 0.125  \\
0.125  & 0.125  \\ 
0.125  & 0.125  \\
0.125  & 0.125  \\ \hline 
0.0625& 0.1875 \\
0.1875 & 0.0625 \\ 
0.1875 & 0.0625 \\
0.0625& 0.1875 \\  
\end{tabularx} \\ \hhline{-||--}
\hsh c=2\hsh   \begin{tabular}{>{$}c<{$}} 
   \hsh b=1\hsh  \begin{tabular}{>{$}c<{$}} 
         \hsh z=\true\hsh  \begin{tabular}{>{$}c<{$}} 
             \hsh y=\true\hsh  \\
             \hsh y=\false\hsh 
             \end{tabular} \\ 
         \hsh z=\false\hsh   \begin{tabular}{>{$}c<{$}} 
             \hsh y=\true\hsh  \\
             \hsh y=\false\hsh 
             \end{tabular}
           \end{tabular} \\ \hline
   \hsh b=2\hsh  \begin{tabular}{>{$}c<{$}} 
         \hsh z=\true\hsh  \begin{tabular}{>{$}c<{$}} 
             \hsh y=\true\hsh  \\
             \hsh y=\false\hsh 
             \end{tabular} \\ 
         \hsh z=\false\hsh  \begin{tabular}{>{$}c<{$}} 
             \hsh y=\true\hsh  \\
             \hsh y=\false\hsh 
             \end{tabular}
           \end{tabular} 
      \end{tabular} & 
\begin{tabularx}{\ghzentrywidth}{MM} 
0.125 & 0.125 \\
0.125 & 0.125 \\ 
0.125 & 0.125 \\
0.125 & 0.125 \\ \hline 
0.0625& 0.1875 \\
0.1875 & 0.0625 \\ 
0.1875 & 0.0625 \\
0.0625& 0.1875 \\  
\end{tabularx} 
& 
\begin{tabularx}{\ghzentrywidth}{MM} 
0.0625& 0.1875 \\
0.1875 & 0.0625 \\ 
0.1875 & 0.0625 \\
0.0625& 0.1875 \\ \hline 
0.125 & 0.125 \\
0.125 & 0.125 \\ 
0.125 & 0.125 \\
0.125 & 0.125\\  
\end{tabularx} 
\\ \hhline{-||--}
\end{tabular}}
\end{table}

\section{The Kullback-Leibler Divergence} \label{app:kl}
This appendix provides in-depth information about the Kullback-Leiber
divergence and its relation to statistical
strength. Appendix~\ref{app:formal} gives the formal connection
between KL divergence and statistical strength.
Appendix~\ref{app:klproperties} discusses some general properties of
KL divergence. Appendix~\ref{app:klvariation} compares it to variation
distance.  Appendix~\ref{app:klintuition} informally explains why KL
divergence is related to statistical strength.
\subsection{Formal Connection between KL Divergence and Statistical Strength}
\label{app:formal}
We consider three methods for statistical hypothesis testing:
\emph{frequentist} hypothesis testing \cite{Rice95}, \emph{Bayesian}
hypothesis \cite{Lee97} testing and \emph{information-theoretic}
hypothesis testing \cite{LiV97,Rissanen89}.  Nearly all
state-of-the-art, theoretically motivated statistical methodology
falls in either the Bayesian or the frequentist
categories. Frequentist hypothesis testing is the most common, the
most taught in statistics classes and is the standard method in, for
example, the medical sciences. Bayesian hypothesis testing is becoming
more and more popular in, for example, econometrics and biological
applications.  While theoretically important, the
information-theoretic methods are less used in practice and are
discussed
mainly because they lead to a very concrete interpretation of
statistical strength in terms of \emph{bits} of information.

We illustrate below that in all three approaches the KL divergence
indeed captures the notion of `statistical strength'.
We consider the general situation with a sample $Z_1, Z_2,\dots$, with
the $Z_i$ independently and identically distributed according
to some $Q_{\svec}$, $Q_{\svec}$ being some distribution over some
finite set $\set{Z}$.  For each $n$, the first $n$ outcomes are
distributed according to the $n$-fold product distribution of
$Q_{\svec}$, which we shall also refer to as $Q_{\svec}$. Hence
$Q_{\svec}(z_1,\dots, z_n) = \prod_{i=1}^n Q_{\svec}(z_i)$. The
independence assumption also induces a distribution over the set
$\set{Z}^{\infty}$ of all infinite sequences\footnote{Readers
familiar with measure theory should note that throughout this paper,
we tacitly assume that $\set{Z}^{\infty}$ is endowed with a suitable
$\svec$-algebra such that all sets mentioned in this paper become
measurable.}  which we shall \emph{also} refer to as $Q_{\svec}$.

 We test $Q_{\svec}$ against a set of distributions $\set{P}_{\svec}$. 
Thus, $Q_{\svec}$ and $\set{P}_{\svec}$ may, but do not
necessarily refer to quantum and local realist theories --- the
statements below hold more generally.

\subsubsection{Frequentist Justification} \label{ssec:frequentist}
In frequentist hypothesis testing, $\set{P}_{\svec}$ is called the
\emph{null-hypothesis} and $Q_{\svec}$ the alternative hypothesis.
Frequentist hypothesis testing can be implemented in a number of
different ways, depending on what \emph{statistical test} one adopts. A
statistical test is a procedure that, when input an arbitrary
sample of arbitrary length, outputs a \emph{decision}. The decision is
either `$Q_{\svec}$ generated the data' or `$\set{P}_{\svec}$
generated the data'. Each test is defined relative to some {\em test
  statistic $T$\/} and {\em critical value\/} $c$. A test statistic
$T$ is a function defined on samples of arbitrary length, that for
each sample outputs a real number. Intuitively, large values of the
test statistic indicate that something has happened which is much more
unlikely under any of the distributions in the null-hypothesis than
under the alternative hypothesis. A function that is often used as a test
statistic is the {\em log-likelihood ratio\/} 
\begin{equation}
T(z_1, \ldots, z_n) := 
\frac{Q_{\svec}(z_1, \ldots, z_n)}{\sup_{P \in \set{P}_{\svec}} P(z_1,
    \ldots, z_n)},
\end{equation}
but many other choices are possible as well.

The critical value $c$ determines
the threshold for the test's decision: if, for the observed data $z_1,
\ldots, z_n$, it holds that $T(z_1, \ldots, z_n) \geq c$, the test
says `$Q_{\svec}$ generates the data'; if $T(z_1, \ldots, z_n) < c$,
the test says `$\set{P}_{\svec}$ generated the data'.

The confidence in a given decision
is determined by a quantity known as \emph{the $p$-value}. This is a
function of the data that was actually observed in the statistical
experiment. 
It only depends on the 
observed value of the test statistic $t_{\text{observed}}
:=  T(z_1,\dots, z_n)$. 
It is defined as
\begin{IEEEeqnarray}{rCl}
\label{eq:bijna}
\freqconf 
& := & 
\sup_{P \in \set{P}_{\svec}} P (T(Z_1, \ldots, Z_n) \geq t_{\text{observed}} ). 
\end{IEEEeqnarray}
Here the $Z_1, \ldots, Z_n$ are  distributed according to  $P$
and thus do not refer to the data that was actually observed in the experiment. Thus, the $p$-value is the
maximum probability, under any distribution in $\set{P}_{\svec}$, that
the  test statistic takes on a value that is at least as extreme as 
its actually observed outcome. Typically, the test is defined such
that the critical value $c$ depends on sample size $n$. It is
set to the value $c_0$ such that the test outputs `$Q_{\svec}$' iff the
$p$-value is smaller than some pre-defined {\em significance level},
typically $0.05$.

Large $p$-values mean small confidence: for example, suppose the test
outputs $Q_{\svec}$ whenever the $p$-value is smaller than
$0.05$. Suppose further that data
are observed with a $p$-value of $0.04$. Then the test says
``$Q_{\svec}$'' but since the $p$-value
is large, this is not that convincing to someone who
considers the possibility that some $P \in \set{P}_{\svec}$ has
generated the data: the large $p$-value indicates that the test may
very well have given the wrong answer. On the other hand, if data are
observed with a $p$-value of $0.001$, this gives a lot more confidence
in the decision output by the test.

We call a test statistic asymptotically optimal for identifying $Q_{\svec}$ if,
under the assumption that $Q_{\svec}$ generated the data, the
$p$-value goes to $0$ at the fastest possible rate. Now let us assume
that $Q_{\svec}$ generates the data, and an optimal test is used. A
well-known result due to Bahadur \cite[Theorem 1]{Bahadur67} says
that, under some regularity conditions on $Q_{\svec}$ and 
$\set{P}_{\svec}$, with $Q_{\svec}$-probability $1$, for all large $n$, 
\begin{IEEEeqnarray}{rCl}\label{eq:freqboem}
\freqconf 
&= & 
2^{-n D(Q_{\svec} \| \set{P}_{\svec}) + o(n)}.
\end{IEEEeqnarray}
where $\lim_{n \rightarrow \infty } o(n)/n = 0$.  We say `the
$p$-value is determined, \emph{to first order in the exponent}, by
$D(Q_{\svec} \| \set{P}_{\svec})$'. Note that what we called the
`actually observed test statistic $t_{\text{observed}}$' in
(\ref{eq:bijna}) has become a random variable in (\ref{eq:freqboem}),
distributed according to $Q_{\svec}$. It turns out that the regularity
conditions, needed for Equation~(\ref{eq:freqboem}) to hold, apply when
$Q_{\svec}$ is instantiated to a quantum theory $Q$ with measurement
setting distributions $\svec$, and $\set{P}_{\svec}$ is instantiated
to the corresponding set of LR theories as defined in
Section~\ref{sec:formal}.

Now imagine that QM, who knows that $Q_{\svec}$ generates the data,
wonders whether to use the experimental setup corresponding to
$\svec_1$ or $\svec_2$. Suppose that $D(Q_{\svec_1} \| \set{P}_{\svec_1}) 
> D(Q_{\svec_1} \| \set{P}_{\svec_2})$. It follows from
Equation~(\ref{eq:freqboem}) that if the experiment corresponding to
$\svec_1$ is performed, the $p$-value will go to $0$ exponentially
faster (in the number of trials) than if the experiment corresponding
to $\svec_2$ is performed. It therefore makes sense to say that `the
statistical strength of the experiment corresponding to $\svec_1$ is
larger than the strength of $\svec_2$'. This provides a frequentist
justification of adopting $D(Q_{\svec} \| \set{P}_{\svec})$ as an
indicator of statistical strength.
\paragraph*{Remark}
Bahadur \cite[Theorem 2]{Bahadur67} also provides a variation of
  Equation~(\ref{eq:freqboem}), which (roughly speaking) says the
  following: suppose $Q_{\svec}$ generates the data. For $\epsilon >
  0$, let $N_{\epsilon}$ be the \emph{minimum number of observations}
  such that, for all $n \geq N_{\epsilon}$, the test rejects 
$\set{P}_{\svec}$ (if $\set{P}_{\svec}$ is not rejected for infinitely
  many $n$, then $N_{\epsilon}$ is defined to be infinite). Suppose
  that an optimal (in the sense we used previously) test is
  used. Then, for small $\epsilon$, $N_{\epsilon}$ is inversely
  proportional to $D(Q_{\svec} \| \set{P}_{\svec})$: with
  $Q_{\svec}$-probability $1$, the smaller $D(Q_{\svec} \| 
  \set{P}_{\svec})$, the larger $N_{\epsilon}$. If a `non-optimal' test is
  used, then $N_{\epsilon}$ can only be larger, never smaller.

The rate at which the $p$-value of a test converges to $0$ is known in
statistics as \emph{Bahadur efficiency}. For an overview of the area,
see \cite{GroeneboomO77}. For an easy introduction to the main ideas,
focusing on `Stein's lemma' (a theorem related to Bahadur's), see
\cite[Chapter 12, Section 8]{BarronC91}.  For an introduction to
Stein's lemma with a physicist audience in mind, see
\cite{Balasubramanian96}.
\subsubsection{Bayesian Justification}
In the Bayesian approach to hypothesis testing
\cite{BernardoS94,Lee97}, when testing $Q_{\svec}$ against 
$\set{P}_{\svec}$, we must first determine an \emph{a priori probability
distribution} over $Q_{\svec}$ and $\set{P}_{\svec}$. This
distribution over distributions is usually just called `the prior'. It
can be interpreted as indicating the prior (i.e., before seeing the
data) `degree of belief' in $Q_{\svec}$ vs.\ $\set{P}_{\svec}$. It is
often used to incorporate prior knowledge into the statistical
decision process.  In order to set up the test as fairly as possible,
QM and LR may agree to use the prior $\Pr(Q_{\svec}) = 
\Pr(\set{P}_{\svec}) = 1/2$ (this should be read as `the prior probability that
$Q_{\svec}$ obtains is equal to the prior probability that some $P \in
\set{P}_{\svec}$ obtains').  Yet as long as $\Pr(Q_{\svec})>0$ and
there is a smooth and positive probability density for
$P_{\svec}\in\set{P}_{\svec}$, the specific values for the priors will be
irrelevant for the result below.

For given prior probabilities and a given sample $z_1,\dots, z_n$,
Bayesian statistics provides a method to compute the \emph{posterior
probabilities} of the two hypotheses, \emph{conditioned} on the
observed data: $\Pr(Q_{\svec})$ is transformed into $\Pr(Q_{\svec} \mid
z_1,\dots, z_n)$. Similarly, $\Pr(\set{P}_{\svec})$ is transformed to
$\Pr(\set{P}_{\svec} \mid z_1,\dots, z_n)$.  One then adopts the
hypothesis $H \in \{ Q_{\svec}, \set{P}_{\svec} \}$ with the larger
posterior probability $\Pr(H \mid z_1,\dots, z_n)$.  The confidence in
this decision is given by the \emph{posterior odds} of $Q_{\svec}$
against $\set{P}_{\svec}$, defined, for given sample $z_1,\dots,
z_n$, as
\begin{IEEEeqnarray}{rCl}
\bayesconf(Q_{\svec},\set{P}_{\svec})
&:=& 
\frac{\Pr(Q_{\svec} \mid z_1,\dots, z_n)}{\Pr(\set{P}_{\svec} \mid z_1,\dots, z_n)}.
\end{IEEEeqnarray}
The larger $\bayesconf$, the larger the confidence. Now suppose that
data are distributed according to $Q_{\svec}$. It can be shown that,
under some regularity conditions on $Q_{\svec}$ and $\set{P}_{\svec}$, 
with $Q_{\svec}$-probability $1$,
\begin{IEEEeqnarray}{rCl}\label{eq:bayesconf}
\bayesconf 
&=&
2^{n D(Q_{\svec} \| \set{P}_{\svec}) + O(\log n)},
\end{IEEEeqnarray}
In our previously introduced terminology, `the Bayesian confidence
(posterior odds) is determined by $(Q_{\svec} \| \set{P}_{\svec})$,
up to first order in the exponent'. We may now reason exactly as in
the frequentist case to conclude that it makes sense to adopt
$D(Q_{\svec} \| \set{P}_{\svec})$ as an indicator of statistical
strength, and that it makes sense for QM to choose the setting
probabilities $\svec$ so as to maximize $D(Q_{\svec} \| \set{P}_{\svec})$.

Equation~(\ref{eq:bayesconf}) is a `folklore result' which `usually'
holds. In Appendix~\ref{app:bayes}, we show that it does indeed holds
with $Q_{\svec}$ and $\set{P}_{\svec}$ defined as nonlocality proofs
and local realist theories, respectively.

\subsubsection{Information-Theoretic Justification}
There exist several approaches to information-theoretic or
compression-based hypothesis testing; see, for example,
\cite{BarronC91,LiV97}.  The most influential of these is the
so-called \emph{Minimum Description Length Principle}
\cite{Rissanen89}. The basic idea is always that the more one can
compress a given sequence of data, the more regularity one has
extracted from the data, and thus, the better one has captured the
`underlying regularities in the data'. Thus, the hypothesis that
allows for the maximum compression of the data should be adopted.

Let us first consider testing a simple hypothesis $Q$ against another
simple hypothesis $P$. Two basic facts of coding theory say that
\begin{enumerate}
\item There exists a uniquely decodeable code with lengths $L_Q$ that
  satisfy, for all $z_1,\dots, z_n \in \set{Z}^n$,
\begin{IEEEeqnarray}{rCl}
L_Q(z_1,\dots, z_n) 
&=& 
\lceil - \log Q(z_1,\dots, z_n) \rceil.
\end{IEEEeqnarray}
The code with lengths $L_Q$ is called the \emph{Shannon-Fano code},
and its existence follows from the so-called \emph{Kraft Inequality},
\cite{CoverT91}.
\item If data $Z_1,\dots, Z_n$ are independently identically
distributed $\sim Q$, then among all uniquely decodeable codes, the
code with length function $L_Q$ has the shortest expected
code-length. That is, let $L$ be the length function of any uniquely
decodeable code over $n$ outcomes, then
\begin{IEEEeqnarray}{rCl}
\Exp_{Q} [L(Z_1,\dots, Z_n)] 
&\geq& 
\Exp_Q [ - \log Q(Z_1,\dots, Z_n)].
\end{IEEEeqnarray}
\end{enumerate}
Thus, under the assumption that $Q$ generated the data, the optimal
(maximally compressing) code to use will be the Shannon-Fano code with
lengths $- \log Q(Z^n)$ (here, as in the remainder of this section, we
ignored the integer requirement for code lengths).  Similarly, under
the assumption that some $P$ with $P \neq Q$ generated the data the
optimal code will be the code with lengths $- \log P(Z^n)$. Thus, from
the information-theoretic point of view, if one wants to find out
whether $P$ or $Q$ better explains the data, one should check whether
the optimal code under $P$ or the optimal code under $Q$ allows for
more compression of the data. That is, one should look at the
difference
\begin{IEEEeqnarray}{rCl}\label{eq:simplebitdiff}
\itconf 
&:=& 
-\log P(z_1,\dots,z_n) - [ - \log Q(z_1,\dots,z_n)].
\end{IEEEeqnarray}
If $\itconf > 0$, then one decides that $Q$ better explains the
data. The confidence in this decision is given by the magnitude of
$\itconf$: the larger $\itconf$, the more extra bits one needs to
encode the data under $P$ rather than $Q$, thus the larger the
confidence in $Q$.

Now suppose that $Q$ actually generates the data. The \emph{expected
  code length difference}, measured in bits, between coding the data
  using the optimal code for $Q$ and coding using the optimal code for
  $P$, is given by
$\Exp_Q [ -\log P(Z^n) - [ - \log Q(Z^n)]] =n D(Q\|P)$.
Thus, the KL divergence can be interpreted as \emph{the expected
additional number of bits needed to encode outcomes generated by $Q$,
if outcomes are encoded using a code that is optimal for $P$ rather
than for $Q$ }.  Thus, the natural `unit' of $D(\cdot \| \cdot)$ is
the `bit', and $D(Q \|P)$ may be viewed as `average amount of
information about $Z$ that is lost if $Z$ is wrongfully regarded as
being distributed by $Q$ rather than $P$'.  By the law of large
numbers, Equation~(\ref{eq:simplebitdiff}) implies that, with
$Q$-probability $1$, as $n \rightarrow \infty$,
\begin{IEEEeqnarray}{rCl}\label{eq:kwaad} 
\frac{1}{n} (\itconf) 
&\rightarrow& 
D(Q \|P).
\end{IEEEeqnarray}
Thus, if $Q$ generates the data, then the information-theoretic
confidence $\itconf$ in decision ``$Q$ explains the data better than
$P$'' is, up to first order, determined by the KL divergence between
$Q$ and $P$: the larger $D(Q\|P)$, the larger the confidence. This
gives an information-theoretic justification of the use of the KL
divergence as an indicator of statistical strength for simple
hypothesis testing. We now turn to composite hypothesis testing.
\paragraph*{Composite Hypothesis Testing}
If one compares $Q_{\svec}$ against a set of hypotheses $\set{P}_{\svec}$, 
then one has to associate $\set{P}_{\svec}$ with a
  code that is `optimal under the assumption that some $P \in \set{P}_{\svec}$ 
generated the data'. It turns out that there exist codes
  with lengths $L_{\set{P}}$ satisfying, for all $z_1,\dots, z_n \in
  \set{Z}^n$,
\begin{IEEEeqnarray}{rCl}
L_{\set{P}_{\svec}}(z_1,\dots, z_n) 
&\leq& 
\inf_{P \in \set{P}_{\svec}} - \log P(z_1,\dots, z_n) + O(\log n).
\end{IEEEeqnarray}
An example of such a code is given in Appendix~\ref{app:mdl}.  The
code $L_{\set{P}_{\svec}}$ is optimal, up to logarithmic terms, for
whatever distribution $P \in \set{P}_{\svec}$ that might actually
generate data. The information theoretic approach to hypothesis
testing now tells us that, to test $Q_{\svec}$ against $\set{P}_{\svec}$, 
we should compute the difference in code lengths
\begin{IEEEeqnarray}{rCl}
\itconf 
& := & 
L_{\set{P}_{\svec}}(z_1,\dots, z_n) - [ - \log Q_{\svec}(z_1,\dots, z_n)].
\end{IEEEeqnarray}
The larger this difference, the larger the confidence that $Q_{\svec}$
rather than $\set{P}_{\svec}$ generated the data. In
Appendix~\ref{app:mdl} we show that, in analogy to
Equation~(\ref{eq:kwaad}), as $n \rightarrow \infty$,
\begin{IEEEeqnarray}{rCl}\label{eq:kwader}
\frac{1}{n}(\itconf) 
&\rightarrow & 
D(Q_{\svec} \| \set{P}_{\svec})
\end{IEEEeqnarray}
Thus, up to sublinear terms, the information-theoretic confidence in
$Q_{\svec}$ is given by $n D(Q_{\svec} \| \set{P}_{\svec})$. This
provides an information-theoretic justification of adopting
$D(Q_{\svec} \| \set{P}_{\svec})$ as an indicator of statistical
strength.
\subsection{Properties of Kullback-Leibler Divergence} \label{app:klproperties}
\subsubsection{General Properties}
Let $\set{P}$ be the set of distributions on $\set{Z}$. We equip
$\set{P}$ with the Euclidean topology by identifying each $P \in
\set{P}$ with its probability vector. Then $D(P \| Q)$ is jointly
continuous in $P$ and $Q$ on the interior of $\set{P}$. It is jointly
\emph{lower semicontinuous} (for a definition see, e.g.,
\cite{Rockafellar70}), but not continuous, on $\set{P}\times \set{P}$. It is also
jointly strictly convex on $\set{P}\times \set{P}$.

Because $Q(z)\log Q(z)=0$ as $Q(z)\downarrow 0$ we can ignore the 
$Q(z)=0$ parts in the summation, and hence 
\begin{IEEEeqnarray}{rCl} \label{eq:kldecomposition}
D(Q\|P) 
& = & 
\sum_{\substack{z \in \set{Z}\\ Q(z) > 0}} {Q(z) [- \log P(z) + \log Q(z)]}.
\end{IEEEeqnarray}
\subsubsection{The Additivity Property}
The KL divergence has the following additivity property. 
Let $\set{X}$ and $\set{Y}$ be finite sample spaces, and let $P$ and $Q$ be
distributions over the product space $\set{X} \times \set{Y}$. Let
$P_X$ ($Q_X$) denote the marginal distribution of $P$ ($Q$) over
$\set{X}$, and for each $x \in \set{X}$, let $P_{Y|x}$ ($Q_{Y|x}$)
denote the conditional distribution over $\set{Y}$ conditioned on $X=
x$, i.e.\ $P_{Y|x}(y) := P(y|x)$ for all $y \in \set{Y}$.  Then
\begin{IEEEeqnarray}{rCl}
\label{eq:additivity}
D(Q \| P)
& =& 
\sum_{x \in \set{X}} Q(x) D(Q_{Y|x} \| P_{Y|x}) + D(Q_X \|P_X) \\ 
& = & \Exp_{Q_X} [D(Q_{Y|X} \| P_{Y|X})] + D(Q_X \| P_X).
\end{IEEEeqnarray}
An important consequence of this property is that the divergence
between the joint distribution of $n$ independent drawings from $Q$ to
that of $n$ independent drawings from $P$, is $n$ times the divergence
for one drawing.  It also implies Equation~(\ref{eq:avkl}) in
Section~\ref{sec:kl}.

\subsection{Kullback-Leibler versus Total Variation Distance} \label{app:klvariation}
In discussions about the strengths of nonlocality proofs, it has
sometimes been claimed that QM should use the filter settings that
give the largest deviation in the Bell inequality.  This would mean
that QM should try to set up the experiment such that the distribution
of outcomes $Q$ is as distant as possible to LR's distribution over
outcomes $P$ where distance is measured by the so-called \emph{total
variation distance}, \cite{Feller68b} between $Q$ and $P$, defined as
$\sum_{z \in \set{Z}} | P(z) - Q(z) |$. 
While it is true that this defines a distance
between probability distributions, it is only one of large number of
possible distances or divergences that can be defined for probability
distributions. But if one is interested in measuring `statistical
distance', total variation is \emph{not} the appropriate distance
measure to use. Instead, one should use the KL divergence. To get some
feel for how different KL and total variation can be, let $\set{Z} =
\{1,2\}$ and consider the following possibilities for $P$ and $Q$:
\begin{enumerate}
\item $P(1) = 0.99$ and $Q(1) = 1$. Then the absolute difference in
probabilities between $P$ and $Q$ is very small ($0.02$); however, if
data are sampled from $P$, then, with high probability, after a few
hundred trials we will have observed at least one $0$. From that point on,
we are \emph{100\% certain} that $P$, and not $Q$, has generated the
data. This is reflected by the fact that $D(P\|Q) = \infty$.
\item Let $P$ and $Q$ be as above but consider $D(Q\|P)$. We have
  $D(Q\|P) = - 1 \cdot \log 0.99 = 0.015$. This illustrates that, if
  $Q$ rather than $P$ generates the data, we typically need an
  enormous amount of data before we can be reasonably sure that $Q$
  indeed generated the data.
\item $P(1) = 0.49, Q(1)= 0.5$. In this case, $D(P\|Q) = 
0.49 \log 0.98 + 0.51 \log 1.02 \approx 0.000289$ and 
$D(Q\|P) = 0.5 (- \log 0.98 - \log 1.02) \approx 0.000289$.
Now the average support per trial in favor of $Q$
under distribution $Q$ is about equal to the average support per trial
in favor of $P$ under $P$. 
\item Note that the KL divergences for the `near uniform' distributions 
with $P(1),Q(1)\approx 0.5$ is much smaller than the divergences for the 
skewed distributions with $P(1),Q(1)\approx 1$, while the total variation 
distance is the same for all these distributions.  
\end{enumerate}
The example stresses the asymmetry of KL divergence as well as its
difference from the absolute deviations between probabilities.

\subsection{Intuition behind it all} \label{app:klintuition}
Here we give some intuition on the relation between KL divergence and
statistical strength. It can be read without any statistical
background.  Let $Z_1, Z_2,\dots$ be a sequence of random variables
independently generated either by some distribution $P$ or by some
distribution $Q$ with $Q \neq P$. Suppose we are given a sample
(sequence of outcomes) $z_1,\dots, z_n$. Perhaps simplest
(though by no means only) way of finding out whether $Q$ or $P$
generated this data is to compare the \emph{likelihood} (in our case,
`likelihood' $=$ `probability') of the data $z_1,\dots,z_n$ according to the two
distributions. That is, we look at the ratio
\begin{IEEEeqnarray}{rCl}  \label{eq:likrat}
  \frac{Q(z_1,\dots,z_n)}{P(z_1,\dots,z_n)} 
&=& 
\frac{\prod_{i=1}^n Q(z_i)} {\prod_{i=1}^n P(z_i)}.
\end{IEEEeqnarray}
Intuitively, if this ratio is larger than $1$, the data is more
typical for $Q$ than for $P$, and we might decide that $Q$ rather than
$P$ generated the data. Again intuitively, the magnitude of the ratio
in Equation~(\ref{eq:likrat}) might give us an idea of the confidence we
should have in this decision.

Now assume that the data are actually generated according to $Q$, i.e.\
`$Q$ is true'. We will study the behavior of the logarithm of the
likelihood ratio in Equation~(\ref{eq:likrat}) under this assumption
(the use of the logarithm is only to simplify the analysis; using
Equation~(\ref{eq:likrat}) directly would have led to the same
conclusions).  The \emph{Law of Large Numbers} \cite{Feller68a} tells
us that, with $Q$-probability $1$, averages of bounded random
variables will converge to their $Q$-expectations. In particular, if
the $Z_i$ take values in a finite set $\set{Z}$, and $P$ and $Q$ are
such that $P(z), Q(z) > 0$ for all $z \in \set{Z}$, then with
$Q$-probability $1$,
\begin{IEEEeqnarray}{rCl}
\frac{1}{n} \sum_{i=1}^n L_i 
&\rightarrow & 
\Exp_Q [L]
\end{IEEEeqnarray}
where $L_i := \log (Q(Z_i)/P(Z_i))$,
and $\Exp_Q [L] = \Exp_Q[L_1] = \cdots = \Exp_Q[L_n]$ is given by
\begin{IEEEeqnarray}{rCcCcCl}
\Exp_Q [L] 
&=& 
\Exp_Q [\log (\frac{Q}{P})] 
&=& 
\sum_{z}{Q(z)}\log(\frac{Q(z)}{P(z)}) 
&=& 
D(Q \| P).
\end{IEEEeqnarray}
Therefore, with $Q$-probability $1$,
\begin{IEEEeqnarray}{rCl}  \label{eq:kllikrat}
  \frac{1}{n} \log \frac{Q(Z_1,\dots,Z_n)}{P(Z_1,\dots,Z_n)}
  &\rightarrow& 
D(Q\|P).
\end{IEEEeqnarray}
Thus, with $Q$-probability $1$, the \emph{average log-likelihood ratio
between $P$ and $Q$ will converge to the KL divergence between $P$ and
$Q$}. This means that the likelihood ratio, which may be viewed as the
amount of evidence for $Q$ vs.\ $P$, is asymptotically determined by
the KL divergence, to first order in the exponent.  For example, let
us test $Q$ first against $P_1$ with $D(Q\|P_1) = \epsilon_1$, and
then against $P_2$ with $D(Q\|P_2) = \epsilon_2 > \epsilon_1$, then,
with $Q$-probability $1$,
\begin{IEEEeqnarray}{rClCrCl}
 \frac{1}{n} \log \frac{Q(Z_1,\dots,Z_n)}{P_1(Z_1,\dots,Z_n)}
  &\rightarrow& \epsilon_1 
& \mbox{~and~} & 
\frac{1}{n} \log \frac{Q(Z_1,\dots,Z_n)}{P_2(Z_1,\dots,Z_n)} 
&\rightarrow& \epsilon_2.
\end{IEEEeqnarray}
This implies that with increasing $n$, the likelihood ratio $Q/P_1$
becomes exponentially smaller than the likelihood ratio $Q/P_2$:
\begin{IEEEeqnarray}{rCl}
\frac{Q(Z_1,\dots,Z_n)}{P_1(Z_1,\dots,Z_n)} &\leq&
\frac{Q(Z_1,\dots,Z_n)}{P_1(Z_2,\dots,Z_n)} \cdot 
\e^{- n (\epsilon_2 - \epsilon_1) + o(n)}.
\end{IEEEeqnarray}
Returning to the setting discussed in this paper, this preliminary
analysis suggests that from QM's point of view (who knows that Q is
true), the most convincing experimental results (highest likelihood
ratio of $Q_{\svec}$ vs.\ $P_{\svec; \mvec}$) are obtained if the KL
divergence between $Q_{\svec}$ and $P_{\svec; \mvec}$ is as large as
possible. If $Q_{\svec}$ is compared against a set $\set{P}_{\svec}$,
then the analysis suggests that the most convincing experimental
results are obtained if the KL divergence between $Q_{\svec}$ and $
\set{P}_{\svec}$ is as large as possible, that is, if $\inf_{P \in
\set{P}_{\svec}} D(Q_{\svec} \| P_{\svec})$ is as large as possible.

\subsection{Bayesian Analysis} \label{app:bayes}
In this appendix we assume some basic knowledge of Bayesian
statistics.  We only give the derivation for the $2\times 2\times 2$
nonlocality proofs. Extension to generalized nonlocality proofs is
straightforward.

Let us identify $H_1 := Q_{\svec}$ and $H_0 := \set{P}_{\svec}$, 
where $Q_{\svec}$ and $\set{P}_{\svec}$ are defined as
quantum and local realist theories respectively, as in
Section~\ref{sec:formal}. We start with a prior $\Pr$ on $H_1$ and
$H_0$, and we assume $0 < \Pr(H_1) < 1$.

Now, \emph{conditioned} on $H_0$ being the case, the actual
distribution generating the data may still be any $P_{\svec;\mvec} \in
H_0$.  To indicate the prior degree of belief in these, we further
need a \emph{conditional prior distribution} $\Pr(\cdot |H_0)$ over
all the distributions in $H_0$. Since $\svec$ is fixed, $H_0 = 
\set{P}_{\svec}$ is parameterized by the set $\mvec$. We suppose the
prior $\Pr(\cdot | H_0)$ is smooth in the sense of having a probability density
function $\priordens$ over $\mvec$, so that for each
(measurable) $A \subset \msim$,
\begin{IEEEeqnarray}{rCl}
\Pr(\{ P_{\svec;\mvec} \;:\; \mvec \in A \} \mid H_0) 
& := &
\int_{\mvec \in A}\priordens(\mvec) \d\mvec.
\end{IEEEeqnarray}
We restrict attention to prior densities $\priordens(\cdot)$ that are
continuous and uniformly bounded away from $0$. By the latter we mean
that there exists $\priordens_{\min} > 0$ such that 
$\priordens(\mvec) > \priordens_{\min}$ for all $\mvec \in \msim$. 
For concreteness one
may take $\priordens$ to be uniform (constant over $\mvec$), although
this will not affect the analysis.

In order to apply Bayesian inference, we further have to define
$\Pr(z_1,\dots, z_n | H_i)$, `the probability of the data given that
$H_i$ is true'. We do this in the standard Bayesian manner:
\begin{IEEEeqnarray}{rCl}\label{eq:cond}
\Pr(z_1,\dots, z_n | H_1) 
& := & 
Q_{\svec}(z_1,\dots, z_n),
\nonumber \\ 
\Pr(z_1,\dots, z_n | H_0) & := & \int_{\mvec \in
\msim} P_{\svec;\mvec}(z_1,\dots, z_n) \priordens(\mvec) \d \mvec.
\end{IEEEeqnarray}
Here each outcome $z_i$ consists of a realized measurement setting and
an experimental outcome in that setting; hence we can write 
$z_i = (a_i,b_i,x_i,y_i) $ for $a_i, b_i \in \{1,2\}$ and 
$x_i, y_i \in \{\true,\false\}$.

Together with the prior over $\{H_1,H_0\}$, Equation~(\ref{eq:cond})
defines a probability distribution over the product space $\{
H_1,H_0\} \times \set{Z}^n$ where $
\set{Z} := \{1,2\} \times \{1,2\} 
\times \{\true,\false\} \times \{\true,\false\}$.
Given experimental data $z_1,\dots, z_n$ and prior distribution $\Pr$,
we can now use Bayes' rule \cite{Feller68a,Lee97} to compute the
\emph{posterior distribution} of $H_i$:
\begin{IEEEeqnarray}{rCl}\label{eq:posterior}
\Pr(H_i | z_1,\dots, z_n) 
&=& 
\frac{\Pr(z_1,\dots, z_n | H_i)  \Pr(H_i)}{\sum_i\Pr(z_1,\dots, z_n | H_i) \Pr(H_i) }
\end{IEEEeqnarray}
According to Bayesian hypothesis testing, we should select the $H_i$
maximizing the posterior probability of Equation~(\ref{eq:posterior}).
The \emph{confidence} in the decision, which we denote by
$\bayesconf$, is given by the posterior odds against $H_0$:
\begin{IEEEeqnarray}{rCl} \label{eq:logodds}
  \bayesconf 
&:=& 
\frac{\Pr(H_1 | z_1,\dots, z_n)}{\Pr(H_0 | z_1,\dots,z_n)}\\ 
& =& 
\frac{\Pr(z_1,\dots, z_n |H_1)\Pr(H_1)}{\Pr(z_1,\dots,z_n | H_0 )\Pr(H_0)}\\ 
&= &
\frac{Q_{\svec}(z_1,\dots, z_n)}{\int_{\mvec \in \msim}
P_{\svec;\mvec}(z_1,\dots,z_n) \priordens(\mvec) \d \mvec} \cdot
\frac{\Pr(H_1)}{\Pr(H_0)}
\end{IEEEeqnarray}
Note that $\bayesconf$ depends on $H_0, H_1$ and the data $z_1,\dots,
z_n$.  The factor on the left of Equation~(\ref{eq:logodds}) is called
the \emph{Bayes factor}, and the factor on the right is called the
\emph{prior odds}. Since the Bayes factor typically increases
exponentially with $n$, the influence of the prior odds on the
posterior odds is negligible for all but the smallest $n$.  Below we
show that, if $H_1$ is true (`QM is right'), then with probability $1$,
\begin{IEEEeqnarray}{rCl}  \label{eq:limodds}
  \frac{1}{n} \log(\bayesconf) 
&\rightarrow& 
\infm{\msim} D(Q_{\svec} \| P_{\svec;\mvec}).
\end{IEEEeqnarray}
Hence Equation~(\ref{eq:bayesconf}) holds:
the confidence $\bayesconf$ will be determined, to first order in the
exponent, by $\infm{\msim} D(Q_{\svec} \|
P_{\svec;\mvec})$. This gives a Bayesian justification of adopting
$D(Q_{\svec} \| \set{P}_{\svec})$ as an indicator of statistical
strength --- provided that we can prove that Equation~(\ref{eq:limodds})
holds.  We proceed to show this.

\begin{proof}
To prove Equation~(\ref{eq:limodds}), we first note that
\begin{IEEEeqnarray}{rCl}
\log P_{\svec;\mvec}(z_1,\dots, z_n)
& =& 
\log P_{\svec;\mvec}((a_1,b_1,x_1,y_1),\dots, (a_n,b_n,x_n,y_n)) \\ 
& = & 
n \cdot \sum_{a,b \in \{1,2\}} \empfreq(a, b) \svec_{ab} \log \empfreq(a,b) \ + \\ 
& & n\cdot \sum_{\substack{a,b \in \{1,2\}\\x,y \in \{\true,\false\}}}
{\empfreq(a,b,x,y) \log \left(\sum_{\substack{x_1,x_2,y_1,y_2\\x_a = x, y_b = y}}
{\mvec_{x_1 x_2 y_1 y_2}}\right)}.
\end{IEEEeqnarray}
Here $\empfreq(a,b)$ is the relative frequency (number of occurrences in the
sample divided by $n$) of experimental outcomes with measurement
setting $(a,b)$ and $\empfreq(a,b,x,y)$ is the relative frequency of experimental
outcomes with measurement setting $(a,b)$ and outcome $X=x, Y=y$.

Let $\tilde{\mvec}$ be any $\mvec$ achieving
$\infm{\msim} D(Q_{\svec} \| P_{\svec;\mvec})$.
By Theorem~\ref{thm:supinf}, such a $\tilde{\mvec}$ must exist, and
$Q_{\svec}$ must be absolutely continuous with respect to $P_{\svec ;
\tilde{\mvec}}(a,b,x,y)$.  It follows that
\begin{IEEEeqnarray}{rCl}
P_{\svec; \tilde{\mvec}}(a,b,x,y) 
&=& 
{\svec}_{ab} \sum_{\substack{x_1,x_2,y_1,y_2\\x_a = x, y_b = y}}{\mvec_{x_1 x_2 y_1 y_2}}
\end{IEEEeqnarray}
may be equal to $0$ \emph{only} if $\svec_{ab} Q_{\svec}(x,y,a,b) =
0$.  From this it follows (with some calculus) that there must be a
constant $c$ and an $\epsilon>0$
such that if $\mvec$ with $|\mvec-\tilde{\mvec}|_1<\epsilon$, then, 
for all $n$ and all sequences $z_1,\dots,z_n$ with 
$Q_{\svec}(z_1,\dots, z_n) > 0$, we have 
\begin{IEEEeqnarray}{rCl}  \label{eq:lipschitz}
  \frac{|\frac{1}{n} \log P_{\svec;\mvec}(z_1,\dots,z_n) -\frac{1}{n} \log
P_{\svec ; \tilde{\mvec}}(z_1,\dots,z_n)|} {|\mvec - \tilde{\mvec}|_1} 
&\leq& 
c
\end{IEEEeqnarray}
and whence $|\log P_{\svec;\mvec}(z_1,\dots,z_n) - \log P_{\svec ;
  \tilde{\mvec}}(z_1,\dots,z_n)| \leq n c | \mvec - \tilde{\mvec}|_1
\leq nc \epsilon$.
For sufficiently large $n$, we find that $\epsilon > n^{-2}$ and then
\begin{IEEEeqnarray}{rCl}  \label{eq:taylor}
\Pr(z_1,\dots, z_n | H_0)
& =& 
\int_{\mvec \in \msim}P_{\svec;\mvec}(z_1,\dots,z_n) \priordens(\mvec ) \d \mvec\\ 
&\geq &
\int_{|\mvec-\tilde{\mvec}|_1 < \epsilon} P_{\svec;\mvec}(z_1,\dots,z_n)
\priordens(\mvec ) \d \mvec\\ 
& \geq& 
\priordens_{\min}\cdot\frac{v}{n^{2k}} \cdot \e^{-c/n} P_{\svec;\tilde{\mvec}}(z_1,\dots,z_n),
\end{IEEEeqnarray}
where $v\cdot n^{-2k}$ is a lower bound on the volume $\int 1 \d \pi$ of 
the set $\{\mvec : |\mvec-\tilde{\mvec}|_1<\epsilon=n^{-2}\}$.
Hence, $- \log \Pr(z_1,\dots,z_n | H_0) \leq -\log P_{\svec;
  \tilde{\mvec}}(z_1,\dots,z_n) +O(\log n)$.  By applying the strong
law of large numbers to $n^{-1} \sum \log P_{\svec;
  \tilde{\mvec}}(Z_i)$, we find that, with $Q_{\svec}$-probability
$1$,
\begin{IEEEeqnarray}{rCl}\label{eq:bayesahoy}
- \frac{1}{n} \log \Pr(Z_1,\dots, Z_n | H_0) 
&\leq& 
\Exp_{Q_{\svec}} [-\log P_{\svec; \tilde{\mvec}}(Z)] + O(\frac{\log n}{n})\\ 
& = &
\infm{\msim} \Exp_{Q_{\svec}} [- \log P_{{\svec},\mvec}(Z)]+ O(\frac{\log n}{n}).
\end{IEEEeqnarray}
This bounds $- \frac{1}{n} \log \Pr(Z^n | H_0)$ from above. We proceed
to bound it from below. Note that for all $n$, $z_1,\dots, z_n$,
\begin{IEEEeqnarray}{rCl}  \label{eq:hallowim}
- \frac{1}{n} \log \int_{\mvec\in \msim} P_{\svec;\mvec}(z_1,\dots,z_n) 
\priordens(\mvec) \d \mvec 
&\geq&
\infm{\msim} -\frac{1}{n} \log P_{\svec ; \mvec}(z_1,\dots,z_n).
\end{IEEEeqnarray}
To complete the proof, we need to relate
\begin{IEEEeqnarray}{rCl}
\infm{\msim} - \frac{1}{n} \log P_{\svec ; \mvec}(z_1,\dots,z_n) 
&=& 
\infm{\msim} - \frac{1}{n} \sum_{i=1}^n \log P_{\svec ; \mvec}(z_i)
\end{IEEEeqnarray} 
(which depends on the data) to
its `expectation version' $\infm{\msim} \Exp_{Q_{\svec}} [ -
\log P_{\svec ; \mvec}(Z)]$. This can be done using a version of the
\emph{uniform law of large numbers} \cite{Vapnik98}. Based on such a
uniform law of large numbers, (for example, \cite[Chapter 5, Lemma
5.14]{Grunwald98b}) one can show that for all distributions $Q$ over
$\set{Z}$, with $Q$-probability $1$, as $n \rightarrow \infty$,
\begin{IEEEeqnarray}{rCl}  \label{eq:empexp}
 \infm{\msim} - \frac{1}{n} \log P_{\svec ;\mvec}(Z_1,\dots,Z_n) 
&\rightarrow& 
\infm{\msim} \Exp_Q [ -\log P_{\svec ;\mvec}(Z)].
\end{IEEEeqnarray}
Together, Equations~(\ref{eq:bayesahoy}), (\ref{eq:hallowim}) and
(\ref{eq:empexp}) show that, with $Q_{\svec}$-probability $1$, as $n
\rightarrow \infty$,
\begin{IEEEeqnarray}{rCl}  \label{eq:supconv}
 - \frac{1}{n} \log \Pr(Z_1,\dots,Z_n | H_0) 
&\rightarrow&
 \infm{\msim} \Exp_{Q_{\svec}} [ - \log P_{\svec ;\mvec}(Z)]
\end{IEEEeqnarray}
Together with the law of large numbers applied to $n^{-1} \sum \log
\Pr(Z_i| H_1)$, we find that
\begin{IEEEeqnarray}{rCl}  \label{eq:bayeskl}
\frac{1}{n} \log\frac{\Pr(Z_1,\dots, Z_n | H_1)}{\Pr(Z_1,\dots, Z_n |H_0)} 
& \rightarrow & 
\infm{\msim} \Exp_{Q_{\svec}} [ \log\frac{Q_{\svec}(Z)}{P_{\svec ; \mvec}(Z)} ].
\end{IEEEeqnarray}
Noting that the right hand side is equal to 
$\infm{\msim}{D(Q_{\svec} \| P_{\svec ;\mvec})}$ and plugging this in into
Equation~(\ref{eq:logodds}), we see that indeed with $H_1$-probability $1$,
Equation~(\ref{eq:limodds}) holds.  
\end{proof}

\subsection{Information Theoretic Analysis} \label{app:mdl}
In this appendix we assume that the reader is familiar with the basics
of information theory.

The code with lengths $L_{\set{P}_{\svec}}$ is simply the
Shannon-Fano code for the Bayesian marginal likelihood $\Pr(z_1,\dots,
z_n \mid \set{P}_{\svec}) = \Pr(z_1,\dots, z_n \mid H_0)$ as defined
in Equation~(\ref{eq:cond}). For each $n$, each $z_1,\dots, z_n \in
\set{Z}^n$, this code achieves lengths (up to 1 bit) $ -\log
\Pr(z_1,\dots, z_n \mid \set{P}_{\svec})$.  The code corresponding to
$Q_{\svec}$ achieves lengths $- \log Q(z_1,\dots, z_n)$. We have
already shown in Appendix~\ref{app:bayes}, Equation~(\ref{eq:bayeskl}),
that, with $Q$-probability $1$, as $n \rightarrow \infty$,
\begin{IEEEeqnarray}{rCl}
\frac{1}{n} \bigl[ -\log \Pr(z_1,\dots, z_n \mid \set{P}_{\svec}) 
- [- \log Q(z_1,\dots, z_n) ] \bigr] 
&\rightarrow& 
D(Q_{\svec} \| \set{P}_{\svec}).
\end{IEEEeqnarray}
Noting that the left hand side is equal to $\itconf/n$,
Equation~(\ref{eq:kwader}) follows.

\section{Proofs of Theorems~\ref{thm:supinf},\ref{thm:spcor} and \ref{thm:2x2}}
\label{app:theoremproofs}
\subsection{Preparation}
The proof of Theorem~\ref{thm:supinf} uses the following lemma, which
is of some independent interest.
\begin{lemma} \label{lem:wim}
Let $(\ssim,\msim, U)$ be the game corresponding to an arbitrary $2$
party, $2$ measurement settings per party nonlocality proof.  For any
$(a_0,b_0) \in \{1,2\}^2$, there exists a $\mvec \in \msim$ such that 
for all $(a,b) \in \{1,2\}^2\setminus\{(a_0,b_0)\}$ we have 
$Q_{ab} = P_{ab ; \mvec}$.
Thus, for any three of the four measurement settings, the probability
distribution on outcomes can be perfectly explained by a local realist
theory.
\end{lemma}
\begin{proof}
We give a detailed proof for the case that the measurement
outcomes are two values $\{\true,\false\}$; the general case can be proved in a
similar way.

Without loss of generality let $(a_0,b_0) = (2,2)$. Now we must prove
that the equation $Q_{ab} = P_{ab ; \mvec}$ holds for the three settings $(a,b)
\in \{ (1,1),(1,2),(2,1) \}$.  Every triple of distributions
$P_{ab;\mvec}$ for these three settings may be represented by a table
of the form of Table~\ref{tb:generic}.
\begin{table}
\centering
\caption{Generic Classical Distribution}
\label{tb:generic}
\Prtable%
{ p_1 & p_2 \\ p_3 & p_4}
{ p_5 & p_6 \\ p_7 & p_8}
{ p_9 & p_{10} \\ p_{11} & p_{12}}
{  &  \\  & }
\end{table}
with $p_1,\dots,p_{12} \geq 0$
and the normalization restrictions 
$p_1+\dots + p_4 = p_5+\dots +p_8 = p_9+\cdots +p_{12}=1$. 
Given any table of this form, we
say that the LR distribution $P_\mvec$ \emph{corresponds} to the 
$p$-table if $P_{11 ; \mvec}(\true,\true) = p_1$, 
$P_{12 ;\mvec}(\false,\true) = p_{10}$ etc., for all $p_i$. 

The no-signalling restriction implies that the realized measurement 
setting on $A$'s side should not influence the probability on $B$'s 
side and vice versa.  Hence, for example,  
$\Pr_{00}(Y=\true) = \Pr_{10}(Y=\true)$, which gives $p_1+p_2=p_5+p_6$.  
In total there are four such no-signaling restrictions:
\begin{IEEEeqnarray}{c}\label{eq:rest}
\left\{\begin{array}{rcl} 
p_1+p_2& = &p_5+p_6\\ 
p_3+p_4 & = & p_7+p_8 \\ 
p_1+p_3 & = & p_9+p_{11}\\ 
p_2+p_4 & = & p_{10}+p_{12}.
\end{array}
\right.
\end{IEEEeqnarray}
We call a table with $p_1,\dots, p_{12} \geq 0$, 
that obeys the normalization restriction on the sub-tables and 
that satisfies Equations~(\ref{eq:rest}) a \emph{$\Gamma$-table}. 
We already showed that each
triple of conditional LR distributions may be represented as a
$\Gamma$-table. In exactly the same way one shows that each triple of
conditional quantum experimentalist distributions 
$Q_{00}$, $Q_{01}$, $Q_{10}$ can be represented as a $\Gamma$-table. It therefore suffices
if we can show that \emph{every} $\Gamma$-table corresponds to some LR
theory $P_{\mvec}$.  We show this by considering the $16$ possible
deterministic theories $T_{\mvala_1 \mvala_2 \mvalb_1 \mvalb_2}$.
Here $T_{\mvala_1 \mvala_2 \mvalb_1 \mvalb_2}$ is defined as the
theory with
$P_{\mvec}(\mvara_1 = \mvala_1, \mvara_2 = \mvala_2, \mvarb_1 =
\mvalb_1, \mvarb_2 = \mvalb_2) = \mvec_{\mvala_1 \mvala_2 \mvalb_1 \mvalb_2} = 1$.
Each deterministic theory $T_{\mvala_1 \mvala_2 \mvalb_1 \mvalb_2}$ 
corresponds to a specific $\Gamma$-table denoted by
$\Gamma_{\mvala_1 \mvala_2 \mvalb_1 \mvalb_2}$.  For example, the
theory $T_{\false \false \true \false}$ gives the 
$\Gamma_{\false \false \true \false}$-table shown in Table~\ref{tb:detexample}.
\begin{table}
\centering
\caption{Example of a Classical, Deterministic Distribution}
\label{tb:detexample}
\Prtable%
{0 &1 \\0 &0 }
{0 &1 \\0 &0 }
{0 &0 \\0 &1 }
{ & \\ & }
\end{table}
We will prove that the set of $\Gamma$-tables is in fact the convex 
hull of the $16$ tables $\Gamma_{\mvala_1 \mvala_2 \mvalb_1 \mvalb_2}$
corresponding to deterministic theories. This shows that any
$\Gamma$-table can be reproduced by a mixture of deterministic
theories. Since every LR theory $\mvec \in \msim$ can be written as
such a mixture, this proves the lemma.

First we observe that a $\Gamma$-table with all entries $0$ or $1$
has to be one of the $16$ deterministic theories.  
Given a $\Gamma$-table that is not a deterministic theory, 
we focus on its smallest nonzero entry
$\Gamma_{ab}=\varepsilon>0$. By the restrictions of
Equations~(\ref{eq:rest}) there exists a deterministic theory $T_k$ such
that the table $(\Gamma - \varepsilon \Gamma_k)/(1-\varepsilon)$ has no 
negative entries.  For example, suppose that the smallest element in 
$\Gamma$ corresponds to $P_\mvec(\mvara_1 = \false, \mvarb_1 = \true)$ 
(denoted as $p_2$ in the first table above).  
By the restrictions of Equation~(\ref{eq:rest}),
either the table $(\Gamma - p_2 \Gamma_{\false \false \true \false})/(1-p_2)$ 
(where $\Gamma_{\false \false \true \false}$ is shown above) or one of the 
three tables $(\Gamma - p_2 \Gamma_{\false \false \true \true})/(1-p_2)$, 
$(\Gamma - p_2 \Gamma_{\false \true \true\false})/(1-p_2)$, 
$(\Gamma - p_2 \Gamma_{\false \true \true\true})/(1-p_2)$ has only nonnegative entries.

Let $\Gamma' := (\Gamma - \varepsilon \Gamma_k)/(1-\varepsilon)$ where $k$ 
is chosen such that $\Gamma'$ has no negative entries. Clearly, either 
$\Gamma'$ describes a deterministic theory with entries $0$ and $1$, 
or $\Gamma'$ is a $\Gamma$-table with number of
nonzero entries one less than that of $\Gamma$.  Hence by applying the
above procedure at most $16$ times, we obtain a decomposition
$\Gamma = \varepsilon_1 \Gamma_{k_1}+\dots + \varepsilon_{16}\Gamma_{k_{16}}$,
which shows that $\Gamma$ lies in the convex hull of the
$\Gamma$-tables corresponding to deterministic theories.
Hence, any such $\Gamma$ can be described as a LR theory.

For measurement settings with more than two outcomes, 
the proof can be generalized in a straightforward manner.  
\end{proof}

\subsection{Proof of Theorem~\ref{thm:supinf}} \label{app:prf:supinf}
\newcounter{theoremtracker}
\setcounter{theoremtracker}{\value{theorem}}
 \setcounter{theorem}{\value{cr:supinf}}
 \addtocounter{theorem}{-1}
\begin{theorem}
Let $Q$ be a given (not necessarily $2\times 2 \times 2$) nonlocality
proof and $\msim$ the corresponding set of local realist theories.
\begin{enumerate}
\item Let $U(\svec,\mvec) := D(Q_{\svec} \| P_{\svec;\mvec})$, then:
\begin{enumerate}
\item For a $2\times 2\times 2$ proof, we have that
\begin{IEEEeqnarray}{rCl}\label{eq:appendixavkl}
U(\svec,\mvec) 
&=& 
\sum_{a,b \in \{1,2\}} \svec_{ab} D(Q_{ab}(\cdot ) \| P_{ab; \mvec}(\cdot ))
\end{IEEEeqnarray} 
Hence, the KL divergence $D(Q_{\svec}\|P_{\svec;\mvec})$ may
alternatively be viewed as the average KL divergence between the
distributions of $(X,Y)$, where the average is over the settings
$(A,B)$. For a generalized nonlocality proof, the analogous
generalization of Equation~(\ref{eq:appendixavkl}) holds.
\item For fixed $\svec$, $U(\svec,\mvec)$ is convex and lower
  semicontinuous on $\msim$, and continuous and differentiable on the
  interior of $\msim$.
\item If $Q$ is absolutely continuous with respect to some fixed
  $\mvec$, then $U(\svec,\mvec)$ is linear in $\svec$.
\end{enumerate}
\item Let
\begin{IEEEeqnarray}{rCl}\label{eq:appendixuinf}
U(\svec) & :=& \infm{\msim} U(\svec,\mvec),
\end{IEEEeqnarray} 
then
\begin{enumerate}
\item For all $\svec \in \ssim$, the infimum in
Equation~(\ref{eq:appendixuinf}) is achieved for some $\mvec^*$.
\item The function $U(\svec)$ is nonnegative, bounded, concave and
continuous on $\svec$.
\item If $Q$ is not a proper nonlocality proof, then $U(\svec) = 0$ 
for all $\svec\in\ssim$. 
If $Q$ is a proper nonlocality proof, then
  $U(\svec) > 0$ for \emph{all} $\svec$ in the interior of $\ssim$.
\item For a 2 party, 2 measurement settings per party nonlocality
  proof, we further have that, even if $Q$ is proper, then still
  $U(\svec) = 0$ for all $\svec$ on the boundary of $\ssim$.
\end{enumerate}
\item Suppose that $\svec$ is in the interior of $\ssim$, then:
\begin{enumerate}
\item Let $Q$ be a $2\times 2\times 2$ nonlocality proof. Suppose that
  $Q$ is non-trivial in the sense that, for some $a,b$, $Q_{ab}$ is
  not a point mass (i.e.\ $0 < Q_{ab}(\mvala,\mvalb) < 1$ for some
  $\mvala,\mvalb$). Then $\mvec^* \in \msim$ achieves the infimum in
  Equation~(\ref{eq:appendixuinf}) if and only if the following 16
  (in)equalities hold:
\begin{IEEEeqnarray}{rCl}\label{eq:appendixem}
\sum_{a,b \in \{1,2\}} \svec_{ab} \frac{ Q_{ab}(\mvala_a, \mvalb_b ) }
{ P_{ab ; \mvec^*}(\mvala_a, \mvalb_b )} 
&=& 1
\end{IEEEeqnarray}
for all $(\mvala_1,\mvala_2,\mvalb_1,\mvalb_2) \in \{ \true,\false \}^4$
such that $\mvec^*_{\mvala_1,\mvala_2,\mvalb_1,\mvalb_2} > 0$, and
\begin{IEEEeqnarray}{rCl}\label{eq:appendixemb}
\sum_{a,b \in \{1,2\}} \svec_{ab} \frac{ Q_{ab}(\mvala_a, \mvalb_b ) }
{ P_{ab ; \mvec^*}(\mvala_a, \mvalb_b )} 
&\leq& 1
\end{IEEEeqnarray}
for all $(\mvala_1,\mvala_2,\mvalb_1,\mvalb_2) \in \{ \true,\false \}^4$
such that $\mvec^*_{\mvala_1,\mvala_2,\mvalb_1,\mvalb_2} = 0$.

For generalized nonlocality proofs, $\mvec^* \in \msim$ achieves
Equation~(\ref{eq:appendixuinf}) if and only if the corresponding
analogues of Equations~(\ref{eq:appendixem}) and (\ref{eq:appendixemb})
both hold.
\item Suppose that $\mvec^*$ and $\mvec^\circ$ both achieve the
infimum in Equation~(\ref{eq:appendixuinf}).  Then, for all 
$\mvala,\mvalb \in \{ \true,\false\}$, $a,b \in \{1,2\}$ with
$Q_{ab}(\mvala,\mvalb) > 0$, we have
$P_{ab; \mvec^*}(\mvala,\mvalb) = P_{ab ; \mvec^\circ}(\mvala, \mvalb)> 0$.
In words, $\mvec^*$ and $\mvec^\circ$ coincide in every measurement
setting for every measurement outcome that has positive probability
according to $Q_{\svec}$, and $Q$ is absolutely continuous with
respect to $\mvec^*$ and $\mvec^\circ$.
\end{enumerate}
\end{enumerate}
\end{theorem}
\setcounter{theorem}{\value{theoremtracker}} 
\begin{proof}
We only give proofs for the $2\times 2\times 2$ case; extension to the
general case is entirely straightforward.  We define
\begin{IEEEeqnarray}{rCl} \label{eq:rewrite}
U((a,b),\mvec)
& := & D(Q_{ab}(\cdot) \|P_{ab; \mvec}(\cdot)) \\ 
& = &
\sum_{\substack{\mvala, \mvalb \in \{\true,\false \}\\ 
Q_{ab}(\mvala, \mvalb) > 0}}{Q_{ab}(\mvala, \mvalb) 
[\log Q_{ab}(\mvala, \mvalb) - \log P_{ab; \mvec}( \mvala, \mvalb) ]}.
\end{IEEEeqnarray}
Note that $U(\svec,\mvec)$ can be written as
$U(\svec,\mvec) =\sum_{a,b \in \{1,2\}} \svec_{ab} U((a,b), \mvec)$.
\paragraph*{Part 1}
Equation~(\ref{eq:appendixavkl}) follows directly from the additivity
property of KL divergence, Equation~(\ref{eq:additivity}).  Convexity is
immediate by Jensen's inequality applied to the logarithm in
Equation~(\ref{eq:rewrite}) and the fact that $P_{ab; \mvec}( \mvala,
\mvalb)$ is linear in $\mvec_{\mvala_1 \mvala_2 \mvalb_1 \mvalb_2}$
for each $(\mvala_1,\mvala_2, \mvalb_1, \mvalb_2) \in \{\true,\false\}^4$. 
If $\mvec$ lies in the interior of $\msim$, then
$P_{ab; \mvec}(\mvala, \mvalb) > 0$ for $a,b \in \{1,2\}$ so that
$U(\svec,\mvec)$ is finite. Continuity and differentiability are then
immediate by continuity and differentiability of $\log x$ for $x > 0$. 
Lower semicontinuity of $U(\svec,\mvec)$ on $\msim$ is implied by
the fact that, on general spaces, $D(Q \| P)$ is jointly lower
semi-continuous in $Q$ and $P$ in the weak topology, as proved by
Posner \cite[Theorem 2]{Posner75}. Part 1(c) is immediate.

\paragraph*{Part 2}
We have already shown that for fixed $\svec$, $U(\svec,\mvec)$ is
lower semicontinuous on $\msim$. Lower semicontinuous functions
achieve their infimum on a compact domain (see for example \cite[page
84]{Ferguson67}), so that for each $\svec$,
Equation~(\ref{eq:appendixuinf}) is achieved for some $\mvec^*$. This
proves (a). To prove (b), note that nonnegativity of $U(\svec)$ is
immediate by nonnegativity of the KL divergence. Boundedness of
$U(\svec)$ follows by considering the uniform distribution
$\mvec^\circ$, with, for all $\mvala_1, \mvala_2, \mvalb_1, \mvalb_2$,
$\mvec^\circ_{\mvala_1 \mvala_2 \mvalb_1 \mvalb_2} =
1/16$. $\mvec^\circ$ is in $\msim$, so that
\begin{IEEEeqnarray}{rCl}
U(\svec) 
&\leq& 
U(\svec,\mvec^\circ)\\ 
& = & 
\sum_{a,b \in \{1,2\}}\svec_{ab} \biggl({
\sum_{\substack{\mvala, \mvalb \; \in \{\true,\false \} \\
Q_{ab}(\mvala, \mvalb) > 0}}{Q_{ab}(\mvala, \mvalb) [\log Q_{ab}(\mvala, \mvalb) + 2]}} 
\biggr) \\ 
&\leq& 
- \sum_{a,b \{1,2\}}\svec_{ab} H(Q_{ab})+ 8,
\end{IEEEeqnarray}
where $H(Q_{ab})$ is the Shannon-entropy of the distribution $Q_{ab}$.
Boundedness of $U(\svec)$ now follows from the fact that $H(Q) \geq 0$
for every distribution $Q$, which is a standard result (see,
e.g. \cite{CoverT91}).

Let $\svec$ be in the interior of $\ssim$ and let $\mvec^* \in \msim$
achieve $\infm{\msim} U(\svec,\mvec)$. Since $U(\svec)$ is
bounded, $Q$ is absolutely continuous with respect to $\mvec^*$
(otherwise $U(\svec) = \infty$, a contradiction).  Thus, $U(\svec)$
satisfies
\begin{IEEEeqnarray}{rCl}\label{eq:appendixuinfb}
U(\svec) 
&=& 
\inf_{\mvec \; : \; Q \ll \mvec}{U(\svec,\mvec)},
\end{IEEEeqnarray}
where $Q \ll \mvec$ means that $Q$ is absolutely continuous with
respect to  $\mvec\in\msim$.
We already proved that if $Q$ is absolutely continuous with
respect to $\mvec^*$, then $U(\svec,\mvec^*)$ is linear in
$\svec$. Thus, by Equation~(\ref{eq:appendixuinfb}), $U(\svec)$ is an
infimum of linear functions, which (by a standard result of convex
analysis, see e.g.\ \cite{Rockafellar70}) is concave. A concave and
bounded function with a convex domain must be continuous on the
interior of this domain (see, e.g., \cite{Rockafellar70}). It remains
to show that $U(\svec)$ is continuous at boundary points of
$\ssim$. Showing this is straightforward by taking limits (but
tedious).  We omit the details.

Now for part (c). If $Q$ is not a proper nonlocality proof, then by
definition there exists a $\mvec_0 \in \msim$ such that, for $a,b \in
\{1,2\}$, we have $Q_{ab} = P_{ab; \mvec_0}$ and hence 
$U(\svec,\mvec_0) = 0$ for all $\svec \in \ssim$.

Now suppose $Q$ is a proper nonlocality proof. Let $\svec$ be in the
interior of $\ssim$.  $\infm{\msim} U(\svec,\mvec)$ is achieved for
some $\mvec^*$. Suppose, by means of contradiction, that
$U(\svec,\mvec^*) = 0$. Since $\svec_{ab} > 0$ for $a,b \in \{1,2\}$,
we must have $Q_{ab} = P_{ab; \mvec^*}$ for $a,b \in \{1,2\}$. But
then $Q$ is not a proper nonlocality proof; contradiction.  For part
(d), if $\svec$ is on the boundary of $\ssim$, then for some $a,b$,
$\svec_{ab} = 0$. It then follows from Lemma~\ref{lem:wim} and the
fact that, for all $P$, $D(P \| P) = 0$ that $U(\svec,\mvec^*) = 0$.

\paragraph*{Part 3}
Part (a) The condition that $Q_{ab}$ is not a point mass for some
$a,b$, implies that all $\mvec^*$ that achieve the infimum must have
$\mvec^*_{\mvala_1 \mvala_2 \mvalb_1 \mvalb_2} < 1$ for all
$\mvala_1,\mvala_2, \mvalb_1, \mvalb_2$,
(otherwise $U(\svec,\mvec^*) = \infty$, which is a contradiction).  
Thus, we assume that $\mvec^* \in \msim_0$, with
$\msim_0$ the set of $\mvec$s that satisfy this ``$< 1$'' restriction.

For $\rho \in [0, \infty)^{16}$, let
\begin{IEEEeqnarray}{rCl}
\overline{\rho}_{\mvala_1 \mvala_2 \mvalb_1 \mvalb_2}(\rho)
& :=&
\frac{\rho_{\mvala_1 \mvala_2 \mvalb_1 \mvalb_2} }{\sum_{\mvala'_1,
\mvala'_2, \mvalb'_1, \mvalb'_2 \in \{\true,\false\}} \rho_{{\mvala'_1
\mvala'_2 \mvalb'_1 \mvalb'_2}}}.
\end{IEEEeqnarray}
In this way, each vector $\rho$ with at least one non-zero component
uniquely defines a local theory $\overline{\rho} \in \msim$, and
\begin{IEEEeqnarray}{rCl}
\biggl\{ \overline{\rho} : \rho \in [0,\infty)^{16} \text{ and } 
\sum_{\mvala_1,\mvala_2, \mvalb_1, \mvalb_2 \in \{\true,\false\}} 
\rho_{\mvala_1 \mvala_2 \mvalb_1 \mvalb_2} > 0 \biggr\} 
&=& 
\msim_0.
\end{IEEEeqnarray}
Let $\rho^*$ be such that $\overline{\rho}^*$ achieves the infimum in
Equation~(\ref{eq:appendixuinf}). Then $Q$ is absolutely continuous 
 with respect to $\overline{\rho}^*$.  One can now show that for each
$(\mvala_1,\mvala_2,\mvalb_1,\mvalb_2) \in \{ \true,\false \}^4$, the
partial derivative
$\partial \; U(\svec,\overline{\rho}) /\partial \rho_{\mvala_1,\mvala_2,\mvalb_1,\mvalb_2}$
evaluated at $\rho=\rho^*$ exists (even if
$\rho^*_{\mvala_1,\mvala_2,\mvalb_1,\mvalb_2} = 0$).  Since
$\overline{\rho}^*$ achieves the infimum, it follows that, for each
$(\mvala_1,\mvala_2,\mvalb_1,\mvalb_2) \in \{ \true,\false \}^4$, we must
have that 
$(\partial /\partial\rho_{\mvala_1,\mvala_2,\mvalb_1,\mvalb_2}) U(\svec,\overline{\rho})$
evaluated at $\rho^*$ is not less than $0$, or, equivalently,
\begin{IEEEeqnarray}{rCl}\label{eq:nounou}
\biggl\{ \frac{\partial U(\svec,\overline{\rho})}{\partial
\rho_{\mvala_1,\mvala_2,\mvalb_1,\mvalb_2}} 
\biggr\}_{\rho = \rho^*} \cdot
\left(\sum_{\mvala_1,\mvala_2,\mvalb_1,\mvalb_2}
\rho_{\mvala_1,\mvala_2,\mvalb_1,\mvalb_2}\right) &\geq& 0
\end{IEEEeqnarray}
with equality if $\rho^*_{\mvala_1,\mvala_2,\mvalb_1,\mvalb_2} > 0$.
Straightforward evaluation of Equation~(\ref{eq:nounou}) gives
Equations~(\ref{eq:appendixem}) and (\ref{eq:appendixemb}).  This shows
that each $\mvec^*$ achieving Equation~(\ref{eq:appendixuinf}) satisfies
Equations~(\ref{eq:appendixem}) and (\ref{eq:appendixemb}).  On the other
hand, each $\mvec^*$ corresponding to a $\rho^*$ with
$\overline{\rho}^* = \mvec^*$ such that Equation~(\ref{eq:nounou}) holds
for each $(\mvala_1,\mvala_2,\mvalb_1,\mvalb_2) \in \{ \true,\false \}^4$
must achieve a local minimum of $U(\svec,\mvec)$ (viewed as a function
of $\mvec$), Since $U(\svec,\mvec)$ is convex, $\mvec^*$ must achieve
the infimum of Equation~(\ref{eq:appendixuinf}).

For part (b), suppose, by way of contradiction, that for at least one
$(\mvala_1, \mvalb_1) \in \{ \true,\false \}^2$, $a_0, b_0 \in \{1,2\}$
with $Q_{a_0 b_0}(\mvala_1, \mvalb_1) > 0$, we have 
$P_{a_0 b_0 ;\mvec^*}(\mvala_1, \mvalb_1) \neq P_{a_0 b_0 ; \mvec^\circ}
(\mvala_1,\mvalb_1)$.  For each $\mvala, \mvalb \in \{\true,\false\}, 
a,b \in\{1,2\}$, we can write
\begin{IEEEeqnarray}{rCl}\label{eq:friday}
P_{ab ; \mvec^*}(\mvala, \mvalb ) 
& = & 
\mvec^*_{k_1}+\mvec^*_{k_2} +
\mvec^*_{k_3} +\mvec^*_{k_4}, \nonumber \\ P_{ab ; \mvec^\circ}(\mvala,
\mvalb) & = & \mvec^\circ_{k_1}+\mvec^\circ_{k_2} + \mvec^\circ_{k_3}
+ \mvec^\circ_{k_4},
\end{IEEEeqnarray}
for some $k_1,\dots, k_4$ depending on $\mvala,\mvalb,a,b$.  Here each
$k_j$ is of the form $\mvala_1 \mvala_2 \mvalb_1 \mvalb_2$ with
$\mvala_i, \mvalb_i \in \{ \true,\false \}$. Now consider $\mvec^+ :=
(1/2) \mvec^* + (1/2) \mvec^\circ$. Clearly $\mvec^+ \in \msim$.  By
Jensen's inequality applied to the logarithm and using
Equation~(\ref{eq:friday}), we have for $a,b \in \{1,2\}$:
$Q_{ab}(\mvala, \mvalb) [\log Q_{ab}(\mvala, \mvalb) - \log P_{ab ;
\mvec^+}(\mvala, \mvalb )] \leq Q_{ab}(\mvala, \mvalb) [\log
Q_{ab}(\mvala, \mvalb) - \frac{1}{2} \log P_{ab ; \mvec^*}(\mvala,
\mvalb) - \frac{1}{2} \log P_{ab ; \mvec^\circ}(\mvala, \mvalb)]$,
where the inequality is strict if 
$\mvala = \mvala_1, \mvalb = \mvalb_1, a = a_0$ 
and $b= b_0$. But then for $a,b \in \{1,2\}$:
$U((a,b),\mvec^+) \leq \frac{1}{2} U((a,b),\mvec^*) + \frac{1}{2}
U((a,b),\mvec^\circ)$, which for $(a,b) = (a_0,b_0)$ must be strict. 
By assumption, $\svec_{a_0 b_0} > 0$. But that implies $
U(\svec,\mvec^+) < U(\svec,\mvec^*) = \infm{\msim}
U(\svec,\mvec)$ and we have arrived at the desired contradiction.
\end{proof}

\subsection{Proofs of Game-Theoretic Theorems}
\subsubsection{Game-Theoretic Preliminaries}
Proposition~\ref{prop:game} gives a few standard game-theoretic
results (partially copied from \cite{Ferguson67}).  We will use these
results at several stages in later proofs.
\begin{prop}
\label{prop:game}
Let $A$ and $B$ be arbitrary sets and let $L: A \times B \rightarrow
\R \cup \{ - \infty, \infty \}$ be an arbitrary function on $A \times
B$.  We have
\begin{enumerate}
\item $\inf_{\beta \in B} \sup_{\alpha \in A} L(\alpha,\beta) \geq
\sup_{\alpha \in A} \inf_{\beta \in B} L(\alpha,\beta)$.
\item Suppose the following conditions hold:
\begin{enumerate}
\item The game $(A,B,L)$ has a value $V \in \R \cup \{ - \infty,
  \infty \}$, that is
$\inf_{\beta \in B} \sup_{\alpha \in A} L(\alpha,\beta) = V = \sup_{\alpha
\in A} \inf_{\beta \in B} L(\alpha,\beta)$.
\item There exists $\alpha^*$ that achieves $\sup_{\alpha \in A}
\inf_{\beta \in B} L(\alpha,\beta)$.
\item There exists $\beta^*$ that achieves $\inf_{\beta \in B}
\sup_{\alpha \in A} L(\alpha,\beta)$.
\end{enumerate}
Then 
$(\alpha^*,\beta^*)$ is a saddle point and $L(\alpha^*,\beta^*) = V$.
\item Suppose there exists a pair $(\alpha^*,\beta^*)$ such that
\begin{enumerate}
\item $\beta^*$ achieves $\inf_{\beta \in B} L(\alpha^*,\beta)$ and
\item $\beta^*$ is an \emph{equalizer strategy}, that is, there exists
  a $K \in \R \cup \{ - \infty, \infty \}$ with for all $\alpha \in
  A$, $L(\alpha,\beta^*) = K$.
\end{enumerate}
Then the game $(A,B,L)$ has value $K$, i.e.\ $ \inf_{\beta \in B}
\sup_{\alpha \in A} L(\alpha,\beta) = \sup_{\alpha \in A} \inf_{\beta
\in B} L(\alpha,\beta)= K$, and $(\alpha^*,\beta^*)$ is a
saddle point.
\end{enumerate}
\end{prop}
\begin{proof}
(1) For all $\alpha' \in A$,
\begin{IEEEeqnarray}{rCl}
\inf_{\beta \in B} \sup_{\alpha \in A} L(\alpha,\beta)& \geq&
\inf_{\beta \in B} L(\alpha',\beta).
\end{IEEEeqnarray}
Therefore, $ \inf_{\beta \in B} \sup_{\alpha \in A} L(\alpha,\beta)
\geq \sup_{\alpha' \in A} \inf_{\beta \in B} L(\alpha',\beta)$.

(2) Under our assumptions,
\begin{IEEEeqnarray}{rCl}
L(\alpha^*,\beta^*) &\leq& \sup_{\alpha \in A} L(\alpha,\beta^*) \\ 
&=&
\inf_{\beta \in B} \sup_{\alpha \in A} L(\alpha,\beta)\\ & =& V \\ &=&
\sup_{\alpha \in A} \inf_{\beta \in B} L(\alpha,\beta)\\ & = &
\inf_{\beta \in B} L(\alpha^*,\beta) \leq L(\alpha^*,\beta^*),
\end{IEEEeqnarray}
so $L(\alpha^*,\beta^*) = V = \inf_{\beta \in B} L(\alpha^*,\beta)$
and $L(\alpha^*,\beta^*) = V =\sup_{\alpha \in A} L(\alpha,\beta^*)$.

(3) To show that the game has a value, by (1) it is sufficient to show
that
$\inf_{\beta \in B} \sup_{\alpha \in A} L(\alpha,\beta) 
\leq
\sup_{\alpha \in A} \inf_{\beta \in B} L(\alpha,\beta)$.
But this is indeed the case:
\begin{IEEEeqnarray}{rCcCcCcCcCl}
\inf_{\beta \in B}\sup_{\alpha \in A}L(\alpha,\beta) 
&\leq&
\sup_{\alpha \in A}L(\alpha,\beta^*)  
&=& 
L(\alpha^*,\beta^*) 
&=& 
K 
&=& 
\inf_{\beta \in B} L(\alpha^*,\beta) 
&\leq&
\sup_{\alpha \in A}\inf_{\beta \in B} L(\alpha,\beta),
\end{IEEEeqnarray}
where the first equalities follow because $\beta^*$ is an equalizer
strategy. Thus, the game has a value equal to $K$. Since 
$\sup_{\alpha}{L(\alpha,\beta^*)} = K$, $\beta^*$ achieves 
$\inf_{\beta}{\sup_{\alpha}{L(\alpha,\beta)}}$. 
Since $\inf_{\beta}{L(\alpha^*,\beta)} = K$, 
$\alpha^*$ achieves $\sup_{\alpha}{\inf_{\beta} L(\alpha,\beta)}$.  
Therefore, $(\alpha^*,\beta^*)$ is a saddle point.
\end{proof}

\subsubsection[Proof of Theorem~\ref{thm:spcor}]{Proof of Theorem~\ref{thm:spcor}, 
the Saddle Point Theorem for Correlated Settings and Generalized
  Nonlocality Proofs}\label{app:prf:spcor}
\setcounter{theoremtracker}{\value{theorem}}
\setcounter{theorem}{\value{cr:spcor}}
 \addtocounter{theorem}{-1}
\begin{theorem}
For every (generalized) nonlocality proof, the correlated game
$(\msim,\ssim,U)$ corresponding to it has a finite value, i.e.\ there
exists $0 \leq V < \infty$ with
\begin{IEEEeqnarray*}{rCl}
V
&=& \infm{\msim} \sups{\ssim} U(\svec,\mvec)\\ 
&=& 
\sups{\ssim}\infm{\msim} U(\svec,\mvec). 
\end{IEEEeqnarray*}
The infimum on the first line is achieved for some 
$\mvec^* \in \msim$; the supremum on the second line is achieved 
for some $\svec^*$ in $\ssim$, so
that $(\mvec^*, \svec^*)$ is a saddle point.
\end{theorem}
\setcounter{theorem}{\value{theoremtracker}} 
\begin{proof}
We use the following well-known minimax theorem due to Ferguson. The
form in which we state it is a straightforward combination of
Ferguson's \cite{Ferguson67} Theorem~1, page~78 and Theorem~2.1,
page~85, specialized to the Euclidean topology.
\begin{theorem}[Ferguson 1967]
\label{thm:ferguson}
Let $(A,B,L)$ be a statistical game where $A$ is a finite set, $B$ is
a convex compact subset of $\R^{k}$ for some $k > 0$ and $L$ is such
that for all $\alpha \in A$,
\begin{enumerate}
\item $L(\alpha,\beta)$ is a convex function of $\beta \in B$.
\item $L(\alpha,\beta)$ is lower semicontinuous in $\beta \in B$.
\end{enumerate}
Let $\set{A}$ be the set of distributions on $A$ and define, for $P
\in \set{A}$, $L(P,\beta) = \Exp_P L(\alpha,\beta) = \sum_{\alpha \in A}
P_{\alpha} L(\alpha,\beta)$.  Then the game $(\set{A},B,L)$ has a
value $V$, i.e.\
\begin{IEEEeqnarray}{rCl}
\sup_{P \in \set{A}} \inf_{\beta \in B} L(P,\beta)
& =& 
\inf_{\beta\in B} \sup_{P \in \set{A}} L(P,\beta),
\end{IEEEeqnarray}
and a minimax $\beta^* \in B$ achieving $\inf_{\beta \in B}
\sup_{{\mathbb \alpha} \in \set{A}} L(\alpha,\beta)$ exists.
\end{theorem}
By Theorem~\ref{thm:supinf}, part (1), 
$U(\svec,\mvec) = D(Q_{\svec} \| P_{\svec; \mvec} )$ is lower
semicontinuous in $\mvec$ for all $\svec \in \ssim$. 
Let us now focus on the case of a $2\times
2\times 2$ game. We can apply Theorem~\ref{thm:ferguson} with 
$A = \{11,12,21,22 \}$, $\set{A} = \ssim$ and $B = \msim$. It follows that
the game $(\ssim,\msim,U)$ has a value $V$, and $\infm{\msim}
\sups{\ssim} U(\svec,\mvec)= V$ is achieved for some $\mvec^* \in
\msim$.  By Theorem~\ref{thm:supinf}, part (2), $0 \leq V < \infty$,
and, since $U(\svec)$ is continuous in $\svec$, there exists some
$\svec^*$ achieving $\sups{\ssim} \infm{\msim} U(\svec,\mvec)$.

The proof for generalized nonlocality proofs is completely analogous;
we omit details.
\end{proof}

\subsubsection[Proof of Theorem~\ref{thm:2x2}]{Proof of 
Theorem~\ref{thm:2x2}, Saddle Points and Equalizer Strategies for 
${2\times 2\times 2}$ Nonlocality Proofs}\label{app:prf:2x2}
\setcounter{theoremtracker}{\value{theorem}}
\setcounter{theorem}{\value{cr:2x2}}
 \addtocounter{theorem}{-1}
\begin{theorem}
Fix any proper nonlocality proof based on 2 parties with 2
measurement settings per party and let $(\ssim,\msim,U)$ and
$(\ssimuc,\msim,U)$ be the corresponding correlated and uncorrelated
games, then:
\begin{enumerate}
\item The correlated game has a saddle point with value $V >0$. Moreover,
\begin{IEEEeqnarray}{rCcCl}\label{eq:appendixucltcor}
\sups{\ssimuc} \infm{\msim} U(\svec,\mvec) 
&\leq& 
\sups{\ssim} \infm{\msim} U(\svec,\mvec)
& =& 
V \\ \label{eq:appendixuciscor}
\infm{\msim} \sups{\ssimuc} U(\svec,\mvec) 
&=& 
\infm{\msim} \sups{\ssim} U(\svec,\mvec) 
&=&
V.
\end{IEEEeqnarray}

\item Let
\begin{IEEEeqnarray}{rCl}
\msim^* 
&:=& 
\{ \mvec : \mvec \text{\ achieves\ } \infm{\msim}\sups{\ssim} U(\svec,\mvec) \},\\
\msimuc^* 
&:=& 
\{ \mvec : \mvec \text{\ achieves\ } \infm{\msim} \sups{\ssimuc} U(\svec,\mvec) \},
\end{IEEEeqnarray}
then
\begin{enumerate}
\item $\msim^*$ is non-empty.
\item $\msim^* = \msimuc^*$.
\item All $\mvec^* \in \msim^*$ are `equalizer strategies',
i.e.\  for all $\svec \in \ssim, U(\svec,\mvec^*) = V$.
\end{enumerate}
\item The uncorrelated game has a saddle point if and only if there
exists $(\mvec^*,\svec^*)$, with $\svec^* \in \ssimuc$, such that
\begin{enumerate}
\item $\mvec^*$ achieves $\infm{\msim} U(\svec^*,\mvec)$.
\item $\mvec^*$ is an equalizer strategy.
\end{enumerate}
If such $(\svec^*,\mvec^*)$ exists, it is a saddle point.
\end{enumerate}
\end{theorem}
\setcounter{theorem}{\value{theoremtracker}} 
\begin{proof}
The correlated game has a value $V$ by Theorem~\ref{thm:spcor} and 
$V > 0$ by Theorem~\ref{thm:supinf}.  
Inequality~\ref{eq:appendixucltcor} is immediate.

Let $U((a,b),\mvec)$ be defined as in the proof of
Theorem~\ref{thm:supinf}, Equation~(\ref{eq:rewrite}).  To prove
Equation~(\ref{eq:appendixuciscor}), note that for every $\mvec \in
\msim$,
\begin{IEEEeqnarray}{rCl}
\sups{\ssimuc} U(\svec,\mvec) 
&=& 
\sups{\ssim} U(\svec,\mvec)\\ 
&=&
 \max_{a,b \in \{1,2\}} U((a,b),\mvec).
\end{IEEEeqnarray}
Thus, Equation~(\ref{eq:appendixuciscor}) and part 2(b) of the theorem
follow.  Part 2(a) is immediate from Theorem~\ref{thm:spcor}. To
prove part 2(c), suppose, by way of contradiction, that there exists
a $\mvec^* \in \msim^*$ that is not an equalizer strategy.
Then the set
$\{ (a,b) \mid U((a,b),\mvec^*) = \max_{a,b \in \{1,2\}}
U((a,b),\mvec^*) \}$
has less than four elements.  By Theorem~\ref{thm:spcor}, there exists
a $\svec^* \in \ssim$ such that $(\svec^*,\mvec^*)$ is a saddle point
in the correlated game.  Since $\svec^* \in \ssim$ achieves 
$\sups{\ssim} U(\svec,\mvec^*)$, it follows that for some 
$a_0,b_0 \in\{1,2\}$, $\svec^*_{a_0 b_0 } = 0$. 
But then $\svec^*$ lies on the boundary of $\ssim$. 
By Theorem~\ref{thm:supinf}, part 2(d), this is
impossible, and we have arrived at the desired contradiction.

It remains to prove part (3). Part (3), `if' follows directly from
Proposition~\ref{prop:game}.  To prove part (3), `only if', suppose
the uncorrelated game has saddle point $(\svec^*, \mvec^*)$. It is
clear that $\mvec^*$ achieves $\infm{\msim} U(\svec^*,\mvec)$. We have
already shown above that $\mvec^*$ is an equalizer strategy.
\end{proof}
\end{document}